\documentclass[a4paper,11pt]{article}
\usepackage[english]{babel}
\usepackage{jheppub}
\pdfoutput=1
\usepackage[table]{xcolor}
\usepackage[T1]{fontenc}
\usepackage{bbold}
\usepackage{float}
\usepackage{amsmath}
\usepackage{empheq}
\usepackage{booktabs}
\usepackage{color}
\usepackage[utf8]{inputenc}
\usepackage{xspace}
\usepackage{scalerel}
\usepackage[most]{tcolorbox}
\usepackage{multirow}
\usepackage{scalefnt}
\usepackage{bold-extra}
\usepackage[shortlabels]{enumitem}
\usepackage[tikz]{bclogo}
\usepackage{subcaption}
\usepackage{cancel}
\usepackage{verbatim}
\usepackage{slashed}
\usetikzlibrary{arrows,shapes}
\usepackage{afterpage}
\usepackage{graphicx,grffile}
\usepackage{xspace}
\usepackage[normalem]{ulem}
\usepackage{makecell}
\usepackage{tabularx,colortbl}
\usepackage{bm}
\usepackage[export]{adjustbox}

\newcommand{\sig}{\sigma}
\newcommand{\lzm}{\left(}
\newcommand{\dzm}{\right)}
\newcommand{\lzs}{\left[}
\newcommand{\dzs}{\right]}
\newcommand{\lzv}{\left\{}
\newcommand{\dzv}{\right\}}
\newcommand{\lzu}{\left|}
\newcommand{\dzu}{\right|}
\newcommand{\cL}{\mathcal{L}}

\newcommand{\cW}{{\mathcal W}}
\newcommand{\cB}{{\mathcal B}}

\newcommand{\cQ}{{\mathcal Q}}

\newcommand{\cC}{{\mathcal C}}
\newcommand{\cS}{{\mathcal S}}
\newcommand{\cG}{{\mathcal G}}
\newcommand{\cU}{{\mathcal U}}
\newcommand{\cX}{{\mathcal X}}
\newcommand{\cY}{{\mathcal Y}}
\newcommand{\cH}{{\mathcal H}}
\newcommand{\cO}{{\mathcal O}}
\newcommand{\re}{{\mathrm{Re}} \,}
\newcommand{\im}{{\mathrm{Im}} \,}
\newcommand{\hermc}{\text{h.c.}}
\newcommand{\ud}[2]{\phantom{}^{#1}\phantom{}_{#2}}
 
\usepackage{xcolor}

\newtcolorbox{empheqboxed}{colback=white!35,
 colframe=black,
 width=\textwidth,
 sharpish corners,
 top=-2mm, 
 bottom=0pt
}

\title{Leading directions in the SMEFT}

\author[]{Admir Greljo,}
\author[]{Ajdin Palavri\' c}

\emailAdd{admir.greljo@unibas.ch}
\emailAdd{ajdin.palavric@unibas.ch}

\affiliation[]{Department of Physics, University of Basel, Klingelbergstrasse 82,  CH-4056 Basel, Switzerland}

\abstract{Short-distance new physics at (or slightly) above the TeV scale should not excessively violate the approximate flavor symmetries of the SM in order to comply with stringent constraints from flavor-changing neutral currents. In this respect, flavor symmetries provide an effective organizing principle for the vast parameter space of the SMEFT. In this work, we classify all possible irreducible representations under $U(3)^5$ flavor symmetry of new heavy spin-0, 1/2, and 1 fields which integrate out to dimension-6 operators at the tree level. For a general perturbative UV model, the resulting flavor-symmetric interactions are very restrictive and, in most cases, predict a single Hermitian operator with a definite sign. These \textit{leading directions} in the SMEFT space deserve particular attention. We derive an extensive set of present experimental constraints by utilizing the existing global SMEFT fits, which incorporate data from top quark, Higgs boson, and electroweak measurements, along with constraints on dilepton and 4-lepton contact interactions. The derived set of bounds comprehensively summarises the present knowledge from indirect searches of flavor-blind new physics mediators.}


\begin{document}

\maketitle

\newpage

\section{Introduction}
\label{sec:intro}

The Standard Model (SM) of particle physics is a remarkably successful theory that withstood decades of experimental scrutiny. However, the SM is not the ultimate theory; there must be new physics (NP) beyond the SM that resolves some of its outstanding issues. The electroweak hierarchy problem, for example, suggests that the theory breaks not far above the electroweak scale, motivating the exploration of the TeV scale, the next short-distance frontier. Despite high expectations, the Large Hadron Collider (LHC) has not yet discovered NP, setting stringent limits on many favored models addressing, e.g., the hierarchy problem. While it is still premature to draw conclusions, the negative experimental results have already shifted the focus toward bottom-up model building. As the LHC is heading towards the precision era, we must explore every possibility without theoretical biases.

A robust approach to broadly explore short-distance NP effects at colliders is provided by the Standard Model Effective Field Theory (SMEFT)~\cite{Buchmuller:1985jz, Grzadkowski:2010es, Brivio:2017vri, Isidori:2023pyp, Giudice:2007fh, Henning:2014wua}. Higher-dimensional operators are added to the SM Lagrangian, built from the SM fields respecting the SM spacetime and gauge symmetries, suppressed by the cutoff scale $\Lambda$ that characterizes the NP scale. These operators encode the effects of NP at energy scales below $\Lambda$. In a bottom-up approach, the SMEFT allows for a model-independent study of NP effects, which can then constrain or motivate particular ultraviolet (UV) models.

However, the SMEFT comes with its own challenges; there is a vast number of theory parameters, a price to pay in order to be very general. The importance of a particular higher-dimensional operator depends on its canonical dimension, which determines the suppression due to the cutoff scale $\Lambda$. The leading baryon and lepton number--conserving effects are due to operators of dimension-6. There are (59) 2499 such operators for one (three) generation(s) of SM fermions~\cite{Alonso:2013hga}. Managing this complexity, which arises due to the flavor degree of freedom, becomes crucial in order to extract meaningful information~\cite{Faroughy:2020ina,Greljo:2022cah}.

There is more to consider when it comes to new physics at the TeV scale. Notably, low-energy experiments have already conducted thorough tests of the (approximate) accidental symmetries of the Standard Model, which predict a suppression in flavor-changing neutral currents and CP violation – a consequence of the SM flavor structure. These indirect tests impose crucial constraints on the higher-dimensional operators of the SMEFT that violate symmetries; see, e.g.,~\cite{Isidori:2010kg, EuropeanStrategyforParticlePhysicsPreparatoryGroup:2019qin, Aebischer:2020dsw, Silvestrini:2018dos, Pruna:2014asa, Feruglio:2015yua}. In other words, TeV-scale new physics with unsuppressed flavor violation is already ruled out, a fact known as the new physics flavor puzzle.

Here we build upon our previous work~\cite{Greljo:2022cah}, in which we investigated the role of flavor symmetries and their breaking patterns in the SMEFT, with a focus on utilizing them to organize the vast parameter space of the theory. In Ref.~\cite{Greljo:2022cah}, we examined a variety of flavor structures, including Minimal Flavor Violation (MFV)~\cite{DAmbrosio:2002vsn} and the $U(2)^5$ flavor symmetry~\cite{Barbieri:2011ci, Kagan:2009bn}, while also exploring a wide array of other possibilities in both the quark and lepton sectors. We identified several symmetry hypotheses that permit new physics at the TeV scale in accordance with indirect tests conducted in flavor physics. As a general guideline, the flavor-symmetric operators primarily influence the flavor-conserving processes, such as those in the top, Higgs, and electroweak sectors. In contrast, the spurion-suppressed operators play a role in flavor-changing transitions. In this respect, approximate flavor symmetries provide an organizing principle simplifying phenomenological studies. For example, the flavor-symmetric basis of the $U(3)^5$ case (leading terms in the MFV) contains only 41 CP even and 6 CP odd operators (see Appendix~\ref{app:basis} for explicit construction). 

In this paper, we take advantage of the aforementioned simplification and move one step beyond the EFT, classifying all perturbative ultraviolet (UV) completions that integrate out to dimension-6 flavor-symmetric SMEFT at the tree level. As a representative example, we focus on the $U(3)^5$ symmetry, deferring the exploration of alternative cases to future work. The resulting set of new fields and interactions extending the SM not only justify the current global fits, which are performed in isolation from flavor data but also hold promise for direct searches at the LHC and future colliders. These hypothetical new particles possess inherent protection against flavor constraints, allowing them to easily have masses in the TeV range with order one interactions with the SM states. In a sense, this approach represents a convenient way to restructure new physics searches at the LHC using a purely bottom-up methodology.

In the seminal work of Ref.~\cite{deBlas:2017xtg}, the authors provided a comprehensive classification of new spin 0, 1/2, and 1 fields that match at tree-level to the dimension-6 SMEFT operators, calculating the explicit contributions to the Wilson coefficients (WCs) -- a crucial endeavor known as the tree-level UV/IR dictionary (see also~\cite{Arzt:1994gp, Einhorn:2013kja, Li:2022abx}). Building on these findings, our objective in this work is to classify all possible irreducible representations (irreps) under the $U(3)^5$ flavor symmetry of the fields in the aforementioned UV/IR dictionary. Imposing the flavor symmetry will tremendously restrict the allowed interactions with the SM fields. By definition, such scenarios will integrate out to the $U(3)^5$ flavor-symmetric SMEFT basis.

In the context of a general perturbative UV model, the resulting flavor-symmetric interactions are highly restrictive. For a given flavor irrep, they predict, in most instances, a single operator, i.e., a linear combination of operators in a chosen basis. These \textit{leading directions} in the SMEFT space are especially intriguing as they represent single-operator scenarios motivated from a UV perspective, and are crucial when experimental collaborations aim to fit a limited set of operators. Currently, fits are performed using an arbitrary EFT basis, such as the Warsaw basis, where one-dimensional fits are predominantly inadequate for interpretation within a UV model. Considering the SMEFT space as a vector space, the leading directions can be viewed as unique vectors with well-defined directions while their lengths are controlled by the heavy mediator mass and couplings.

Upon establishing the flavor-symmetric UV/IR dictionary, we proceed in this work to derive a comprehensive set of experimental constraints on each irreducible scenario. These constraints are obtained from the available global SMEFT fits of the Higgs and electroweak sectors, top quark, dilepton, and 4-lepton processes. As a result, we effectively summarize existing limits on flavor-blind new physics mediators derived from indirect (EFT) searches.

The structure of this paper is as follows: In Section~\ref{sec:theory}, we delve into the classification of new tree-level mediators under the $U(3)^5$ flavor symmetry, generating dimension-6 operators in the SMEFT through flavor-symmetric interactions. In Section~\ref{sec:pheno}, we explore the phenomenology associated with each identified case and derive the current experimental limits from the available EFT analyses. We offer our concluding remarks in Section~\ref{sec:conc}, while additional technical details can be found in the Appendices.

\section{The tree-level UV/IR dictionary: Flavor representations}
\label{sec:theory}

The SM fermion representations under $(SU(3)_C, SU(2)_L)_{U(1)_Y}$ gauge group are $q \sim (\bm3,\bm2)_{\frac{1}{6}}$, $\ell \sim (\bm1,\bm2)_{-\frac{1}{2}}$, $u \sim (\bm3,\bm1)_{\frac{2}{3}}$, $d \sim (\bm3,\bm1)_{-\frac{1}{3}}$, and $e\sim (\bm1,\bm1)_{-1}$. The $SU(2)_L$ doublets (singlets) are left (right) Weyl fermions. Every gauge representation comes in three identical copies (flavors). As a result, the SM Lagrangian without Yukawa interactions enjoys a large global flavor symmetry,
\begin{equation}
U(3)^5 \equiv U(3)_q \times U(3)_\ell \times U(3)_u \times U(3)_d \times U(3)_e ~,
\end{equation}
The corresponding fields transform as
\begin{equation}
    q\sim\bm3_q~, \quad \ell\sim\bm3_\ell~, \quad u\sim\bm3_u~, \quad d\sim\bm3_d~, \quad e\sim\bm3_e~,
\end{equation}
where we use a shorthanded notation: $\bm3_f$ is a $(\bm3)_{1}$ under $(SU(3)_f)_{U(1)_f}$ and a singlet under other group factors. The Yukawa couplings $y_{u,d,e}$ act as spurious fields that formally transform in a nontrivial representation of the flavor group with its background value inducing flavor symmetry breaking,
\begin{equation}
    y_u \sim  (\bm3_u,\bar{\bm3}_q)~, \quad y_d \sim  (\bm3_d,\bar{\bm3}_q)~, \quad y_e \sim  (\bm3_e,\bar{\bm3}_\ell)~.
\end{equation}
With this, the prescription for building an MFV structure~\cite{DAmbrosio:2002vsn} of the SMEFT is straightforward~\cite{Faroughy:2020ina, Greljo:2022cah}: postulate the $U(3)^5$ symmetry in the higher-dimensional operators build from the SM fields and spurions. The expansion is organized based on the number of spurion insertions, rendering the series valuable due to the smallness of Yukawa matrix entries.\footnote{The top Yukawa stands out, with $(y_u)_{33} \approx 1$. For this case, one can either assume an additional expansion parameter, $\epsilon_u \times y_u$, where $\epsilon_u \ll 1$ (linear MFV), or consider resummation to all orders, resulting in a $U(2)$ subgroup (general MFV)~\cite{Kagan:2009bn}. We will take the first approach.} The construction of the MFV SMEFT basis order by order in the spurion expansion is detailed in Sections 2.4 and 3.7 in Ref.~\cite{Greljo:2022cah}. In Appendix~\ref{app:basis}, we summarise the main findings up to the linear order $\mathcal{O}(y_{u,d,e})$. The focus of this work is on the \textit{flavor-symmetric} basis (without spurion insertions) shown in the first two tables of Appendix~\ref{app:basis}.

Moving forward, we presume that the subsequent layer beyond the SM is a weakly coupled, perturbative quantum field theory encompassing arbitrary new spin-0, 1/2, and 1 fields. With their masses $M_X \gg v_{\text{EW}}$, we first concentrate on their leading, renormalizable interactions (up to canonical dimension 4). Our goal is to identify and study all possible ways to generate the dimension-6 operator in the $U(3)^5$ flavor-symmetric basis when matching the UV theory on the SMEFT at the tree level (leading order).

We build on Ref.~\cite{deBlas:2017xtg}, which derived a comprehensive tree-level UV/IR dictionary. The BSM fields, which integrate out at the tree level to dimension-6 SMEFT operators, are organized in representations of the SM gauge group. Their Lagrangians ($\mathcal{L}_{{\rm BSM}}$) can be found in Appendix A of~\cite{deBlas:2017xtg}. Field rotations and rescalings are performed to guarantee diagonal and canonical kinetic terms, along with diagonal mass terms in the symmetric electroweak phase. In Appendix D of~\cite{deBlas:2017xtg}, the authors report a complete set of tree-level matching contributions to the dimension-6 SMEFT WCs in terms of the new masses and couplings present in $\mathcal{L}_{{\rm BSM}}$.

Finally, what UV physics integrates out to the MFV SMEFT basis at dimension 6? Imposing a $U(3)^5$ global flavor symmetry on $\mathcal{L}_{{\rm BSM}}$ will: 
\begin{itemize}
\item organize the new fields in irreps of the flavor group,
\item impose mass degeneracy of the states within nontrivial flavor irreps, and,
\item define the form of the flavor coupling tensors and drastically reduce the number of theory parameters.
\end{itemize}
Focusing only on new physics interactions with the canonical dimension $\le 4$, only up to two SM fermions can be present.\footnote{In this study, we disregard matching contributions arising from terms $\mathcal{L} \supset V_\mu D^\mu S$, where $V_\mu$ represents a vector, and $S$ denotes a scalar. These terms are absent in the unitary gauge of a renormalizable UV completion and do not contribute at the tree level~\cite{deBlas:2017xtg}.} Therefore, the BSM fields can either transform as trivial (singlets) or nontrivial (triplets, sextets, octets) representations under the corresponding $U(3)$ flavor factors. In Appendix~\ref{app:sm1}, we write down all options for a single field extension of the SM. For each Lorentz and gauge representation, we specify the Lagrangian, identify possible flavor representations and their coupling tensors, and match every scenario to the SMEFT. The main results of Appendix~\ref{app:sm1} are summarised in Tables~\ref{tab:sm1-spin0}, \ref{tab:sm1-spin12}, \ref{tab:sm1-spin1}, \ref{tab:singlets}, and \ref{tab:singlets-except}.

Each nontrivial flavor irrep generates a single dim-6 SMEFT operator at the tree level. The predicted operators are Hermitian, and the related Wilson coefficients are real with a well-defined sign. Since $\mathcal{L}_{{\rm BSM}}$ is $U(3)^5$-invariant, the derived operators constitute linear combinations of those in the $U(3)^5$ flavor-universal basis found in Appendix~\ref{app:basis} (initial two tables).\footnote{One can also get operators in the flavor-universal + $\mathcal{O}(y_{u,d,e})$ basis due to the SM Yukawa interactions in the matching even though $\mathcal{L}_{{\rm BSM}}$ contains no Yukawa spurions.} Overall, there are 25 distinct cases for new scalars (Table~\ref{tab:sm1-spin0}), 14 for fermions (Table~\ref{tab:sm1-spin12}), and 25 for vectors (Table~\ref{tab:sm1-spin1}). We associate the term \textit{leading direction} to such linear combinations of the SMEFT operators, which are multiplied only by a single real parameter with a definite sign. 

Significant simplification transpires, even for trivial flavor irreps, upon enforcing $U(3)^5$ symmetry on $\mathcal{L}_{\text{BSM}}$. Flavor singlets can only be either spin 0 or spin 1. In total, 12 such instances are shown in Tables~\ref{tab:singlets} and \ref{tab:singlets-except}. The former table presents nine straightforward cases, six expressible by a single parameter and three cases comprising a direction plus a free Wilson coefficient for the $\mathcal{O}_\phi$ operator. Remarkably, only three exceptional vector fields necessitate three or more parameters (at most seven) for describing the tree-level matching to dimension-6 SMEFT (Table~\ref{tab:singlets-except}).

In a UV theory featuring multiple new fields (flavor irreps), besides simply aggregating their WCs, nontrivial matching contributions may arise from diagrams involving several BSM fields. All such instances are charted in Appendix~\ref{app:smN}. They involve either two or three new scalars and always match to a single dimension-6 operator at the tree level, $\mathcal{O}_\phi$.

Finally, for completeness, Appendix~\ref{app:nrint} explores dimension-5 interactions of new fields with the SM fields in $\mathcal{L}_{\text{BSM}}$. These nonrenormalizable interactions may emerge due to the existence of a higher scale, ultimately leading to dimension-6 SMEFT. As expected, the presence of a flavor symmetry fosters simplifications in comparison to the general scenario.

\begin{table}[th!]
\centering
\scalebox{1.00}{
\begin{tabular}{ccccc}
\toprule
\textbf{Field}&\textbf{Irrep}&\textbf{Normalization}&\textbf{Operator}
\\
\midrule
$\cS_1\sim(\bm1,\bm1)_1$
&$\bm3_\ell$
&$\phantom{-}|y_{\cS_1}|^2/M_{\cS_1}^2$
&$\cO_{\ell\ell}^D-\cO_{\ell\ell}^E$
\\[6pt]
$\cS_2\sim(\bm1,\bm1)_2$
&$\bar{\bm6}_e$
&$\phantom{-}|y_{\cS_2}|^2/(2M_{\cS_2}^2)$
&$\cO_{ee}$
\\
\midrule
\multirow{3}{*}{\rule{0pt}{5ex}$\varphi\sim(\bm1,\bm2)_{\frac{1}{2}}$}
&$(\bar{\bm3}_e,\bm3_\ell)$
&$-|y_\varphi^e|^2/(2M_\varphi^2)$
&$\cO_{\ell e}$
\\[4pt]
&$(\bar{\bm3}_d,\bm3_q)$
&$-|y_\varphi^d|^2/(6M_\varphi^2)$
&$\cO_{qd}^{(1)}+6\cO_{qd}^{(8)}$
\\[4pt]
&$(\bar{\bm3}_q,\bm3_u)$
&$-|y_\varphi^u|^2/(6M_\varphi^2)$
&$\cO_{qu}^{(1)}+6\cO_{qu}^{(8)}$
\\
\midrule
$\Xi_1\sim(\bm1,\bm3)_1$&$\bar{\bm6}_\ell$
&$\phantom{-}\lzu y_{\Xi_1} \dzu^2/(2M_{\Xi_1}^2)$
&$\cO_{\ell\ell}^D+\cO_{\ell\ell}^E$
\\
\midrule
\multirow{4}{*}{\rule{0pt}{6.7ex}$\omega_1\sim(\bm3,\bm1)_{-\frac{1}{3}}$}
&$(\bm3_q,\bm3_\ell)$
&$\phantom{-}|y_{\omega_1}^{q\ell}|^2/(4M_{\omega_1}^2)$
&$\cO^{(1)}_{\ell q}-\cO^{(3)}_{\ell q}$
\\[5pt]
& $(\bm3_e,\bm3_u)$
&$\phantom{-}|y_{\omega_1}^{eu}|^2/(2M_{\omega_1}^2)$
&$\cO_{eu}$
\\[5pt]
& $\bar{\bm6}_q$
&$\phantom{-}|y_{\omega_1}^{qq}|^2/(4M_{\omega_1}^2)$
&$\cO_{qq}^{(1)D}-\cO_{qq}^{(3)D}+\cO_{qq}^{(1)E}-\cO_{qq}^{(3)E}$
\\[5pt]
& $(\bar{\bm3}_d,\bar{\bm3}_u)$
&$\phantom{-}|y_{\omega_1}^{du}|^2/(3M_{\omega_1}^2)$
&$\cO_{ud}^{(1)}-3\cO_{ud}^{(8)}$
\\
\midrule
$\omega_2\sim(\bm3,\bm1)_{\frac{2}{3}}$
&$\bm3_d$
&$\phantom{-}|y_{\omega_2}|^2/M_{\omega_2}^2$
&$\cO_{dd}^D-\cO_{dd}^E$
\\
\midrule
\multirow{2}{*}{\rule{0pt}{4ex}$\omega_4\sim(\bm3,\bm1)_{-\frac{4}{3}}$} 
&$(\bm3_e,\bm3_d)$
&$\phantom{-}|y_{\omega_4}^{ed}|^2/(2M_{\omega_4}^2)$
&$\cO_{ed}$
\\[5pt]
&$\bm3_u$
&$\phantom{-}|y_{\omega_4}^{uu}|^2/M_{\omega_4}^2$
&$\cO_{uu}^D-\cO_{uu}^E$ 
\\
\midrule
$\Pi_1\sim(\bm3,\bm2)_{\frac{1}{6}}$
&$(\bar{\bm3}_\ell,\bm3_d)$
&$-|y_{\Pi_1}|^2/(2M_{\Pi_1}^2)$
&$\cO_{\ell d}$
\\
\midrule
\multirow{2}{*}{\rule{0pt}{3.5ex}$\Pi_7\sim(\bm3,\bm2)_{\frac{7}{6}}$}
&$(\bar{\bm3}_\ell,\bm3_u)$
&$-\lzu y_{\Pi_7}^{\ell u} \dzu^2/(2M_{\Pi_7}^2)$
&$\cO_{\ell u}$
\\[4pt]
&$(\bar{\bm3}_e,\bm3_q)$
&$-|y_{\Pi_7}^{qe}|^2/(2M_{\Pi_7}^2)$
&$\cO_{qe}$ 
\\
\midrule
\multirow{2}{*}{\rule{0pt}{3.5ex}$\zeta\sim(\bm3,\bm3)_{-\frac{1}{3}}$}
&$(\bm3_q,\bm3_\ell)$
&$\phantom{-}|y_{\zeta}^{q\ell}|^2/(4M_\zeta^2)$
&$3\cO^{(1)}_{\ell q}+\cO^{(3)}_{\ell q}$
\\[5pt]
&$\bm3_q$
&$\phantom{-}|y_\zeta^{qq}|^2/(2M_\zeta^2)$
&$3\cO_{qq}^{(1)D}+\cO_{qq}^{(3)D}-3\cO_{qq}^{(1)E}-\cO_{qq}^{(3)E}$
\\
\midrule
\multirow{2}{*}{\rule{0pt}{4ex}
$\Omega_1\sim(\bm6,\bm1)_{\frac{1}{3}}$}
&$(\bm3_u,\bm3_d)$
&$\phantom{-}|y^{ud}_{\Omega_1}|^2/(6M_{\Omega_1}^2)$
&$2\cO^{(1)}_{ud}+3\cO^{(8)}_{ud}$
\\[7pt]
&$\bar{\bm3}_q$
&$\phantom{-}|y_{\Omega_1}^{qq}|^2/(4M_{\Omega_1}^2)$
&$\cO_{qq}^{(1)D}-\cO_{qq}^{(3)D}-\cO_{qq}^{(1)E}+\cO_{qq}^{(3)E}$ 
\\
\midrule
$\Omega_2\sim(\bm6,\bm1)_{-\frac{2}{3}}$
&$\bm6_d$
&$\phantom{-}|y_{\Omega_2}|^2/(4M_{\Omega_2}^2)$
&$\cO_{dd}^D+\cO_{dd}^E$
\\[6pt]
$\Omega_4\sim(\bm6,\bm1)_{\frac{4}{3}}$
&$\bm6_u$
&$\phantom{-}|y_{\Omega_4}|^2/(4M_{\Omega_4}^2)$
&$\cO_{uu}^D+\cO_{uu}^E$
\\[6pt]
$\Upsilon\sim(\bm6,\bm3)_{\frac{1}{3}}$
&$\bm6_q$
&$\phantom{-}|y_\Upsilon|^2/(8M_\Upsilon^2)$
&$3\cO_{qq}^{(1)D}+\cO_{qq}^{(3)D}+3\cO_{qq}^{(1)E}+\cO_{qq}^{(3)E}$
\\
\midrule
\multirow{2}{*}{\rule{0pt}{4ex}    
$\Phi\sim(\bm8,\bm2)_{\frac{1}{2}}$}
&$(\bar{\bm3}_q,\bm3_u)$
&$-|y^{qu}_\Phi|^2/(18M_\Phi^2)$
&$4\cO_{qu}^{(1)}-3\cO_{qu}^{(8)}$
\\[4pt]
&$(\bar{\bm3}_d,\bm3_q)$
&$-|y^{dq}_\Phi|^2/(18M_\Phi^2)$
&$4\cO_{qd}^{(1)}-3\cO_{qd}^{(8)}$ 
\\
\bottomrule
\end{tabular}
}
\caption{\textbf{New scalars} (nontrivial flavor irreps): The first column presents the names of the spin-0 fields alongside their respective charges under the SM gauge group. The second column delineates possible representations under the $U(3)^5$ flavor symmetry. The SMEFT operators generated upon integrating out a given field are detailed across the third and fourth columns. For clarity, we separately display normalization and the generated operators, noting that the SMEFT Lagrangian can be reconstructed as $\cL_{\text{SMEFT}}\supset \text{Normalization}\times\text{Operator}$. Further details can be found in Appendix~\ref{app:sm1-spin0}.}\label{tab:sm1-spin0}
\end{table}

\begin{table}[h!]
\centering
\begin{tabular}{cccc}
\toprule
\textbf{Field} & \textbf{Irrep} & \textbf{Normalization} & \textbf{Operator} \\
\midrule
$N\sim(\bm1,\bm1)_0$
&$\bm3_\ell$
&$\phantom{-}|\lambda_N|^2/(4M_N^2)$
&$\cO_{\phi\ell}^{(1)} -\cO_{\phi\ell}^{(3)}$
\\[4pt]
$E\sim(\bm1,\bm1)_{-1}$
& $\bm3_\ell$
&$-|\lambda_E|^2/(4M_E^2)$
&$\cO_{\phi\ell}^{(1)} + \cO_{\phi\ell}^{(3)}-\lzs2y_e^*\cO_{e\phi}+\hermc\dzs$
\\[4pt]
$\Delta_1\sim(\bm1,\bm2)_{-\frac{1}{2}}$
&$\bm3_e$
&$\phantom{-}|\lambda_{\Delta_1}|^2/(2M_{\Delta_1}^2)$
&$ \cO_{\phi e}+\lzs y_e^*\cO_{e\phi}+\hermc \dzs$
\\[6pt]
$\Delta_3\sim(\bm1,\bm2)_{-\frac{3}{2}}$
& $\bm3_e$
&$-|\lambda_{\Delta_3}|^2/(2M_{\Delta_3}^2)$
&$\cO_{\phi e}-\lzs y_e^*\cO_{e\phi}+\hermc \dzs$
\\[6pt]
$\Sigma\sim(\bm1,\bm3)_0$
&$\bm3_\ell$
&$\phantom{-}|\lambda_\Sigma|^2/(16M_\Sigma^2)$
&$3 \cO_{\phi\ell}^{(1)} +\cO_{\phi\ell}^{(3)}+\lzs 4y_e^*\cO_{e\phi}+\hermc\dzs$
\\[5pt]
$\Sigma_1\sim(\bm1,\bm3)_{-1}$ 
&$\bm3_\ell$
&$\phantom{-}|\lambda_{\Sigma_1}|^2/(16M_{\Sigma_1}^2)$
&$\cO_{\phi\ell}^{(3)} -3 \cO_{\phi\ell}^{(1)}+\lzs 2y_e^*\cO_{e\phi}+\hermc \dzs$
\\
\midrule
$U\sim(\bm3,\bm1)_{\frac{2}{3}}$
& $\bm3_q$
&$\phantom{-}\lzu\lambda_{U}\dzu^2/(4M_{U}^2)$
&$\cO_{\phi q}^{(1)}-\cO_{\phi q}^{(3)}+\lzs 2y_u^*\cO_{u\phi}+\hermc \dzs$
\\[5pt]
$D\sim(\bm3,\bm1)_{-\frac{1}{3}}$
&$\bm3_q$
&$-\lzu\lambda_{D}\dzu^2/(4M_{D}^2)$
&$\cO_{\phi q}^{(1)} + \cO_{\phi q}^{(3)}-\lzs 2y_d^*\cO_{d\phi}+\hermc \dzs$
\\
\midrule
\multirow{2}{*}{$\rule{0pt}{4.4ex} Q_1\sim(\bm3,\bm2)_{\frac{1}{6}}$}
&$\bm3_u$
&$-|\lambda^u_{Q_1}|^2/(2M_{Q_1}^2)$
&$\cO_{\phi u}-\lzs y_u^*\cO_{u\phi}+\hermc \dzs$
\\[6pt]
&$\bm3_d$
&$\phantom{-}|\lambda^d_{Q_1}|^2/(2M_{Q_1}^2)$
&$\cO_{\phi d}+\lzs y_d^*\cO_{d\phi}+\hermc \dzs$
\\
\midrule
$Q_5\sim(\bm3,\bm2)_{-\frac{5}{6}}$
&$\bm3_d$
&$-|\lambda_{Q_5}|^2/(2M_{Q_5}^2)$
&$\cO_{\phi d}-\lzs y_d^*\cO_{d\phi}+\hermc \dzs$
\\[7pt]
$Q_7\sim(\bm3,\bm2)_{\frac{7}{6}}$
&$\bm3_u$
&$\phantom{-}|\lambda_{Q_7}|^2/(2M_{Q_7}^2)$
&$\cO_{\phi u}+\lzs y_u^*\cO_{u\phi}+\hermc \dzs$
\\[7pt]
$T_1\sim(\bm3,\bm3)_{-\frac{1}{3}}$
&$\bm3_q$
&$\phantom{-}|\lambda_{T_1}|^2/(16M_{T_1}^2)$
&$\cO_{\phi q}^{(3)}-3\cO_{\phi q}^{(1)}+\lzs 2y_d^*\cO_{d\phi}+4y_u^*\cO_{u\phi}+\hermc \dzs$ 
\\[6pt]
$T_2\sim(\bm3,\bm3)_{\frac{2}{3}\phantom{}}$
&$\bm3_q$
&$\phantom{-}|\lambda_{T_2}|^2/(16M_{T_2}^2)$
&$\cO_{\phi q}^{(3)}+3\cO_{\phi q}^{(1)}+\lzs 4y_d^*\cO_{d\phi}+2y_u^*\cO_{u\phi}+\hermc \dzs$
\\
\bottomrule
\end{tabular}
\caption{\textbf{New fermions}: The first column presents the names of the spin-$\frac{1}{2}$ fields alongside their respective charges under the SM gauge group. The second column delineates possible representations under the $U(3)^5$ flavor symmetry. The SMEFT operators generated upon integrating out a given field are detailed across the third and fourth columns. For clarity, we separately display normalization and the generated operators, noting that the SMEFT Lagrangian can be reconstructed as $\cL_{\text{SMEFT}}\supset \text{Normalization}\times\text{Operator}$. The flavor indices in the $\cO(y_{e,d,u})$ terms are suppressed to reduce clutter. See Appendix \ref{app:sm1-spin12} for more details.}\label{tab:sm1-spin12}
\end{table}

\begin{table}[h!]
\centering
\begin{tabular}{cccc}
\toprule
\textbf{Field} & \textbf{Irrep} & \textbf{Normalization} &\textbf{Operator}\\
\midrule
\multirow{5}{*}{$\cB\sim(\bm1,\bm1)_0$}
&$\bm8_\ell$
&$-(g_\cB^\ell)^2/(12M_\cB^2)$
&$3\cO_{\ell\ell}^E-\cO_{\ell\ell}^D$
\\
&$\bm8_e$
&$-(g_\cB^e)^2/(6M_\cB^2)$
&$\cO_{ee}$
\\
&$\bm8_q$
&$-(g_\cB^q)^2/(12M_\cB^2)$
&$3\cO_{qq}^{(1)E}-\cO_{qq}^{(1)D}$
\\
&$\bm8_u$
&$-(g_\cB^u)^2/(12M_\cB^2)$
&$3\cO_{uu}^E-\cO_{uu}^D$
\\
&$\bm8_d$
&$-(g_\cB^d)^2/(12M_\cB^2)$
&$3\cO_{dd}^{E}-\cO_{dd}^D$
\\
\midrule
$\cB_1\sim(\bm1,\bm1)_1$
&$(\bar{\bm3}_d,\bm3_u)$
&$-|g_{\cB_1}^{du}|^2/(3M_{\cB_1}^2)$
&$\cO_{ud}^{(1)}+6\cO_{ud}^{(8)}$
\\
\midrule
\multirow{2}{*}{$\cW\sim(\bm1,\bm3)_0$}
&$\bm8_q$
&$-(g_{\cW}^q)^2/(48M_{\cW}^2)$
&$3\cO_{qq}^{(3)E}-\cO_{qq}^{(3)D}$
\\
&$\bm8_\ell$
&$\phantom{-}(g_\cW^\ell)^2/(48M_\cW^2)$
&$5\cO_{\ell\ell}^E-7\cO_{\ell\ell}^D$
\\
\midrule
$\cL_3\sim(\bm1,\bm2)_{-\frac{3}{2}}$
&$(\bm3_e,\bm3_\ell)$
&$\lzu g_{\cL_3} \dzu^2/{M_{\cL_3}^2}$
&$\cO_{\ell e}$
\\
\midrule
\multirow{2}{*}{\rule{0pt}{3ex}    
$\cU_2\sim(\bm3,\bm1)_{\frac{2}{3}}$}
&$(\bar{\bm3}_e,\bm3_d)$
&$-|g_{\cU_2}^{ed}|^2/{M_{\cU_2}^2}$
&$\cO_{ed}$
\\
&$(\bar{\bm3}_\ell,\bm3_q)$
&$-|g_{\cU_2}^{\ell q}|^2/{(2M_{\cU_2}^2)}$
&$\cO_{\ell q}^{(1)}+\cO_{\ell q}^{(3)}$
\\
\midrule
$\cU_5\sim(\bm3,\bm1)_{\frac{5}{3}}$
&$(\bar{\bm3}_e,\bm3_u)$
&$-|g_{\cU_5}|^2/{M_{\cU_5}^2}$
&$\cO_{eu}$
\\
\midrule
\multirow{2}{*}{$\cQ_1\sim(\bm3,\bm2)_{\frac{1}{6}}$}
& $(\bm3_u,\bm3_\ell)$
& $\phantom{-}|g_{\cQ_1}^{u\ell}|^2/M_{\cQ_1}^2$ & $\cO_{\ell u}$
\\
& $(\bar{\bm3}_d,\bar{\bm3}_q)$
& $2|g_{\cQ_1}^{dq}|^2/(3M_{\cQ_1}^2)$
& $\cO_{qd}^{(1)}-3\cO_{qd}^{(8)}$
\\
\midrule
\multirow{3}{*}{$ \cQ_5\sim(\bm3,\bm2)_{-\frac{5}{6}}$}
& $(\bm3_d,\bm3_\ell)$
& $|g_{\cQ_5}^{d\ell}|^2/{M_{\cQ_5}^2}$ 
& $\cO_{\ell d}$ 
\\
&$(\bm3_e,\bm3_q)$
& $|g_{\cQ_5}^{eq}|^2/{M_{\cQ_5}^2}$
& $\cO_{qe}$
\\
& $(\bar{\bm3}_u,\bar{\bm3}_q)$ 
& $2|g_{\cQ_5}^{uq}|^2/(3M_{\cQ_5}^2)$
& $\cO_{qu}^{(1)}-3\cO_{qu}^{(8)}$
\\
\midrule
$\cX\sim(\bm3,\bm3)_{\frac{2}{3}}$
& $\rule{0pt}{3ex}  (\bar{\bm3}_\ell,\bm3_q)$
& $-\lzu g_\cX \dzu^2/(8M_\cX^2)$ 
& $3\cO_{\ell q}^{(1)}-\cO_{\ell q}^{(3)}$
\\
\midrule
$\cY_1\sim(\bar{\bm6},\bm2)_{\frac{1}{6}}$ 
& $(\bar{\bm3}_d,\bar{\bm3}_q)$ 
& $\lzu g_{\cY_1} \dzu^2/{(3M_{\cY_1}^2)}$ 
& $2\cO_{qd}^{(1)}+3\cO_{qd}^{(8)}$
\\
\midrule
$\cY_5\sim(\bar{\bm6},\bm2)_{-\frac{5}{6}}$ 
& $  (\bar{\bm3}_u,\bar{\bm3}_q)$ 
& $\lzu g_{\cY_5} \dzu^2/(3M_{\cY_5}^2)$ 
& $2\cO_{qu}^{(1)}+3\cO_{qu}^{(8)}$
\\
\midrule
\multirow{3}{*}{$\cG\sim(\bm8,\bm1)_0$}
& $\bm8_q$
& $-(g_\cG^q)^2/(144M_\cG^2)$ 
& $11\cO_{qq}^{(1)D}-9\cO_{qq}^{(1)E}+9\cO_{qq}^{(3)D}-3\cO_{qq}^{(3)E}$ 
\\
& $\bm8_u$ 
& $\phantom{-}(g_\cG^u)^2/(36M_\cG^2)$
& $3\cO_{uu}^E-5\cO_{uu}^D$
\\
& $\bm8_d$ 
& $\phantom{-}(g_\cG^d)^2/(36M_\cG^2)$ 
& $3\cO_{dd}^E-5\cO_{dd}^D$
\\
\midrule
$\cG_1\sim(\bm8,\bm1)_1$
& $(\bar{\bm3}_d,\bm3_u)$
& $\phantom{-}\lzu g_{\cG_1} \dzu^2/(9M_{\cG_1}^2)$
& $-4\cO_{ud}^{(1)}+3\cO_{ud}^{(8)}$
\\
\midrule
$\cH\sim(\bm8,\bm3)_0$
& $\bm8_q$ 
& $-(g_\cH)^2/(576 M_\cH^2)$
& $27\cO_{qq}^{(1)D}-9\cO_{qq}^{(1)E}-7\cO_{qq}^{(3)D}-3\cO_{qq}^{(3)E}$\\
\bottomrule
\end{tabular}
\caption{\textbf{New vectors} (nontrivial flavor irreps): The first column presents the names of the spin-1 fields alongside their respective charges under the SM gauge group. The second column delineates possible representations under the $U(3)^5$ flavor symmetry. The SMEFT operators generated upon integrating out a given field are detailed across the third and fourth columns. For clarity, we separately display normalization and the generated operators, noting that the SMEFT Lagrangian can be reconstructed as $\cL_{\text{SMEFT}}\supset \text{Normalization}\times\text{Operator}$. See Appendix \ref{app:sm1-spin1} for more details.}\label{tab:sm1-spin1}
\end{table}

\begin{table}[h]
\centering
\scalebox{0.91}{
\begin{tabular}{cccc}
\toprule
\textbf{Field}&\textbf{Irrep}&\textbf{Normalization}&\textbf{Operator}\\
\midrule
\multirow{1}{*}{$\varphi\sim(\bm1,\bm2)_{\frac{1}{2}}$}
&$\bm1$
&$\lzu \lambda_\varphi \dzu^2/M_\varphi^2$
&$\cO_\phi$
\\[4pt]
$\Theta_1\sim(\bm1,\bm4)_{\frac{1}{2}}$
&$\bm1$
&$\lzu \lambda_{\Theta_1} \dzu^2/(6M_{\Theta_1}^2)$
&$\cO_\phi$\\[4pt]
$\Theta_3\sim(\bm1,\bm4)_{\frac{3}{2}}$
&$\bm1$
&$\lzu \lambda_{\Theta_3} \dzu^2/(2M_{\Theta_3}^2)$
&$\cO_\phi$\\
\midrule
$\cS\sim(\bm1,\bm1)_0$
&$\bm1$
&$-\kappa_\cS^2/(2M_\cS^4)$
&$\cO_{\phi\Box}-\bar\cC_\cS\cO_\phi$
\\[3pt]
$\Xi\sim(\bm1,\bm3)_0$
&$\bm1$
&$\kappa_\Xi^2/(2M_\Xi^4)$
&$-4\cO_{\phi D}+\cO_{\phi\Box}+\bar\cC_\Xi\cO_\phi+2\lzs \sum_{f}y_f^*\cO_{f\phi} +\hermc \dzs$
\\[4pt]
$\Xi_1\sim(\bm1,\bm3)_1$
&$\bm1$
&$\lzu \kappa_{\Xi_1} \dzu^2/M_{\Xi_1}^4$
&$4\cO_{\phi D}+2\cO_{\phi\Box}+\bar\cC_{\Xi_1}\cO_\phi+2\lzs \sum_{f}y_f^*\cO_{f\phi} +\hermc \dzs$
\\[2pt]
\midrule
$\cB_1\sim(\bm1,\bm1)_1$
&$\bm1$
&$-|g_{\cB_1}^\phi|^2/(2M_{\cB_1}^2)$
&$4(\lambda_\phi+C_{\phi4}^{\cB_1})\cO_\phi-2\cO_{\phi D}+\cO_{\phi\Box}+\lzs \sum_{f}y_f^*\cO_{f\phi} +\hermc\dzs$
\\[4pt]
$\cW_1\sim(\bm1,\bm3)_1$
&$\bm1$
&$-|g_{\cW_1}|^2/(8M_{\cW_1}^2)$
&$4(\lambda_\phi+C_{\phi4}^{\cW_1})\cO_\phi+2\cO_{\phi D}+\cO_{\phi\Box}+\lzs \sum_{f}y_f^*\cO_{f\phi} +\hermc \dzs$
\\[4pt]
$\cH\sim(\bm8,\bm3)_0$
&$\bm1$
&$(g_\cH)^2/(96M_\cH^2)$
&$2\cO_{qq}^{(3)D}+3\cO_{qq}^{(3)E}-9\cO_{qq}^{(1)E}$
\\
\bottomrule
\end{tabular}
}
\caption{\textbf{Flavor singlets}: First six rows are scalars (spin-0) while the last three are vectors (spin-1). The table format is the same as for Tables~\ref{tab:sm1-spin0}, \ref{tab:sm1-spin12} and \ref{tab:sm1-spin1}. The $f$ index in the $\cO(y_f)$ terms goes over all three right-handed fields, i.e., $f=\{e,u,d\}$. The flavor indices are suppressed to reduce clutter. Parameters $C^X_{\phi4}$ are fixed in terms of the normalisation, while $\bar\cC_X$ are independent. See Appendices~\ref{app:sm1-spin0} and \ref{app:sm1-spin1} for details.} \label{tab:singlets}
\end{table}

\begin{table}[h]
\centering
\scalebox{0.95}{
\begin{tabular}{cccc}
\toprule
\textbf{Field}&\textbf{Irrep}&\textbf{\# of parameters}&\textbf{Operators}\\
\midrule
\multirow{3}{*}{$\cB\sim(\bm1,\bm1)_0$}
&\multirow{3}{*}{$\bm1$}
&\multirow{3}{*}{$5\mathbb{R}+1\mathbb{C}$}
&$\cO_{\ell\ell}^D$, $\cO_{qq}^{(1)D}$, $\cO_{\ell q}^{(1)}$, $\cO_{ee}$, $\cO_{dd}^D$, $\cO_{uu}^D$, $\cO_{ed}$, $\cO_{eu}$, $\cO_{ud}^{(1)}$\\
&&&$\cO_{\ell e}$, $\cO_{\ell d}$, $\cO_{\ell u}$, $\cO_{qe}$, $\cO_{qu}^{(1)}$, $\cO_{qd}^{(1)}$, $\cO_{\phi\Box}$, $\cO_{\phi D}$, $\cO_{\phi u}$\\
&&&$\cO_{\phi d}$, $\cO_{\phi e}$, $\cO_{\phi\ell}^{(1)}$, $\cO_{\phi q}^{(1)}$, $\cO_{e\phi}$, $\cO_{d\phi}$, $\cO_{u\phi}$
\\
\midrule
\multirow{2}{*}{$\cW\sim(\bm1,\bm3)_0$}
&\multirow{2}{*}{$\bm1$}
&\multirow{2}{*}{$2\mathbb{R}+1\mathbb{C}$}
&$\cO_{\ell\ell}^D-2\cO_{\ell\ell}^E$, $\cO_{qq}^{(3)D}$, $\cO_{\ell q}^{(3)}$, $\cO_\phi$, $\cO_{\phi D}$,  
\\
&&&$\cO_{\phi \Box}$, $\cO_{\phi\ell}^{(3)}$, $\cO_{\phi q}^{(3)}$, $\cO_{e\phi}$, $\cO_{d\phi}$, $\cO_{u\phi}$
\\
\midrule
\multirow{2}{*}{$\cG\sim(\bm8,\bm1)_0$}
&\multirow{2}{*}{$\bm1$}
&\multirow{2}{*}{$3\mathbb{R}$}
&$\cO_{dd}^D-3\cO_{dd}^E$, $\cO_{uu}^D-3\cO_{uu}^E$, $\cO_{qq}^{(3)E}$, $\cO_{qu}^{(8)}$, $\cO_{qd}^{(8)}$,\\
&&&$2\cO_{qq}^{(1)D}-3\cO_{qq}^{(1)E}$, $\cO^{(8)}_{ud}$
\\
\bottomrule
\end{tabular}
}
\caption{\textbf{Flavor singlets} (exceptions): Three vector (spin-1) fields match at tree-level to dimension-6 SMEFT operators shown in the last column. The corresponding WCs can be parameterised by a number of parameters indicated in the third column. See Appendix~\ref{app:sm1-spin1} for details.}\label{tab:singlets-except}
\end{table}

\clearpage
\section{Phenomenology}
\label{sec:pheno}

As articulated in Section~\ref{sec:theory}, imposing $U(3)^5$ symmetry on the interaction Lagrangian between an NP field and SM fields, in the vast majority of considered cases, leads to a single Hermitian operator with a definite sign, once an NP field has been integrated out and matched to the SMEFT at leading order. These leading SMEFT directions capture the leading order IR phenomenology of high-energy single-field BSM extensions defined in Appendix~\ref{app:sm1}. We collected the NP fields denoting their possible representations under the flavor group, along with the corresponding dimension-6 SMEFT operator they generate, in Tables~\ref{tab:sm1-spin0}, \ref{tab:sm1-spin12}, \ref{tab:sm1-spin1}, and \ref{tab:singlets}. This section delves into the phenomenological impacts of all proposed scenarios, along with a compilation of the current experimental bounds. The three exceptional scenarios in Table~\ref{tab:singlets-except} are left for future work.

The section is organized in two parts. In the first part, we briefly describe the global fits employed for the numerical analysis. Following that, we present the derived constraints on the WCs, which can alternatively be viewed as limits on mediator masses, given that the associated couplings in $\mathcal{L}_{{\rm BSM}}$ are set to unity for convenience. These results are logically sorted into five categories, each dictated by different sectors of dominant phenomenology. 

\subsection{Global fits}
\label{subsec:global_fits}

Global SMEFT fits serve as a highly beneficial intermediate step for confronting short-distance new physics scenarios with existing data. The outcomes of these fits are confidence level intervals within the WCs parameter space. These intervals are subsequently transformed into the UV model's parameter space using matching relations. Global fits are dependent on data harvested from a multitude of experiments, encompassing high-energy collider processes, as well as precision measurements at low energies. The recent growing interest in SMEFT has led to considerable progress in this field, with a plethora of SMEFT data interpretations now accessible in the literature, see e.g.~\cite{Falkowski:2015jaa, Falkowski:2016cxu, Ellis:2018gqa, Ellis:2020unq, Aoude:2020dwv, Falkowski:2017pss, Breso-Pla:2023tnz, Falkowski:2019hvp, Gonzalez-Alonso:2016etj, Cirigliano:2009wk, Efrati:2015eaa, Buckley:2015lku, Englert:2016aei, Hartland:2019bjb, Brivio:2019ius, Durieux:2018tev, vanBeek:2019evb, Bissmann:2020mfi, Bruggisser:2021duo, Ethier:2021bye, Baglio:2017bfe, Panico:2017frx, Grojean:2018dqj, Gomez-Ambrosio:2018pnl, Dedes:2020xmo, Pomarol:2013zra, deBlas:2016ojx, deBlas:2017wmn, Falkowski:2014tna, Krauss:2016ely, Alte:2017pme, Hirschi:2018etq, Goldouzian:2020wdq, Greljo:2017vvb, Greljo:2021kvv, Greljo:2022jac, Kassabov:2023hbm, Grunwald:2023nli, Bellafronte:2023amz, Aoude:2022aro, Altmannshofer:2023bfk, deBlas:2022hdk}. However, it is important to note that each of these fits has inherent limitations concerning the breadth of data included and the span of operators considered. In what follows, we give a brief description of global fits that we have identified as being particularly applicable to our numerical analysis. 

\subsubsection*{$W/Z$ vertex corrections, 4-lepton and 2-quark-2-lepton operators}

The flavor-diagonal 4-fermion operators, which consist of two or four leptons, along with flavor-diagonal vertex corrections for $W$ and $Z$, are concurrently fitted to a comprehensive dataset in~\cite{Falkowski:2017pss} (updated in~\cite{Breso-Pla:2023tnz}). Notably, the authors present results for $U(3)^5$ flavor-symmetric operators, which are directly applicable to our cases, covering a substantial number of the leading directions. The data utilized in this fit comprise of:
\begin{itemize}
    \item\textbf{Neutrino scattering}: CHARM \cite{CHARM:1986vuz, CHARM:1987pwr}, CDHS \cite{Blondel:1989ev} and CCFR \cite{CCFR:1997zzq} measured the ratios of neutral- and charged-current (anti-)neutrino cross sections on nuclei.  
    \item\textbf{Atomic parity violation (APV)}: Experiments reported in~\cite{Dzuba:2012kx, Wood:1997zq, Qweak:2013zxf, Edwards:1995zz, Vetter:1995vf, Qweak:2018tjf} probe the parity-violating nature of the couplings between electrons and quarks. In addition to the weak-charge measurements, these are also constrained from the deep-inelastic scattering of polarized electron~\cite{PVDIS:2014cmd, Beise:2004py} and from the asymmetry measurements using muons~\cite{Argento:1982tq}.   
    
    \item\textbf{Low-energy flavor}: Kaon, pion and tau decays into muons and electrons~\cite{FlaviaNetWorkingGrouponKaonDecays:2008hpm, FlaviaNetWorkingGrouponKaonDecays:2010lot, Cirigliano:2011ny, ParticleDataGroup:2014cgo, Gonzalez-Alonso:2016etj}.
    
    \item\textbf{Quark pair production in $e^+e^-$ collisions}: The cross section and forward-backward asymmetry in $e^+ e^-\to q \bar q$ production~\cite{ALEPH:2013dgf, ALEPH:2006bhb, VENUS:1993pob, TOPAZ:2000evx} including also strange, charm and bottom quarks in the final state.
    
    \item\textbf{Lepton pair production in $e^+e^-$ collisions}: The polarization and forward-backward asymmetry in $e^+ e^-\to\ell^+\ell^-$ production~\cite{ALEPH:2013dgf,VENUS:1997cjg}. 
    
    \item\textbf{$W$ and $Z$ pole observables}: Various cross sections, decay rates, ratios, and asymmetries shown in Table~1 of~\cite{Efrati:2015eaa, Falkowski:2019hvp}.
    
\end{itemize}
Moreover, the compilation incorporates $\nu_\mu e$ scattering, low-energy parity-violating $e^+e^-$ scattering, and neutrino trident production, among others. These are comprehensively encapsulated in Table~4 of~\cite{Falkowski:2017pss}. The fit detailed in~\cite{Breso-Pla:2023tnz} encompasses data on coherent elastic neutrino-nucleus scattering from the COHERENT experiment~\cite{COHERENT:2020iec, COHERENT:2021xmm}. Additional updates of~\cite{Falkowski:2017pss} are thoroughly documented in Appendix~C.2.\,of~\cite{Breso-Pla:2023tnz} to which we refer for more details. Important RGE effects are also included, as discussed in Section 2.3 of~\cite{Falkowski:2017pss}. Our numerical results rely on the latest fit~\cite{Breso-Pla:2023tnz}. Nevertheless, we have checked that the earlier fit~\cite{Falkowski:2017pss} leads to only slightly reduced bounds.

In practice, we investigate the $\chi^2$ function of the $U(3)^5$ dimension-6 SMEFT fit of~\cite{Breso-Pla:2023tnz} separately for each leading direction from Section~\ref{sec:theory}.  We express WCs in terms of the mass of a given NP mediator while setting the coupling constant to unity. Consequently, $\chi^2$ becomes a function of a single real and positive parameter, which then serves to determine the allowed mass range.

It is also illuminating to indicate which WCs from the MFV basis (Appendix~\ref{app:basis}) are constrained in this global fit:
\begin{equation}
    \begin{alignedat}{2}
        \text{4F}&:\quad\lzv \cO_{\ell\ell}^D,\cO_{\ell\ell}^E,\cO_{\ell e}, \cO_{ee}, \cO_{\ell q}^{(1)}, \cO_{\ell q}^{(3)},\cO_{e q}, \cO_{\ell u}, \cO_{\ell d}, \cO_{eu}, \cO_{ed}  \dzv,\\
        \text{2F2H}&:\quad\lzv \cO_{\phi\ell}^{(1)},\cO_{\phi\ell}^{(3)},\cO_{\phi q}^{(1)},\cO_{\phi q}^{(3)},\cO_{\phi u}, \cO_{\phi d}, \cO_{\phi e} \dzv,
    \end{alignedat}
    \nonumber
\end{equation}
where 4F denotes the set of the 4-fermion, and 2F2H is the set of 2-fermion-2-Higgs operators. For easier comparison and reproduction of the analysis, we point out that the set of $\cO_{\ell\ell}^D$ and $\cO_{\ell\ell}^E$ operators can also be replaced by $\cO_{\ell\ell}^D$ and $\cO_{\ell\ell}^{(3)}$, which are related via $SU(2)$ identities
\begin{equation}\label{eq:ll-ll3-map}
    \cO_{\ell\ell}^E = \frac{1}{2}\cO_{\ell\ell}^D+\frac{1}{2}\cO_{\ell\ell}^{(3)},\qquad
    \cO_{\ell\ell}^{(3)}=(\bar\ell_i\gamma^\mu\sigma^a \ell^i)(\bar\ell_j\gamma_\mu\sigma^a\ell^j).
\end{equation}
This map enables easier extraction of the $c_{\ell\ell}$ and $c_{\ell\ell}^{(3)}$ WCs used in~\cite{Breso-Pla:2023tnz, Falkowski:2017pss}.\footnote{The overall normalization of the $\cO_{ee}$, $\cO_{\ell\ell}$ and $\cO_{\ell\ell}^{(3)}$ operators in~\cite{Breso-Pla:2023tnz, Falkowski:2017pss} differs by an overall factor of 1/2, so once the map given by Eq.~(\ref{eq:ll-ll3-map}) has been applied, this factor also has to be taken care of upon extraction of the $c_{\ell\ell}$ and $c_{\ell\ell}^{(3)}$ Wilson coefficients.}

\subsubsection*{Top quark fit}

Although the fit in the previous section probes a significant number of WCs, many are still left unconstrained. The largest such set contains 4-quark operators, which are generated in a considerable amount of cases (see Tables~\ref{tab:sm1-spin0} and \ref{tab:sm1-spin1}). The global fit presented in Ref.~\cite{Ethier:2021bye}, proves to be valuable when constraining WCs of the 4-quark operators thanks to the top quark data from the LHC. Notably, most of our 4-quark operators, due to the $U(3)^5$ flavor symmetry, give rise to scattering processes with top quarks in the final state and valence quarks in the initial state.

The ATLAS and CMS experiments measure various aspects of the top-quark production: inclusive top-quark pair production~\cite{ATLAS:2015lsn, CMS:2015rld, CMS:2017iqf, ATLAS:2016fbc, CMS:2016asd, CMS:2016oae, CMS:2018htd, CMS:2017xio, ATLAS:2016pal, ATLAS:2019czt, Amoroso:2020lgh, Bailey:2019yze}, associated top-quark pair production~\cite{CMS:2017xnm, CMS:2017ocm, CMS:2015uvn, CMS:2017ugv, ATLAS:2015qtq, ATLAS:2016wgc, CMS:2019rvj, ATLAS:2020hpj, ATLAS:2018fwl, CMS:2019eih, ATLAS:2019fwo, CMS:2019too}, inclusive single top-quark production~\cite{CMS:2014mgj, CMS:2014ika, ATLAS:2017rso, ATLAS:2015jmq, CMS:2016xoq, ATLAS:2016qhd, CMS:2016xnv, CMS:2016lel, CMS:2019jjp} and associated single top-quark production with weak bosons~\cite{ATLAS:2015igu, CMS:2014fut, ATLAS:2016ofl, CMS:2018amb, CMS:2017wrv, ATLAS:2017dsm, ATLAS:2020cwj, ATLAS:2020bhu, CMS:2018sgc}. Besides the top data, the fit in~\cite{Ethier:2021bye} also includes the Higgs and diboson data from the LHC.

The flavor symmetry of the quark sector imposed in Ref.~\cite{Ethier:2021bye} is $U(2)_q\times U(2)_u\times U(3)_d$. Under this assumption, the SMEFT parameter space is enlarged compared to the $U(3)^5$-symmetric scenario. The latter symmetry, assumed in this work, sets relations between operators used in~\cite{Ethier:2021bye}. The translation boils down to rewriting the $U(3)^5$-symmetric operators in terms of decomposed $q$ and $u$ fields, which under the $U(2)_q\times U(2)_u\times U(3)_d$ flavor symmetry decompose as $q \sim q^{1,2}\oplus q^3$ and $u\sim u^{1,2}\oplus u^3$, where indices $1,2$ denote the light, while 3 denotes the heavy quark flavor~\cite{Faroughy:2020ina, Greljo:2022cah}. Mapping from the Warsaw basis onto the basis used in~\cite{Ethier:2021bye}\footnote{Often referred to as the SMEFT top operator basis, see~\cite{Aguilar-Saavedra:2018ksv}.} is given in Table 2.4. For completeness, the set of WCs relevant to our numerical analysis is
\begin{equation}
    \begin{alignedat}{2}
        \text{4Q}:\quad\{& \texttt{cQQ1},\texttt{cQQ8},\texttt{cQt1},\texttt{cQt8},\texttt{ctt1},\texttt{c81qq},\texttt{c11qq},\texttt{c83qq},\texttt{c13qq},\texttt{c8qt},\texttt{c1qt},\texttt{c8ut},\\&\texttt{c1ut},\texttt{c8qu},\texttt{c1qu},\texttt{c8dt},\texttt{c1dt},\texttt{c8qd},\texttt{c1qd}\},
    \end{alignedat}
    \nonumber
\end{equation}
where 4Q denotes the set of 4-quark operators, following the notation of Table~2.6 of~\cite{Ethier:2021bye}. 

In the numerical analysis, we consider only one leading SMEFT direction at a time, taking the corresponding individual $\cO(\Lambda^{-4})$ bounds at $95\%$ CL on the WCs from Table~5.4 of~\cite{Ethier:2021bye}.\footnote{Since we are setting limits on single-field BSM extensions, the $\cO(\Lambda^{-4})$ intervals give a more reliable estimate than the $\cO(\Lambda^{-2})$.} The only correlation we take into account is the correlation between \texttt{cQQ1} and \texttt{cQQ8} operators, which has a value of $\rho=-0.9$. In fact, a comparison between the individual (the third column) and marginalized (the fourth column) intervals shows similarity in their ranges. This implies that the exclusion of other correlations from our analysis does not exert a substantial influence on the final outcomes.

To sum up, the two fits described in this section cover the vast majority of cases. Besides these two, we will also use a few results on WCs from Refs.~\cite{Ellis:2020unq, Durieux:2022hbu} applicable to flavor singlet cases, as well as, the limits on semileptonic contact interactions from~\cite{Allwicher:2022gkm, Allwicher:2022mcg, Greljo:2022jac}.

\subsection{Results}
\label{subsec:results}

The $U(3)^5$ flavor assumption is restrictive enough to suggest a classification motivated by phenomenology where the BSM mediators can be grouped based on a sector they contribute to. These five classes are 4-lepton, 2-quark-2-lepton, 4-quark contact interactions, $W$ and $Z$ vertex corrections, and oblique/Higgs corrections. For each class, we give a brief description and report the summary table, which contains the $95\%$ CL bounds obtained using the global fits described above.

\subsection*{Class I: 4-lepton phenomenology}

Four spin-0 and four spin-1 fields belong to this class; see Tables~\ref{tab:sm1-spin0} and \ref{tab:sm1-spin1}. Using the SMEFT fit from~\cite{Breso-Pla:2023tnz}, we derive the lower bound on the mediator mass assuming the corresponding coupling constant is equal to one, see Table~\ref{table:4-lepton_pheno}. The obtained limits are all in the multi-TeV range, varying significantly depending on the gauge and flavor representation. 

\begin{table}[t]
\centering
\begin{minipage}[t]{0.46\linewidth}
\begin{adjustbox}{valign=t}
\begin{tabular}{cccccc}
\toprule
& \textbf{Scalars} &\\
\textbf{Field} & \textbf{Irrep} & \textbf{$\bm{M}$ [TeV]}\\
\midrule
$\cS_1\sim(\bm1,\bm1)_1$&$\bm3_\ell$&8.8\\
$\cS_2\sim(\bm1,\bm1)_2$&$\bar{\bm6}_e$&3.5\\
$\varphi\sim(\bm1,\bm2)_{\frac{1}{2}}$&$(\bm3_\ell,\bar{\bm3}_e)$&4.4\\
$\Xi_1\sim(\bm1,\bm3)_1$&$\bar{\bm6}_\ell$&11.1
\\[1.3pt]
\bottomrule
\end{tabular}
\end{adjustbox}
\end{minipage}
\begin{minipage}[t]{0.45\linewidth}
\begin{adjustbox}{valign=t}
\begin{tabular}{cccccc}
\toprule
& \textbf{Vectors} &\\
\textbf{Field} & \textbf{Irrep} & \textbf{$\bm{M}$ [TeV]} \\
\midrule
$\rule{0pt}{2.5ex}\cB\sim(\bm1,\bm1)_0$&$\bm8_\ell$&4.6\\
$\cB\sim(\bm1,\bm1)_0$&$\bm8_e$&2.7\\
$\cW\sim(\bm1,\bm3)_0$&$\bm8_\ell$&5.8\\
$\cL_3\sim(\bm1,\bm2)_{-\frac{3}{2}}$&$(\bm3_\ell,\bm3_e)$&4.5\\
\bottomrule
\end{tabular}
\end{adjustbox}
\end{minipage}
\caption{\textbf{4-lepton phenomenology (Class I)}: The first two columns indicate gauge and flavor representations of the new scalars (left panel) and vectors (right panel). The third column shows the lower bounds at $95\%$ CL on the mediator masses (couplings set to unity) for each case. See Tables \ref{tab:sm1-spin0} and \ref{tab:sm1-spin1} for the linear combinations of the SMEFT operators generated and Appendices \ref{app:sm1-spin0} and \ref{app:sm1-spin1} for more details on the flavor structure of the corresponding coupling tensors.}
\label{table:4-lepton_pheno}
\end{table}

\subsection*{Class II: 2-quark-2-lepton phenomenology}

Seven spin-0 and seven spin-1 fields integrate out to a 2-quark-2-lepton contact interaction. Similarly to the 4-lepton class, here we also use the fit from~\cite{Breso-Pla:2023tnz} (described in the previous section) to extract lower bounds on the mediator masses, assuming the couplings are set to unity. In addition, independent constraints on 2-quark-2-lepton contact interactions can be obtained from the high-mass Drell-Yan production at the LHC~\cite{Greljo:2017vvb, Dawson:2018dxp, Cirigliano:2012ab, deBlas:2013qqa, Gonzalez-Alonso:2016etj, Faroughy:2016osc, Cirigliano:2018dyk, Greljo:2018tzh, Bansal:2018eha, Angelescu:2020uug, Farina:2016rws, Alioli:2017nzr, Raj:2016aky, Schmaltz:2018nls, Brooijmans:2020yij, Ricci:2020xre, Fuentes-Martin:2020lea, Alioli:2017ces, Alioli:2018ljm, Alioli:2020kez, Panico:2021vav, Sirunyan:2021khd, ATLAS:2021pvh, Marzocca:2020ueu, Afik:2019htr, Alves:2018krf, Greljo:2021kvv, Boughezal:2023nhe}. For the latter, we use the likelihood tools reported in~\cite{Allwicher:2022gkm, Allwicher:2022mcg, Greljo:2022jac}. Finally, Table~\ref{table:2q2l_pheno} presents the bounds from both low-$p_T$ (denoted as $M^{{\rm LE}}$) and high-$p_T$ ($M^{{\rm DY}}$) fits. (The summary in Figure~\ref{fig:plot} shows the best of the two bounds.) Interestingly, the two sets of complementary constraints are rather competitive with each other.  The bounds are in the multi-TeV range in all cases and vary significantly on the representation.  For more details on the phenomenology of MFV leptoquarks, see, e.g.,~\cite{Greljo:2022jac, Aloni:2017ixa, Davidson:2010uu}.


\begin{table}[t]
\centering
\begin{minipage}[t]{0.49\linewidth}
\begin{adjustbox}{valign=t}
\scalebox{0.78}{
\begin{tabular}{cccccc}
\toprule
\multicolumn{4}{c}{\textbf{Scalars}}\\
\textbf{Field}&\textbf{Irrep}& \textbf{$\bm{M^{\text{LE}}}$ [TeV]} &\textbf{$\bm{M^{\text{DY}}}$ [TeV]}\\
\midrule
$\omega_1\sim(\bm3,\bm1)_{-\frac{1}{3}}$&$(\bm3_q,\bm3_\ell)$&10.0&8.8
\\
$\omega_1\sim(\bm3,\bm1)_{-\frac{1}{3}}$&$(\bm3_u,\bm3_e)$&4.7&7.5
\\
$\omega_4\sim(\bm3,\bm1)_{-\frac{4}{3}}$& $(\bm3_d,\bm3_e)$&3.6&5.1
\\
$\Pi_1\sim(\bm3,\bm2)_{\frac{1}{6}}$&$(\bm3_d,\bar{\bm3}_\ell)$&3.7&2.8
\\
$\Pi_7\sim(\bm3,\bm2)_{\frac{7}{6}}$&$(\bm3_u,\bar{\bm3}_\ell)$&3.5&6.2
\\
$\Pi_7\sim(\bm3,\bm2)_{\frac{7}{6}}$&$(\bm3_q,\bar{\bm3}_e)$&3.4&5.7
\\
$\zeta\sim(\bm3,\bm3)_{-\frac{1}{3}}$&$(\bm3_q,\bm3_\ell)$&4.3&5.3
\\[3.5pt]
\bottomrule
\end{tabular}
}
\end{adjustbox}
\end{minipage}
\begin{minipage}[t]{0.49\linewidth}
\begin{adjustbox}{valign=t}
\scalebox{0.78}{
\begin{tabular}{cccccc}
\toprule
\multicolumn{4}{c}{\textbf{Vectors}}
\\
\textbf{Field}&\textbf{Irrep}& \textbf{$\bm{M^{\text{LE}}}$ [TeV]} &\textbf{$\bm{M^{\text{DY}}}$ [TeV]}\\
\midrule
$\rule{0pt}{2.5ex}\cU_2\sim(\bm3,\bm1)_{\frac{2}{3}}$&$(\bm3_d,\bar{\bm3}_e)$&3.7&5.6
\\
$\cU_2\sim(\bm3,\bm1)_{\frac{2}{3}}$&$(\bm3_q,\bar{\bm3}_\ell)$&14.4&8.3
\\
$\cU_5\sim(\bm3,\bm1)_{\frac{5}{3}}$&$(\bm3_u,\bar{\bm3}_e)$&3.5&12.4
\\
$\cQ_1\sim(\bm3,\bm2)_{\frac{1}{6}}$&$(\bm3_u,\bm3_\ell)$&4.0&7.5
\\
$\cQ_5\sim(\bm3,\bm2)_{-\frac{5}{6}}$&$(\bm3_d,\bm3_\ell)$&3.4&5.1
\\
$\cQ_5\sim(\bm3,\bm2)_{-\frac{5}{6}}$&$(\bm3_q,\bm3_e)$&7.7&6.6
\\
$\cX\sim(\bm3,\bm3)_{\frac{2}{3}}$&$(\bm3_q,\bar{\bm3}_\ell)$&3.1&8.7
\\[2pt]
\bottomrule
\end{tabular}
}
\end{adjustbox}
\end{minipage}
\caption{\textbf{2-quark-2-lepton phenomenology (Class II)}: The first two columns indicate gauge and flavor representations of the new scalars (left panel) and vectors (right panel). The third and fourth columns contain the lower bounds at $95\%$ CL on the mediator masses (couplings set to unity) obtained by the low-energy experiments ($M^{\text{LE}}$) and the Drell-Yan production at the LHC ($M^{\text{DY}}$), respectively. For the induced SMEFT operators, consult the Tables~\ref{tab:sm1-spin0} and \ref{tab:sm1-spin1} and Appendices~\ref{app:sm1-spin0} and \ref{app:sm1-spin1} for more details.}
\label{table:2q2l_pheno}
\end{table}

\subsection*{Class III: 4-quark phenomenology}

There are 14 scalars and 15 vectors in this class. The LHC measurements of jets and top quarks provide leading probes of these operators. Employing the top quark fit from~\cite{Ethier:2021bye} as discussed in Section~\ref{subsec:global_fits}, we set limits on the corresponding directions as shown in Table~\ref{table:4-quark_pheno}. They are significantly weaker in comparison with the previous two classes clustering around the TeV scale. 

Note that there are four fields (two scalars and two vectors) that generate a 4-quark operator containing only down quarks ($\cO_{dd}^D$ and $\cO_{dd}^E$, see Appendix~\ref{app:basis}) and, as such, they can not be bounded from~\cite{Ethier:2021bye}. The corresponding fields are $\omega_2\sim\bm3_d$, $\Omega_2\sim\bm6_d$, $\cB\sim\bm8_d$ and $\cG\sim\bm8_d$ (see Tables~\ref{tab:sm1-spin0} and \ref{tab:sm1-spin1}). 

These operators, as well as, all other 4-quark operators, can be constrained from the dijet production at the LHC~\cite{Krauss:2016ely, Alioli:2017jdo, Alte:2017pme, Hirschi:2018etq, Goldouzian:2020wdq, Bruggisser:2022rhb, CMS:2018ucw}. Unfortunately, the limits on the WCs derived in these analyses are not directly applicable to our study. For instance, the ATLAS and CMS searches for contact interactions in dijet production~\cite{ATLAS:2017eqx, CMS:2018ucw} consider a very limited set of operators (see Section 4 of~\cite{CMS:2018ucw} or Eq.~2 in~\cite{ATLAS:2017eqx}) which cannot be applied to a general dimension-6 SMEFT even with the most restrictive flavor structure in the quark sector, $U(3)^3$. We urge the experimental collaboration to correct for this in future interpretations of the dijet data and include (at least) the relevant operators from the flavor-symmetric basis of Appendix~\ref{app:basis}:
\begin{equation}
	\lzv \cO_{dd}^D\,,\,\cO_{dd}^E\,,\,\cO_{uu}^D\,,\,\cO_{uu}^E\,,\,\cO_{ud}^{(1)},\,\cO_{ud}^{(8)},\,\cO_{qu}^{(1)},\,\cO_{qu}^{(8)},\,\cO_{qd}^{(1)},\,\cO_{qd}^{(8)},\,\cO_{qq}^{(1)D},\,\cO_{qq}^{(3)D},\,\cO_{qq}^{(1)E},\,\cO_{qq}^{(3)E}\dzv.
 \nonumber
\end{equation}
Interestingly, Table 2 of~\cite{CMS:2018ucw} suggests that the dijet angular distributions might provide competitive (even stronger) bounds than the top quark production. The recast of the dijet analyses is beyond the scope of this paper. We leave the construction of the SMEFT likelihood of the dijet data for future work.


\begin{table}[t]
\centering
\begin{minipage}[t]{0.45\linewidth}
\begin{adjustbox}{valign=t}
\begin{tabular}{cccccc}
\toprule
& \textbf{Scalars} &\\
\textbf{Field} & \textbf{Irrep} & \textbf{$\bm{M}$ [TeV]}\\
\midrule
\multirow{2}{*}{$\rule{0pt}{2.8ex}\varphi\sim(\bm1,\bm2)_{\frac{1}{2}}$}&$(\bm3_q,\bar{\bm3}_d)$&0.9
\\
&$(\bar{\bm3}_q,\bm3_u)$&1.5\\
\midrule
\multirow{2}{*}{$\omega_1\sim(\bm3,\bm1)_{-\frac{1}{3}}$}&$\bar{\bm6}_q$&1.7
\\
&$(\bar{\bm3}_u,\bar{\bm3}_d)$&1.0
\\
\midrule
$\omega_4\sim(\bm3,\bm1)_{-\frac{4}{3}}$&$\bm3_u$&1.7
\\
$\zeta\sim(\bm3,\bm3)_{-\frac{1}{3}}$&$\bm3_q$&3.0
\\
\midrule
\multirow{2}{*}{$\rule{0pt}{2.9ex}\Omega_1\sim(\bm6,\bm1)_{\frac{1}{3}}$}&$\rule{0pt}{2.9ex}(\bm3_u,\bm3_d)$&1.1
\\
&$\rule{0pt}{2.9ex}\bar{\bm3}_q$&2.2
\\
\midrule
$\Omega_4\sim(\bm6,\bm1)_{\frac{4}{3}}$&$\bm6_u$&1.4
\\
$\Upsilon\sim(\bm6,\bm3)_{\frac{1}{3}}$&$\bm6_q$&1.9
\\
\midrule
\multirow{2}{*}{$\Phi\sim(\bm8,\bm2)_{\frac{1}{2}}$}&$(\bar{\bm3}_q,\bm3_u)$&1.3
\\
&$(\bm3_q,\bar{\bm3}_d)$&0.8
\\[5.8pt]
\bottomrule
\end{tabular}
\end{adjustbox}
\end{minipage}
\begin{minipage}[t]{0.45\linewidth}
\begin{adjustbox}{valign=t}
\begin{tabular}{cccccc}
\toprule
& \textbf{Vectors} &\\
\textbf{Field} & \textbf{Irrep} & \textbf{$\bm{M}$ [TeV]}\\
\midrule
\multirow{2}{*}{$\cB\sim(\bm1,\bm1)_0$}&$\bm8_q$&1.0
\\
&$\bm8_u$&0.7
\\
\midrule
$\cB_1\sim(\bm1,\bm1)_1$&$(\bm3_u,\bar{\bm3}_d)$&1.3
\\
$\cW\sim(\bm1,\bm3)_0$&$\bm8_q$&0.7
\\
$\cQ_1\sim(\bm3,\bm2)_{\frac{1}{6}}$&$\rule{0pt}{2.8ex}(\bar{\bm3}_q,\bar{\bm3}_d)$&1.3
\\
$\cQ_5\sim(\bm3,\bm2)_{-\frac{5}{6}}$&$(\bar{\bm3}_q,\bar{\bm3}_u)$&2.2
\\
$\cY_1\sim(\bar{\bm6},\bm2)_{\frac{1}{6}}$&$(\bar{\bm3}_q,\bar{\bm3}_d)$&1.4
\\
$\cY_5\sim(\bar{\bm6},\bm2)_{-\frac{5}{6}}$&$(\bar{\bm3}_q,\bar{\bm3}_u)$&2.1
\\
\midrule
\multirow{2}{*}{$\cG\sim(\bm8,\bm1)_0$}&$\bm8_q$&0.9
\\
&$\bm8_u$&0.7
\\
\midrule
$\cG_1\sim(\bm8,\bm1)_1$&$(\bm3_u,\bar{\bm3}_d)$&1.1
\\
\midrule
\multirow{2}{*}{$\cH\sim(\bm8,\bm3)_0$}&$\bm8_q$&0.4
\\
&$\bm1$&0.5
\\
\bottomrule
\end{tabular}
\end{adjustbox}
\end{minipage}
\caption{\textbf{4-quark phenomenology (Class III)}: The first two columns indicate gauge and flavor representations of the new scalars (left panel) and vectors (right panel). The third column shows the lower bounds at $95\%$ CL on the mediator masses (couplings set to unity) for each case. See Tables~\ref{tab:sm1-spin0} and \ref{tab:sm1-spin1} for the linear combinations of the SMEFT operators generated and Appendices~\ref{app:sm1-spin0} and \ref{app:sm1-spin1} for more details.}
\label{table:4-quark_pheno}
\end{table}

\subsection*{Class IV: $W/Z$ vertex corrections}

The dominant effect when integrating out a new heavy spin-$\frac{1}{2}$ field with an $U(3)^5$ flavor-symmetric Lagrangian is the universal correction of the $W$ and $Z$ couplings to fermions, see Table~\ref{tab:sm1-spin12}. The bounds on these fields, extracted using the fit reported in~\cite{Breso-Pla:2023tnz}, are presented in Table~\ref{table:vertex_pheno}. The obtained lower limits on the mediator masses, for a coupling equal to one, sit in the multi-TeV range. Note that some cases also predict a correlated universal relative correction of the SM Higgs couplings to fermions. However, the experimental limits on the latter are not competitive.

\begin{table}[t]
\centering
\begin{minipage}[t]{0.48\linewidth}
\begin{adjustbox}{valign=t}
\begin{tabular}{cccccc}
\toprule
&\textbf{Leptons}&\\
\textbf{Field}&\textbf{Irrep}&\textbf{$\bm{M}$ [TeV]}\\
\midrule
$N\sim(\bm1,\bm1)_0$&$\bm3_\ell$&9.0
\\
$E\sim(\bm1,\bm1)_{-1}$&$\bm3_\ell$&8.0
\\
$\Delta_1\sim(\bm1,\bm2)_{-\frac{1}{2}}$&$\bm3_e$&6.1
\\
$\Delta_3\sim(\bm1,\bm2)_{-\frac{3}{2}}$&$\bm3_e$&8.6
\\
$\Sigma\sim(\bm1,\bm3)_0$ &$\bm3_\ell$&5.9
\\
$\Sigma_1\sim(\bm1,\bm3)_{-1}$&$\bm3_\ell$&4.3
\\
\bottomrule
\end{tabular}
\end{adjustbox}
\end{minipage}
\begin{minipage}[t]{0.45\linewidth}
\begin{adjustbox}{valign=t}
\begin{tabular}{cccccc}
\toprule
&\textbf{Quarks}&\\
\textbf{Field}&\textbf{Irrep}&\textbf{$\bm{M}$ [TeV]}\\
\midrule
$U\sim(\bm3,\bm1)_{\frac{2}{3}}$&$\bm3_q$&4.2
\\
$D\sim(\bm3,\bm1)_{-\frac{1}{3}}$&$\bm3_q$&4.9
\\
$Q_1\sim(\bm3,\bm2)_{\frac{1}{6}}$&$\bm3_u$&3.7
\\
$Q_1\sim(\bm3,\bm2)_{\frac{1}{6}}$&$\bm3_d$&4.6
\\
$Q_5\sim(\bm3,\bm2)_{-\frac{5}{6}}$&$\bm3_d$&2.3
\\
$Q_7\sim(\bm3,\bm2)_{\frac{7}{6}}$&$\bm3_u$&3.3
\\
$T_1\sim(\bm3,\bm3)_{-\frac{1}{3}}$&$\bm3_q$&4.6
\\
$T_2\sim(\bm3,\bm3)_{\frac{2}{3}}$&$\bm3_q$&4.6
\\
\bottomrule
\end{tabular}
\end{adjustbox}
\end{minipage}
\caption{\textbf{Vertex corrections (Class IV)}: The first two columns indicate gauge and flavor representations of the new leptons (left panel) and quarks (right panel). The third column shows the lower bounds at $95\%$ CL on the mediator masses (couplings set to unity) for each case. Consult Table \ref{tab:sm1-spin12} for the linear combinations of the SMEFT operators generated and Appendix \ref{app:sm1-spin12} for more details regarding the flavor structure of the couplings.}
\label{table:vertex_pheno}
\end{table}

\subsection*{Class V: Oblique corrections and Higgs phenomenology}

We denote the last class as ``Oblique/Higgs'', as it contains flavor-singlet scalars and vectors of Table~\ref{tab:singlets} (except for the last entry) contributing to operators $\cO_\phi$, $\cO_{\phi D}$, $\cO_{\phi\Box}$, and $\cO_{(e,u,d)\phi}$. The three scalars, $\varphi$, $\Theta_1$, and $\Theta_3$, match at the tree level only onto $\cO_\phi$ operator, thus, we constrain those from the Higgs self-coupling~\cite{CMS:2020gsy, ATLAS:2021tyg, Durieux:2022hbu}.\footnote{Our entire analysis is performed at the tree level. Since the Higgs self-coupling is poorly known experimentally, the custodial symmetry breaking at the one loop and beyond dimension 6 provides stronger limits~\cite {Durieux:2022hbu}. Organizing fields by the custodial symmetry would make the case for the Higgs self-coupling~\cite{Durieux:2022hbu}.} 

The remaining five fields generate a combination of operators. The bound on $\cO_{\phi D}$ from oblique EW corrections is by far the most stringent~\cite{Ellis:2020unq, Durieux:2022hbu} and is used to set the lower mass limits on $\Xi$, $\Xi_1$, $\cB_1$ and $\cW_1$ fields. For the first two fields, in principle, there is an independent parameter for $\cO_\phi$, which we ignore given the poor constraints on the Higgs self-coupling. The $\cS$ scalar, however, generates at the tree level only $\cO_{\phi\Box}$ and $\cO_\phi$. The experimental limit on the former is stronger than on the latter. Both operators, in this case, come with an independent WC, and we report bounds only on $\cO_{\phi\Box}$ using~\cite{Ellis:2020unq}.

We collect the results in Table~\ref{table:singlets_pheno}. For dimensionful couplings $\kappa_X$, we assume $\kappa_X = M_X$, while the dimensionless couplings are set to unity. Oblique corrections give very strong limits in the multi-TeV range (the left panel of Table~\ref{table:singlets_pheno}). This is followed by the Higgs signal strengths on $\cS$. Finally, the Higgs self-coupling provides the weakest bounds questioning even the existence of a valid BSM interpretation of the corresponding SMEFT limit. A more detailed analysis of similar simplified models can be found in Section~7 of~\cite{Ellis:2020unq}. For the impact of the $m_W$ tension, ignored in our discussion, see~\cite{Bagnaschi:2022whn}.

\begin{table}[t]
\centering
\begin{minipage}[t]{0.45\linewidth}
\begin{adjustbox}{valign=t}
\begin{tabular}{cccccc}
\toprule
\textbf{Field} & \textbf{Irrep} & \textbf{$\bm{M}$ [TeV]} \\
\midrule
$\Xi\sim(\bm1,\bm3)_0$
&$\bm1$&9.3
\\
$\Xi_1\sim(\bm1,\bm3)_1$
&$\bm1$&25.3
\\
\midrule
$\cB_1\sim(\bm1,\bm1)_1$
&$\bm1$&12.7
\\
$\cW_1\sim(\bm1,\bm3)_1$
&$\bm1$&3.3
\\
\bottomrule
\end{tabular}
\end{adjustbox}
\end{minipage}
\begin{minipage}[t]{0.45\linewidth}
\begin{adjustbox}{valign=t}
\begin{tabular}{cccccc}
\toprule
\textbf{Field} & \textbf{Irrep} & \textbf{$\bm{M}$ [TeV]}\\
\midrule
\multirow{1}{*}{$\varphi\sim(\bm1,\bm2)_{\frac{1}{2}}$}
&$\bm1$&0.5
\\[2pt]
$\Theta_1\sim(\bm1,\bm4)_{\frac{1}{2}}$
&$\bm1$&0.2
\\[2pt]
$\Theta_3\sim(\bm1,\bm4)_{\frac{3}{2}}$
&$\bm1$&0.3
\\
$\cS\sim(\bm1,\bm1)_0$
&$\bm1$&0.7\\
\bottomrule
\end{tabular}
\end{adjustbox}
\end{minipage}
\caption{\textbf{Oblique/Higgs (Class V)}: The first two columns indicate gauge and flavor representations of the new scalars and vectors (only $\cB_1$ and $\cW_1$). The third column shows the lower bounds at $95\%$ CL on the mediator masses (dimensionless couplings set to unity while dimensionful set to the mass) for each case. See Table~\ref{tab:singlets} for the linear combinations of the SMEFT operators generated and Appendices~\ref{app:sm1-spin0} and \ref{app:sm1-spin1} for more details.}
\label{table:singlets_pheno}
\end{table}


\begin{figure}[h!]
\includegraphics[width=\linewidth, height=21.5cm]{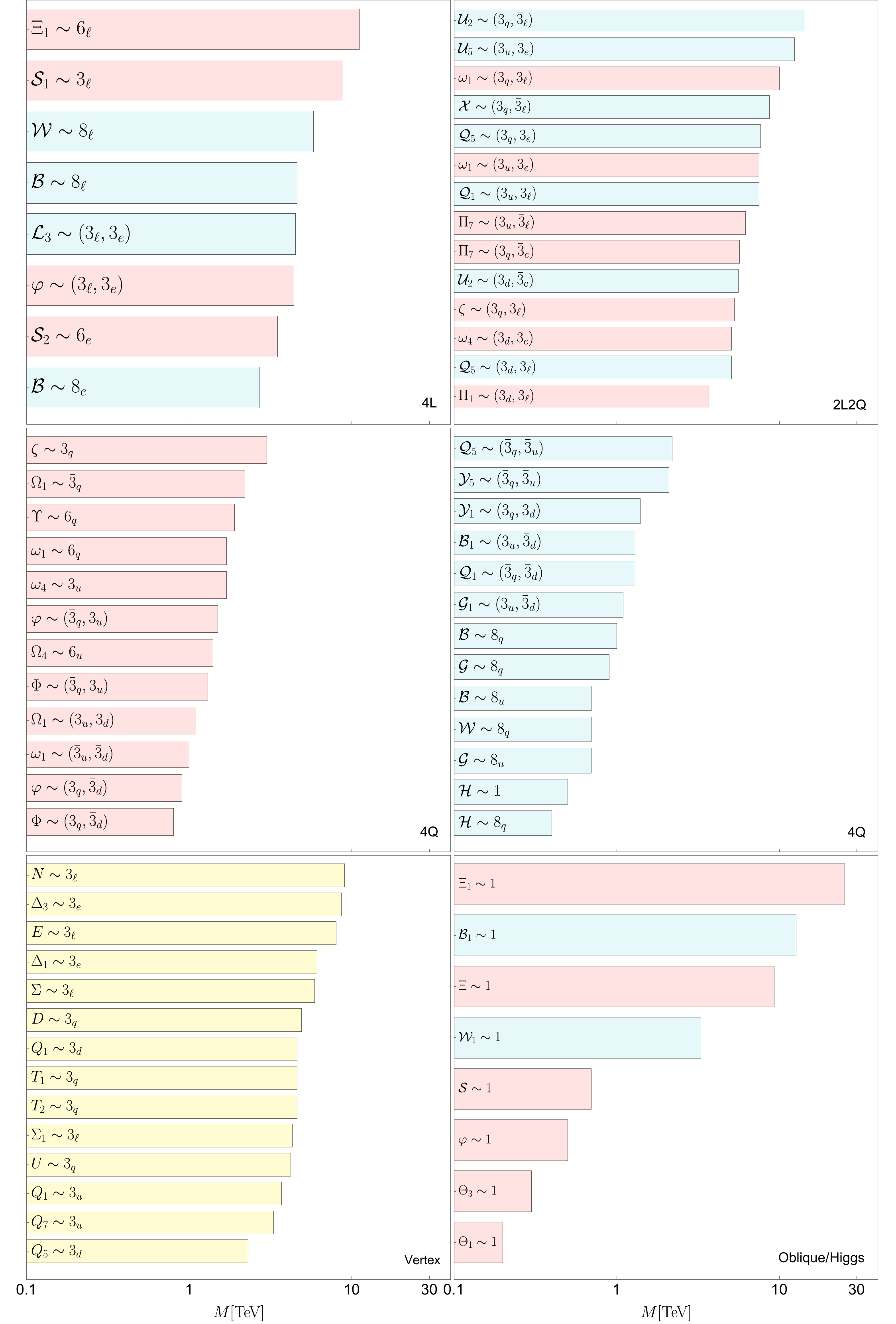}
\caption{Summary of the current experimental limits at $95\%$ CL from indirect searches (using global EFT fits) on flavor-blind new physics mediators, i.e., irreps of $U(3)^5$ flavor symmetry (see Section~\ref{sec:theory}), with spin 0 (red), $\frac{1}{2}$ (yellow) and 1 (blue), which integrate out to dimension-6 SMEFT at the tree level. See Section~\ref{sec:pheno} for details.}
\label{fig:plot}
\end{figure}

\clearpage

\section{Conclusions}
\label{sec:conc}

Among all endeavors in particle physics, the TeV scale exploration continues to be the most compelling in terms of potential for revolutionary discoveries. The LHC at CERN is collecting data at ever-increasing statistics paving the way for precision studies of the TeV dynamics. To this purpose, the SM effective field theory (SMEFT) offers a very powerful theoretical framework capable of describing various short-distance NP effects in a systematic and comprehensive manner. The idea is that a new heavy mediator, which has thus far eluded direct searches for new resonances, might induce subtle deviations from the SM processes captured by the higher-dimensional operators. 

One of the main obstacles to conducting comprehensive SMEFT analyses is a large number of independent parameters, with 2499 leading operators (baryon and lepton number conserving at dimension 6). A promising approach to organizing the vast parameter space of the SMEFT is provided by flavor symmetries, as we argued in~\cite{Greljo:2022cah}. Flavor symmetries lead to a considerable reduction in the number of independent parameters. However, the interplay between flavor symmetries and the TeV scale new physics is more than pragmatic. The stringent bounds from low-energy tests of flavor-changing neutral currents imply some form of an approximate flavor symmetry must be obeyed by the TeV-scale dynamics.

Building upon~\cite{Greljo:2022cah, deBlas:2017xtg}, in this work, we conduct a comprehensive study of all spin-0, $\frac{1}{2}$ and 1 new heavy mediator fields that integrate out at the tree-level to generate the dimension-6 SMEFT, demanding that their mass and interaction terms with the SM fields are fully invariant under $U(3)^5$ flavor symmetry. With this requirement at hand, we proceed by examining each NP mediator field and determining its possible transformations under the flavor group by classifying irreducible representations. In the subsequent step, we devote our attention to analyzing what happens once these flavor irreps are integrated out and matched onto the appropriate MFV basis (Appendix~\ref{app:basis}). Remarkably, in the vast majority of single-field extensions, the matching procedure generates a single Hermitian operator with a definite sign, which we conveniently identify as a leading direction. The full classification of the NP mediators is presented in Section~\ref{sec:theory}, more precisely in Tables~\ref{tab:sm1-spin0}--\ref{tab:singlets-except}, while the exhaustive results, along with the flavor structure of the couplings are presented in Appendix~\ref{app:sm1}.

The astounding simplicity of the leading directions opens doors toward numerical analysis, where we heavily rely on the well-established global SMEFT fits. In Section~\ref{sec:pheno}, we categorize the leading directions into five distinct phenomenologically motivated classes and put bounds on the corresponding WCs. The results are reported in Tables~\ref{table:4-lepton_pheno}--\ref{table:singlets_pheno} in Section~\ref{subsec:results}. The exclusion limits at $95\%$ CL on mediator masses assuming couplings set to unity are compactly presented in Figure~\ref{fig:plot}. This plot comprehensively summarises the EFT limits on flavor-blind short-distance new physics, which are complementary to those obtained from direct searches. The resulting lower limits on the mediator masses vary significantly on the gauge and flavor representations, and in most cases, sit in the $[1 - 10]$\,TeV range (for couplings set to unity) with a few notable exceptions. Figure~\ref{fig:plot} can serve as a benchmark for comparison with other short-distance new physics hypotheses featuring a more elaborate flavor structure. 

The analysis presented in the main part of this work assumes only one NP mediator field representation is active at a time. This assumption, however, can easily be relaxed. For most cases, adding more than one field is, essentially, a trivial sum of different contributions to the WCs at the tree level. Still, there are a few occasions when the introduction of a second or third NP field could have a non-trivial effect on the matching relations. Such combinations and their impacts on the SMEFT WCs have been outlined in Appendix~\ref{app:smN}. The flavor symmetry also serves to simplify this sector, suggesting that a non-trivial tree-level matching exists solely for the operator $\cO_\phi$.  Lastly, Appendix~\ref{app:nrint} includes the allowed dimension-5 interactions of the NP mediators with the SM fields and their matching contribution to the WCs, enlarging the scope beyond renormalizable UV completions.

In winding up, let us emphasize that the approach synthesizing the bottom-up model building and flavor symmetries opens a plethora of interesting directions for future work. A natural extension would be to consider decomposing the obtained flavor representations under a smaller flavor symmetry such as $U(2)^5$ or any other proposed in~\cite{Greljo:2022cah}. Particularly instructive would be repeating the exercise for the third-family dominated interactions and comparing with Figure~\ref{fig:plot}. Even staying within the $U(3)^5$ case, one can consider going beyond the leading order in the MFV spurion expansion and studying the effects of the spurion insertions in the interaction terms. For example, new flavor representations of mediators fields are introduced at higher spurion orders where an interplay with low-energy flavor physics experiments is expected. An additional intriguing issue concerns the UV completions for the newly introduced vector fields, particularly for those with nontrivial flavor representations. Given that vector fields are either composite resonances or gauge bosons, we anticipate added complexity. The approximation of a single-field extension is an oversimplification for vectors. 

There are also several important aspects on the phenomenological side worth mentioning, which could serve as promising future directions. As already pointed out in Section~\ref{subsec:results}, including RGE~\cite{Alonso:2013hga, Jenkins:2013wua, Jenkins:2013zja, Machado:2022ozb}, one-loop effects~\cite{Guedes:2023azv, Cohen:2020qvb, Fuentes-Martin:2022jrf, Dawson:2022ewj, Carmona:2021xtq, Fuentes-Martin:2022vvu, Fuentes-Martin:2020udw, Dawson:2022bxd} and going beyond the dimension-6 operators~\cite{Ellis:2023zim, Corbett:2023qtg, Degrande:2023iob} would be of great importance for improving the phenomenological analyses behind the results presented in Figure~\ref{fig:plot}. In addition, a complementary study focusing on the direct resonance searches for all proposed single-field extensions would allow us to compare against the EFT bounds obtained in this work. Lastly, as indicated in Section~\ref{sec:pheno}, proper inclusion of the dijet data in the SMEFT global likelihood toolkit would be of major relevance.

\section*{Acknowledgements}

We thank Mart\'in Gonz\'alez-Alonso for providing us with the correlation matrix derived in Ref.~\cite{Breso-Pla:2023tnz}. We also thank Nud\v zeim Selimovi\'c, Aleks Smolkovi\v c, Jakub \v Salko and Anders Eller Thomsen for useful discussions.  This work received funding from the Swiss National Science Foundation (SNF) through the Eccellenza Professorial Fellowship ``Flavor Physics at the High Energy Frontier'' project number 186866. AG is also partially supported by the European Research Council (ERC) under the European Union’s Horizon 2020 research and innovation program, grant agreement 833280 (FLAY).


\appendix

\section{Notation}
\label{app:notation}

Here we summarize the notation used throughout the paper. It is closely aligned to the notation used in Ref.~\cite{deBlas:2017xtg}, especially regarding the SM and BSM Lagrangians, yet there are some minor differences we point out. We first give some general remarks and then introduce the notation relevant to the gauge and flavor sectors separately.

\subsubsection*{General remarks}
\begin{itemize}
    \item The SM fermions are denoted by $q \sim (\bm3,\bm2)_{\frac{1}{6}}$, $\ell \sim (\bm1,\bm2)_{-\frac{1}{2}}$, $u \sim (\bm3,\bm1)_{\frac{2}{3}}$, $d \sim (\bm3,\bm1)_{-\frac{1}{3}}$, $e\sim (\bm1,\bm1)_{-1}$ and the Higgs field by $\phi\sim(\bm1,\bm2)_{\frac{1}{2}}$.\footnote{$\tilde\phi=i\sigma_2\phi^*\sim(\bm1,\bm2)_{-\frac{1}{2}}$.} Weak doublets (singlets) are (left) right chiral fermions. The chirality label of the SM fields is suppressed. Representations of BSM fields under the SM gauge group are denoted in the same format, i.e. $X_{\text{NP}}\sim(SU(3)_C,SU(2)_L)_{U(1)_Y}$. 
    \item Letter $c$ in the superscript ($\Box^c$) denotes charge conjugation.
    \item $\sigma_{\mu\nu}=\frac{i}{2}[\gamma_\mu,\gamma_\nu]$ and the dual tensor is denoted by $\tilde A_{\mu\nu}=\frac{1}{2}\varepsilon_{\mu\nu\rho\sigma}A^{\rho\sigma}$. Derivatives appearing in the $\psi^2\phi^2 D$ class of SMEFT operators are defined as $\overset{\text{\footnotesize$\leftrightarrow$}}{D}_\mu= D_\mu-\overset{\text{\footnotesize$\leftarrow$}}{D}_\mu$ and $\overset{\text{\footnotesize$\leftrightarrow$}}{D^a_\mu}=\sigma_a D_\mu-\overset{\text{\footnotesize$\leftarrow$}}{D}_\mu \sigma_a $ where $\sigma_a$ are the Pauli matrices.
\end{itemize}

\subsubsection*{Gauge sector}
\begin{itemize}
    \item $\alpha,\beta,\gamma=1,2,3$ ($\Box^{\alpha,\beta,\gamma}$) denote the $SU(3)_C$ indices of the fundamental representation and $\varepsilon_{\alpha\beta\gamma}$ is the totally antisymmetric tensor in color space. $T_A=\frac{1}{2}\lambda_A$ are generators in the fundamental representation of $SU(3)_C$, with $\lambda_A$ being the Gell-Mann matrices and $f_{ABC}$ the structure constants ($A=1,\ldots,8$). Interaction Lagrangians for color sextet fields employ the symmetric product denoted by $\psi_1^{(\alpha|}\psi_2^{|\beta)}=\frac{1}{2}(\psi_1^\alpha\psi_2^\beta+\psi_1^\beta\psi_2^\alpha)$.
    \item The $SU(2)_L$ triplet components are denoted by $a,b,c=1,2,3$ ($\Box^{a,b,c}$), while $I,J,K=\frac{3}{2},\frac{1}{2},-\frac{1}{2},-\frac{3}{2}$ denote the components of the quadruplets ($\Box^{I,J,K}$). When it comes to contractions of $SU(2)_L$ indices, $f_{abc}=\frac{i}{\sqrt2}\varepsilon_{abc}$ is used to obtain the isospin-1 product out of two triplets. Also, taking a product of a doublet with an isospin-1 field gives a quadruplet, where the matrices $C_{a\beta}^I$ are used and, similarly, the singlet can be obtained as a product of two quadruplets by means of $\varepsilon_{IJ}$.\footnote{Let us point out that in a very small subset of cases, where the quadruplet fields ($\Theta_1$ and $\Theta_3$) are present, Greek indices $\alpha$, $\beta$, $\gamma$ stand for the fundamental $SU(2)_L$ index (see interaction Lagrangians of $\Theta_1$ and $\Theta_3$ in Appendix~\ref{app:sm1-spin0-singlets} and \ref{app:smN}). In all other cases, the contractions of the fundamental $SU(2)_L$ indices (either direct or through $i\sigma_2$) are suppressed.} We refrain from presenting the full expressions for $C_{a\beta}^I$ and $\varepsilon_{IJ}$ matrices, which can be found in Appendix A of Ref.~\cite{deBlas:2017xtg}. Letter $T$ in the superscript ($\Box^T$) indicates that the $SU(2)_L$ indices of a given field have been transposed. 
\end{itemize}

\subsubsection*{Flavor sector}
\begin{itemize}
    \item In the interaction Lagrangian for the BSM fields before assuming $U(3)^5$ flavor symmetry, the generations are denoted by $i,j,r$. Once the flavor symmetry is assumed, the flavor indices present in the definitions of the flavor irreps are of the form $i_f,j_f,k_f,l_f$, where $f=\lzv q,u,d,\ell,e \dzv$ serves as an additional specifier, and it denotes which $U(3)_f$ group is being referred to. 
    \item Fields transforming as singlets under the flavor group carry no flavor indices. Fundamentals ($\bm3_f$) and anti-fundamentals ($\bar{\bm3}_f$) of $U(3)_f$ are designated by one upper ($\Box^{i_f}$) and one lower flavor index ($\Box_{i_f}$), respectively. Likewise, fields transforming as $U(3)_{f_1}\times U(3)_{f_2}$ bifundamentals are labeled in an analogous way: $(\bm3_{f_1},\bm3_{f_2})$ as $\Box^{i_{f_1}j_{f_2}}$, $(\bar{\bm3}_{f_1},\bar{\bm3}_{f_2})$ as $\Box_{i_{f_1}j_{f_2}}$, $(\bar{\bm3}_{f_1},\bm3_{f_2})$ as $\Box^{j_{f_2}}_{i_{f_1}}$ and $(\bm3_{f_1},\bar{\bm3}_{f_2})$ as $\Box^{i_{f_1}}_{j_{f_2}}$. $U(3)_f$ sextets ($\bm6_f$) are denoted as $\Box^{i_f j_f}$, while antisextets ($\bar{\bm6}_f$) as $\Box_{i_f j_f}$ and, lastly, flavor octets ($\bm8_f$) as $\Box\ud{i_f}{j_f}$. 
    \item The contractions of flavor indices in the interaction Lagrangian of the NP fields can be carried out in several ways depending on which irreps enter the interaction, and this analysis directly determines the flavor structure of the couplings. For NP fields transforming trivially (singlets) under the $U(3)^5$, the coupling tensor is either given as a single parameter without any flavor indices (scalars and vectors) or is proportional to $\delta^{i_f}_{j_f}$ (vectors). For $U(3)^5$ triplets, the coupling is either proportional to $\varepsilon^{i_f j_f k_f}$ (scalars) or $\delta^{i_f}_{j_f}$ (fermions). The $U(3)^5$ bifundamental representations (scalars and vectors) have coupling tensors proportional to $\delta^{i_{f_1}}_{j_{f_1}}\delta^{k_{f_2}}_{l_{f_2}}$. Scalar sextets are, au contraire, characterized by the coupling tensor, which is proportional to $\delta^{i_f}_{j_f}\delta^{k_f}_{l_f}+\delta^{i_f}_{l_f}\delta^{k_f}_{j_f}$. This structure originates from the symmetrization over the flavor indices, i.e., from the terms of the form $\psi_1^{(i_f|}\psi_2^{|j_f)}$, which are entering the interaction Lagrangian of the flavor sextets. Ultimately, the coupling tensor of the vector octets is proportional to $(T^{A_f})\ud{i_f}{j_f}$. See Appendix~\ref{app:sm1} for more concrete examples.  
    \item Along with designating the flavor irreps under $U(3)^5$ and the coupling tensors, this notation has also been applied to the Yukawa matrices ($y_{u,d,e}$) appearing in the matching (Appendix~\ref{app:sm1}) and, additionally, to the SMEFT operators arising at $\cO(y_{u,d,e})$ (see Appendix~\ref{app:basis}). Notice that for the two- and four-fermion operators in the MFV basis arising at the leading order in spurion expansion, the flavor indices are taken to be $i,j$, without an additional specifier $f$, since the contractions are fully apparent.     
\end{itemize}


\clearpage
\section{The MFV basis}
\label{app:basis}

Starting from the SMEFT dimension-6 operators in the Warsaw bases~\cite{Grzadkowski:2010es} and imposing $U(3)^5$ flavor symmetry minimally broken by the SM Yukawa interactions treated as spurions, one derives \textit{the MFV basis} order by order in the spurion counting~\cite{Faroughy:2020ina,Greljo:2022cah}. The leading order terms establish the \textit{flavor-symmetric} basis, which is explicitly formulated in the first two tables presented below. It consists of 41 CP even and 6 CP odd operators; see Table~1 of~\cite{Greljo:2022cah}.  The summation over flavor indices $i, j = 1, 2, 3$ is assumed. The abbreviations $D$ and $E$, denoting direct and exchange, respectively, are employed to indicate whether the flavor contraction occurs within the brackets or not. The third table displays the first correction at the linear order in the spurion expansion. 

Operators labeled as \textit{loop-generated} lack a tree-level completion within a renormalizable UV theory. These belong to categories $X^3$, $X^2 \phi^2$, and $\psi^2 X \phi$ and play no role in this work.

\subsubsection*{Operators involving fermions}

\begin{table}[h]
\centering
\begin{tabular}{ccccc}
\toprule
\textbf{Class} & \textbf{Label} & \textbf{Operator} & \textbf{Label} &\textbf{Operator} \\
\midrule
\multirow{4}{*}{\rule{0pt}{3ex}$(\bar L L)(\bar L L)$}&$\cO_{\ell\ell}^D$&$(\bar\ell_{i}\gamma^\mu\ell^{i})(\bar\ell_{j}\gamma_\mu\ell^{j})$&$\cO_{\ell q}^{(1)}$&$(\bar\ell_{i}\gamma^\mu \ell^{i})(\bar q_{j}\gamma_\mu q^{j})$\\
&\rule{0pt}{2.8ex}$\cO_{\ell\ell}^E$&$(\bar\ell_{i}\gamma^\mu\ell^{j})(\bar\ell_{j}\gamma_\mu\ell^{i})$&$\cO_{\ell q}^{(3)}$&$(\bar\ell_{i}\gamma^\mu\sig^a \ell^{i})(\bar q_{j}\gamma_\mu\sig^a q^{j})$\\
&\rule{0pt}{2.8ex}$\cO_{qq}^{(1)D}$&$(\bar q_{i}\gamma^\mu q^{i})(\bar q_{j}\gamma_\mu q^{j})$&$\cO_{qq}^{(3)D}$&$(\bar q_{i}\gamma^\mu \sig^a q^{i})(\bar q_{j}\gamma_\mu \sig^a q^{j})$\\
&\rule{0pt}{2.8ex}$\cO_{qq}^{(1)E}$&$(\bar q_{i}\gamma^\mu q^{j})(\bar q_{j}\gamma_\mu q^{i})$& $\cO_{qq}^{(3)E}$&$(\bar q_{i}\gamma^\mu \sig^a q^{j})(\bar q_{j}\gamma_\mu \sig^a q^{i})$\\
\midrule
\multirow{5}{*}{\rule{0pt}{4.6ex}$(\bar R R)(\bar R R)$} &\rule{0pt}{2.2ex}$\cO_{ee}$             & $(\bar e_{i} \gamma^\mu e^{i})(\bar e_{j} \gamma_\mu e^{j})$                   & $\cO_{dd}^D$         & $(\bar d_{i}\gamma^\mu d^{i})(\bar d_{j}\gamma_\mu d^{j})$                        \\
&\rule{0pt}{2.8ex}$\cO_{uu}^D$           & $(\bar u_{i}\gamma^\mu u^{i})(\bar u_{j}\gamma_\mu u^{j})$                     & $\cO_{dd}^E$         & $(\bar d_{i}\gamma^\mu d^{j})(\bar d_{j}\gamma_\mu d^{i})$                        
\\
&\rule{0pt}{2.8ex}$\cO_{uu}^E$           & $(\bar u_{i}\gamma^\mu u^{j})(\bar u_{j}\gamma_\mu u^{i})$                     & $\cO_{ud}^{(1)}$     & $(\bar u_{i}\gamma^{\mu} u^{i})(\bar d_{j}\gamma_{\mu}  d^{j})$                   
\\
&\rule{0pt}{2.8ex}$\cO_{eu}$             & $(\bar e_{i}\gamma^\mu e^{i})(\bar u_{j}\gamma_\mu u^{j})$                     & $\cO_{ud}^{(8)}$     & $(\bar u_{i}\gamma^{\mu}T^A u^{i})(\bar d_{j}\gamma_{\mu} T^A d^{j})$             
\\
&\rule{0pt}{2.8ex}$\cO_{ed}$             & $(\bar e_{i}\gamma^\mu e^{i})(\bar d_{j}\gamma_\mu d^{j})$                     &                      &                                                                                           
\\
\midrule
\multirow{4}{*}{\rule{0pt}{4.2ex}$(\bar LL)(\bar RR)$}   &$\cO_{\ell e}$         & $(\bar\ell_{i}\gamma^\mu \ell^{i})(\bar e_{j}\gamma_\mu e^{j})$          & $\cO_{qu}^{(1)}$     & $(\bar q_{i}\gamma^\mu q^{i})(\bar u_{j}\gamma_\mu u^{j})$                        
\\
&\rule{0pt}{2.8ex}$\cO_{q e}$            & $(\bar q_{i}\gamma^\mu q^{i})(\bar e_{j}\gamma_\mu e^{j})$                     & $\cO_{qu}^{(8)}$     & $(\bar q_{i}\gamma^\mu T^A q^{i})(\bar u_{j}\gamma_\mu T^A u^{j})$                
\\
&\rule{0pt}{2.8ex}$\cO_{\ell u}$         & $(\bar \ell_{i}\gamma^\mu \ell^{i})(\bar u_{j}\gamma_\mu u^{j})$         & $\cO_{qd}^{(1)}$     & $(\bar q_{i}\gamma^\mu q^{i})(\bar d_{j}\gamma_\mu d^{j})$                        
\\
&\rule{0pt}{2.8ex}$\cO_{\ell d}$         & $(\bar \ell_{i}\gamma^\mu \ell^{i})(\bar d_{j}\gamma_\mu d^{j})$         & $\cO_{qd}^{(8)}$     & $(\bar q_{i}\gamma^\mu T^A q^{i})(\bar d_{j}\gamma_\mu T^A d^{j})$                
\\
\midrule
\multirow{4}{*}{\rule{0pt}{4.2ex}$\psi^2\phi^2 D$}       &$\cO_{\phi\ell}^{(1)}$ & $(\phi^\dag i\overset{\text{\footnotesize$\leftrightarrow$}}{D}_\mu \phi)(\bar\ell_i\gamma^\mu\ell^i)$                 & $\cO_{\phi e}$       & $(\phi^\dag i\overset{\text{\footnotesize$\leftrightarrow$}}{D}_\mu \phi)(\bar e_i\gamma^\mu e^i)$                        
\\
&$\cO_{\phi\ell}^{(3)}$ & $(\phi^\dag i\overset{\text{\footnotesize$\leftrightarrow$}}{D^a_\mu} \phi)(\bar\ell_i\gamma^\mu\sig^a\ell^i)$         & $\cO_{\phi u}$       & $(\phi^\dag i \overset{\text{\footnotesize$\leftrightarrow$}}{D}_\mu \phi)(\bar u_i\gamma^\mu u^i)$                        \\
&$\cO_{\phi q}^{(1)}$   & $(\phi^\dag i\overset{\text{\footnotesize$\leftrightarrow$}}{D}_\mu \phi)(\bar q_i\gamma^\mu q^i)$                     & $\cO_{\phi d}$       & $(\phi^\dag i\overset{\text{\footnotesize$\leftrightarrow$}}{D}_\mu \phi)(\bar d_i\gamma^\mu d^i)$                        \\
&$\cO_{\phi q}^{(3)}$   & $(\phi^\dag i\overset{\text{\footnotesize$\leftrightarrow$}}{D^a_\mu} \phi)(\bar q_i\gamma^\mu\sig^a q^i)$             &                      &                                                                                           
\\[0.11cm]
\bottomrule
\end{tabular}
\end{table}

\clearpage
\subsubsection*{Bosonic only operators}

\begin{table}[h]
\centering
\begin{tabular}{ccccc}
\toprule
\textbf{Class} & \textbf{Label} & \textbf{Operator} & \textbf{Label} &\textbf{Operator} \\
\midrule
\multirow{2}{*}{$\underset{{\scalebox{0.7}{\text{Loop generated}}}}{X^3}$}       &$\cO_W$               &$\varepsilon_{abc} W_\mu ^{a\,\nu}W_\nu^{b\,\rho}W_\rho^{c\,\mu}$        & $\cO_G$                & $f_{ABC} G_\mu ^{A\,\nu}G_\nu^{B\,\rho}G_\rho^{C\,\mu}$        \\
&\rule{0pt}{2.8ex}$\cO_{\tilde W}$      & $\varepsilon_{abc} \tilde W_\mu ^{a\,\nu}W_\nu^{b\,\rho}W_\rho^{c\,\mu}$ & $\cO_{\tilde G}$       & $f_{ABC} \tilde G_\mu ^{A\,\nu}G_\nu^{B\,\rho}G_\rho^{C\,\mu}$ 
\\
\midrule
$\phi^6$                     &$\cO_\phi$            & $(\phi^\dag\phi)^3$                                                      &                        &                                                             \\
\midrule
 $\phi^4 D^2$                 &$\cO_{\phi\Box}$      & $(\phi^\dag\phi)\Box(\phi^\dag\phi)$                                     & $\cO_{\phi D}$         & $(\phi^\dag D_\mu \phi)[(D^\mu\phi)^\dag\phi]$             \\
\midrule
\multirow{4}{*}{\rule{0pt}{4.5ex}$\underset{{\scalebox{0.7}{\text{Loop generated}}}}{X^2\phi^2}$} &$\cO_{\phi B}$        & $(\phi^\dag\phi)B_{\mu\nu}B^{\mu\nu}$                                    & $\cO_{\phi WB}$        & $(\phi^\dag\sig^a\phi)W^a_{\mu\nu}B^{\mu\nu}$                  
\\
&\rule{0pt}{2.8ex}$\cO_{\phi \tilde B}$ & $(\phi^\dag\phi)\tilde B_{\mu\nu}B^{\mu\nu}$                             & $\cO_{\phi \tilde WB}$ & $(\phi^\dag\sig^a\phi)\tilde W^a_{\mu\nu}B^{\mu\nu}$           
\\
&\rule{0pt}{2.8ex}$\cO_{\phi W}$        & $(\phi^\dag\phi)W^a_{\mu\nu}W^{a\,\mu\nu}$                               & $\cO_{\phi G}$         & $(\phi^\dag \phi)G^A_{\mu\nu}G^{A\,\mu\nu}$                    
\\
&\rule{0pt}{2.8ex}$\cO_{\phi \tilde W}$ & $(\phi^\dag\phi)\tilde W^a_{\mu\nu}W^{a\,\mu\nu}$                        & $\cO_{\phi \tilde G}$  & $(\phi^\dag \phi)\tilde G^A_{\mu\nu}G^{A\,\mu\nu}$ 
\\[0.11cm]
\bottomrule
\end{tabular}
\end{table}

\subsubsection*{Operators at $\cO(y_{u,d,e})$}

\begin{table}[h]
\centering
\begin{tabular}{ccccc}
\toprule
\textbf{Class} & \textbf{Label} & \textbf{Operator} & \textbf{Label} &\textbf{Operator} \\
\midrule
\multirow{2}{*}{$\psi^2\phi^3$} &\rule{0pt}{2ex}    
$[\cO_{e\phi}]_{i_\ell}^{j_e}$\phantom{a} & $(\phi^\dag\phi)(\bar\ell_{i_\ell}\phi e^{j_e})$           &                             &                                                                    
\\
&\rule{0pt}{2.8ex}$[\cO_{d\phi}]^{j_d}_{i_q}$    & $(\phi^\dag\phi)(\bar q_{i_q}\phi d^{j_d})$                & $[\cO_{u\phi}]^{j_u}_{i_q}$ & $(\phi^\dag\phi)(\bar q_{i_q}\tilde\phi u^{j_u})$ 
\\
\midrule
\multirow{4}{*}{\rule{0pt}{4.4ex}$\underset{{\scalebox{0.7}{\text{Loop generated}}}}{\psi^2 X\phi}$} &$[\cO_{eB}]^{j_e}_{i_\ell}$    & $(\bar\ell_{i_\ell}\sig^{\mu\nu}e^{j_e})\phi B_{\mu\nu}$   & $[\cO_{eW}]^{j_e}_{i_\ell}$ & $(\bar\ell_{i_\ell}\sig^{\mu\nu}e^{j_e})\sig^a \phi W^a_{\mu\nu}$  
\\
&\rule{0pt}{2.8ex}$[\cO_{uB}]^{j_u}_{i_q}$       & $(\bar q_{i_q}\sig^{\mu\nu} u^{j_u})\tilde\phi B_{\mu\nu}$ & $[\cO_{uW}]^{j_u}_{i_q}$    & $(\bar q_{i_q}\sig^{\mu\nu}u^{j_u})\sig^a \tilde\phi W^a_{\mu\nu}$ 
\\
&\rule{0pt}{2.8ex}$[\cO_{dB}]^{j_d}_{i_q}$       & $(\bar q_{i_q}\sig^{\mu\nu} d^{j_d})\phi B_{\mu\nu}$       & $[\cO_{dW}]^{j_d}_{i_q}$    & $(\bar q_{i_q}\sig^{\mu\nu}d^{j_d})\sig^a \phi W^a_{\mu\nu}$       
\\
&\rule{0pt}{2.8ex}$[\cO_{uG}]^{j_u}_{i_q}$       & $(\bar q_{i_q} \sig^{\mu\nu} T^A u^{j_u})\tilde\phi G^A_{\mu\nu}$    & $[\cO_{dG}]^{j_d}_{i_q}$    & $(\bar q_{i_q}\sig^{\mu\nu}T^Ad^{j_d}) \phi G^A_{\mu\nu}$ 
\\[0.11cm]
\bottomrule
\end{tabular}
\end{table}



\section{Matching: SM + 1 field}
\label{app:sm1}

This Appendix provides an exhaustive list of extending the SM by a single Lorentz, gauge, and $U(3)^5$ flavor representation, resulting in a dimension-6 operator in the SMEFT at the tree level. Presuming that an underlying UV theory is both perturbative and renormalizable, we construct the Lagrangian for each field up to a mass dimension of $\le 4$. This formulation excludes operators of the type $\mathcal{L}\supset V_\mu D^\mu S$, as elaborated in Section~\ref{sec:theory}. Following this, we identify possible flavor representations that determine the coupling tensors. Ultimately, we derive the tree-level matching contributions to the dimension-6 SMEFT for each representation. The organization of this Appendix is based on spin (0, $\frac{1}{2}$ and 1) and color (singlets, triplets, sextets, and octets).

\subsection{Spin 0}
\label{app:sm1-spin0}

\subsubsection{Color singlets}
\label{app:sm1-spin0-singlets}

\subsubsection*{$\boldsymbol{\mathcal S\sim(1,1)_0}$}
\begin{itemize}
	\item Lagrangian:
	\begin{equation}
		\begin{alignedat}{2}
			-\cL_S^{(\le4)}&\supset 
		(\kappa_\cS)_r\cS_r\phi^\dag\phi
		+(\lambda_\cS)_{rs}\cS_r\cS_s\phi^\dag\phi
		+(\kappa_{\cS^3})_{rst}\cS_r\cS_s\cS_t
		.
		\nonumber
		\end{alignedat}
	\end{equation}
	\item Flavor irreps:
	\begin{alignat}{3}
		S&\sim\bm1:
		&\qquad
		\kappa_\cS,\quad\lambda_\cS,\quad\kappa_{\cS^3}.\nonumber
	\end{alignat}
	\item Matching:
			\begin{equation}
		\begin{alignedat}{2}
			\cL_{\text{SMEFT}}&\supset
			-\frac{\kappa_\cS^2}{2M_\cS^4}\cO_{\phi\Box}
			+\frac{\kappa_\cS^2}{2M_\cS^4}\lzm \frac{2\kappa_{\cS^3}\kappa_\cS}{M_\cS^2} -2\lambda_\cS \dzm\cO_\phi.
		\nonumber
		\end{alignedat}
	\end{equation}
\end{itemize}

\subsubsection*{$\boldsymbol{\mathcal S_1\sim(1,1)_1}$}
\begin{itemize}
	\item Lagrangian:
	\begin{equation}
		-\cL_S^{(\le4)}\supset (y_{\mathcal S_1})_{rij}\mathcal S_{1r}^\dag \bar\ell_{i}i\sig_2\ell_{j}^c+\hermc.\nonumber
	\end{equation}
	\item Flavor irreps:
	\begin{alignat}{3}
		S_1^{i_\ell}&\sim\bm3_\ell:
		&\qquad
		[y_{\cS_1}]^{i_\ell j_\ell k_\ell}&=y_{\cS_1}\varepsilon^{i_\ell j_\ell k_\ell}.\nonumber
	\end{alignat}
	\item Matching:
	\begin{equation}
		\cL_{\text{SMEFT}}\supset \frac{|y_{\cS_1}|^2}{M_{\cS_1}^2}\lzm\cO_{\ell\ell}^D-\cO_{\ell\ell}^E \dzm.
		\nonumber
	\end{equation}
\end{itemize}

\subsubsection*{$\boldsymbol{\mathcal S_2\sim(1,1)_2}$}
\begin{itemize}
	\item Lagrangian:
	\begin{equation}
		-\cL_S^{(\le4)}\supset (y_{\mathcal S_2})_{rij}\mathcal S_{2r}^\dag \bar e_{i}e_{j}^c+\hermc.\nonumber
	\end{equation}
	\item Flavor irreps:
	\begin{alignat}{3}
		(S_{2})_{i_e j_e}&\sim\bar{\bm6}_e:
		&\qquad [y_{\cS_2}]^{i_e k_e}_{j_e l_e}&=\frac{y_{\cS_2}}{2}\lzm \delta^{i_e}_{j_e}\delta^{k_e}_{l_e}+\delta^{i_e}_{l_e}\delta^{k_e}_{j_e} \dzm.\nonumber
	\end{alignat}
	\item Matching:
	\begin{equation}
		\begin{alignedat}{2}
			\cL_{\text{SMEFT}}&\supset \frac{\lzu y_{\cS_2} \dzu^2}{2M_{\cS_2}^2}\cO_{ee}.
		\nonumber
		\end{alignedat}
	\end{equation}
\end{itemize}

\subsubsection*{$\boldsymbol{\varphi\sim(1,2)_{\frac{1}{2}}}$}
\begin{itemize}
	\item Lagrangian:
	\begin{equation}
		-\cL_S^{(\le4)}\supset
		(y^e_\varphi)_{rij}\varphi^\dag_r\bar e_{i}\ell_{j}
		+
		(y^d_\varphi)_{rij}\varphi^\dag_r\bar d_{i}q_{j}
		+
		(y^u_\varphi)_{rij}\varphi^\dag_ri\sig_2\bar q^T_{i}u_{j}
		+
		(\lambda_\varphi)_r(\varphi_r^\dag\phi)(\phi^\dag\phi)
		+\hermc.\nonumber
	\end{equation}
	\item Flavor irreps:
	\begin{alignat}{3}
		&\varphi^{i_\ell}_{j_e}\sim(\bar{\bm3}_e,\bm3_\ell):
		&\qquad
		&[y_\varphi^e]^{i_ek_\ell}_{j_el_\ell}=y_{\varphi}^e\delta^{i_e}_{j_e}\delta^{k_\ell}_{l_\ell},\nonumber
		\\
		&\varphi^{i_q}_{j_d}\sim(\bar{\bm3}_d,\bm3_q):
		&\qquad 
		&[y_\varphi^d]^{i_d k_q}_{j_dl_q}=y_{\varphi}^d\delta^{i_d}_{j_d}\delta^{k_q}_{l_q},\nonumber
		\\
		&\varphi^{i_u}_{j_q}\sim(\bar{\bm3}_q,\bm3_u):
		&\qquad 
		&[y_\varphi^u]^{i_q k_u}_{j_ql_u}=y_{\varphi}^u\delta^{i_q}_{j_q}\delta^{k_u}_{l_u},\nonumber
		\\
		&\varphi \sim\bm1:
		&\qquad &\lambda_\varphi.\nonumber
	\end{alignat}
\item Matching:
\begin{itemize}
\item $\varphi^{i_\ell}_{j_e}\sim(\bar{\bm3}_e,\bm3_\ell)$:
		\begin{equation}
		\cL_{\text{SMEFT}}\supset 
		-\frac{|y_\varphi^e|^2}{2M_\varphi^2}\cO_{\ell e}.
		\nonumber
	\end{equation}
\item $\varphi^{i_q}_{j_d}\sim(\bar{\bm3}_d,\bm3_q)$:
		\begin{equation}
		\cL_{\text{SMEFT}}\supset 
		-\frac{|y_\varphi^d|^2}{6M_\varphi^2}\lzm \cO_{qd}^{(1)}+6\cO_{qd}^{(8)} \dzm.
		\nonumber
	\end{equation}
\item $\varphi^{i_u}_{j_q}\sim(\bar{\bm3}_q,\bm3_u)$:
		\begin{equation}
		\cL_{\text{SMEFT}}\supset 
		-\frac{|y_\varphi^u|^2}{6M_\varphi^2}\lzm \cO_{qu}^{(1)}+6\cO_{qu}^{(8)} \dzm.
		\nonumber
	\end{equation}
\item $\varphi\sim\bm1$:
		\begin{equation}
		\begin{alignedat}{2}
			\cL_{\text{SMEFT}}&\supset
			\frac{|\lambda_\varphi|^2}{M_\varphi^2}\cO_\phi. 
		\nonumber
		\end{alignedat}
	\end{equation}
\end{itemize}
\end{itemize}

\subsubsection*{$\boldsymbol{\Xi\sim(1,3)_{0}}$}
\begin{itemize}
	\item Lagrangian:
	\begin{equation}
	\begin{alignedat}{2}
		-\cL_S^{(\le4)}&\supset 
		(\kappa_\Xi)_r\phi^\dag \Xi_r^a\sig^a\phi
		+(\lambda_\Xi)_{rs}(\Xi^a_r\Xi^a_s)(\phi^\dag\phi).
		\nonumber
	\end{alignedat}
\end{equation}
	\item Flavor irreps:
	\begin{alignat}{3}
		\Xi &\sim\bm1:
		&\qquad
		\kappa_\Xi,\quad\lambda_\Xi		\nonumber.
	\end{alignat}
\item Matching:
		\begin{equation}
		\begin{alignedat}{2}
			\cL_{\text{SMEFT}}&\supset 
			\frac{\kappa_\Xi^2}{2M_\Xi^4}\lzm -4\cO_{\phi D}+\cO_{\phi\Box} \dzm
			+\frac{\kappa_\Xi^2}{2M_\Xi^4}\lzs 8(\lambda_\phi+C^\Xi_{\phi4})-2\lambda_\Xi\dzs \cO_\phi
			\\&
			+\frac{\kappa_\Xi^2}{M_\Xi^4} \Bigg\{ [ y_e^*]^{j_\ell}_{i_e}[\cO_{e\phi}]^{i_e}_{j_\ell}+[ y_d^*]^{j_q}_{i_d}[\cO_{d\phi}]^{i_d}_{j_q}+[ y_u^*]^{j_q}_{i_u}[\cO_{u\phi}]^{i_u}_{j_q} +\hermc \Bigg\},
		\nonumber
		\end{alignedat}
	\end{equation}
	where 
	\begin{equation}
		C_{\phi4}^\Xi=\frac{\kappa_\Xi^2}{2M_\Xi^2}-\frac{2\mu^2_\phi \kappa_\Xi^2}{M_\Xi^4}.\nonumber
	\end{equation}
\end{itemize}

\subsubsection*{$\boldsymbol{\Xi_1\sim(1,3)_{1}}$}
\begin{itemize}
	\item Lagrangian:
	\begin{equation}
		\begin{alignedat}{2}
			-\cL_S^{(\le4)}&\supset
			\frac{1}{2}(\lambda_{\Xi_1})_{rs}(\Xi_{1r}^{a\dag}\Xi_{1s}^a)(\phi^\dag\phi)
		+\frac{1}{2}(\lambda'_{\Xi_1})_{rs}f_{abc}(\Xi_{1r}^{a\dag}\Xi_{1s}^b)(\phi^\dag\sig^c\phi)
		\\&
		+\lzs (\kappa_{\Xi_1})_r \Xi_{1r}^{a\dag}(\tilde\phi^\dag\sig^a\phi)+
		(y_{\Xi_1})_{rij}\Xi_{1r}^{a\dag}\bar\ell_{i}\sig^ai\sig_2\ell_{j}^c+\hermc \dzs.
			\nonumber
		\end{alignedat}
	\end{equation}
	\item Flavor irreps:
		\begin{alignat}{3}
		&(\Xi_{1})_{i_\ell j_\ell} 
		\sim\bar{\bm6}_\ell:
		&\qquad &[y_{\Xi_1}]^{i_\ell k_\ell}_{j_\ell l_\ell}=\frac{y_{\Xi_1}}{2}\lzm \delta^{i_\ell}_{j_\ell}\delta^{k_\ell}_{l_\ell}+\delta^{i_\ell}_{l_\ell}\delta^{k_\ell}_{j_\ell} \dzm,\nonumber
		\\
		&\Xi_{1}
		\sim\bm1:
		&\qquad &\lambda_{\Xi_1},\quad\lambda'_{\Xi_1},\quad\kappa_{\Xi_1}\nonumber.
		\nonumber
	\end{alignat}
	\item Matching:
	\begin{itemize}
	\item $(\Xi_{1})_{i_\ell j_\ell}\sim\bar{\bm6}_\ell$:
		\begin{equation}
		\cL_{\text{SMEFT}}\supset 
		\frac{\lzu y_{\Xi_1} \dzu^2}{2M_{\Xi_1}^2}\lzm \cO_{\ell\ell}^D+\cO_{\ell\ell}^E \dzm.
		\nonumber
	\end{equation}
	\item $\Xi_{1}\sim\bm1$:
		\begin{equation}
		\small
		\begin{alignedat}{2}
			\cL_{\text{SMEFT}}&\supset 
			\frac{\lzu \kappa_{\Xi_1} \dzu^2}{M_{\Xi_1}^4}\lzm 4\cO_{\phi D}+2\cO_{\phi\Box} \dzm
			+
			\frac{\lzu \kappa_{\Xi_1} \dzu^2}{M_{\Xi_1}^4}\lzs 8(\lambda_\phi+C^{\Xi_1}_{\phi4})
			-2\lambda_{\Xi_1}
			+\sqrt 2 \lambda'_{\Xi_1}\dzs \cO_{\phi}\\&
			+\frac{2\lzu \kappa_{\Xi_1} \dzu^2}{M_{\Xi_1}^4}\Bigg\{ [y_e^*]^{j_\ell}_{i_e}[\cO_{e\phi}]^{i_e}_{j_\ell}+[y_d^*]^{j_q}_{i_d}[\cO_{d\phi}]^{i_d}_{j_q}+[y_u^*]^{j_q}_{i_u}[\cO_{u\phi}]^{i_u}_{j_q} +\hermc \Bigg\},
		\nonumber
		\end{alignedat}
	\end{equation}
	where
	\begin{equation}
		C^{\Xi_1}_{\phi4}=\frac{2\lzu \kappa_{\Xi_1} \dzu^2}{M_{\Xi_1}^2}-\frac{4\mu_\phi^2 \lzu \kappa_{\Xi_1} \dzu^2}{M_{\Xi_1}^4}.\nonumber
	\end{equation}
	\end{itemize}
\end{itemize}

\subsubsection*{$\boldsymbol{\Theta_1\sim(1,4)_{\frac{1}{2}}}$}
\begin{itemize}
	\item Lagrangian:
	\begin{equation}
	\begin{alignedat}{2}
		-\cL_S^{(\le4)}&\supset 
		(\lambda_{\Theta_1})_r (\phi^\dag \sig^a \phi)C_{a\beta}^I\tilde\phi_\beta\varepsilon_{IJ}\Theta_{1r}^J +\hermc.
		\nonumber
	\end{alignedat}
\end{equation}
	\item Flavor irreps:
		\begin{alignat}{3}
		&\Theta_1
		\sim\bm1:
		&\qquad \lambda_{\Theta_1}
		\nonumber
		.
	\end{alignat}
\item Matching:
		\begin{equation}
		\cL_{\text{SMEFT}}\supset 
		\frac{\lzu \lambda_{\Theta_1} \dzu^2}{6M_{\Theta_1}^2} \cO_\phi.
		\nonumber
	\end{equation}
\end{itemize}

\subsubsection*{$\boldsymbol{\Theta_3\sim(1,4)_{\frac{3}{2}}}$}
\begin{itemize}
	\item Lagrangian:
	\begin{equation}
	\begin{alignedat}{2}
		-\cL_S^{(\le4)}&\supset 
		(\lambda_{\Theta_3})_r(\phi^\dag\sigma^a\tilde\phi)C_{a\beta}^I\tilde\phi_\beta \varepsilon_{IJ}\Theta_{3r}^J +\hermc.
		\nonumber
	\end{alignedat}
\end{equation}
	\item Flavor irreps:
		\begin{alignat}{3}
		&\Theta_3
		\sim\bm1:
		&\qquad \lambda_{\Theta_3}
		\nonumber
		.
	\end{alignat}
\item Matching:
		\begin{equation}
		\cL_{\text{SMEFT}}\supset
		\frac{\lzu \lambda_{\Theta_3} \dzu^2}{2M_{\Theta_3}^2}\cO_\phi .
		\nonumber
	\end{equation}
\end{itemize}

\subsubsection{Color triplets}

\subsubsection*{$\boldsymbol{\omega_1\sim(3,1)_{-\frac{1}{3}}}$}
\begin{itemize}
	\item Lagrangian:
	\begin{equation}
		\begin{alignedat}{2}
			-\cL_S^{(\le4)}&\supset
			(y_{\omega_1}^{q\ell})_{rij}\omega_{1r}^{\dag}\bar q^c_{i}i\sig_2\ell_{j}
			+
			(y_{\omega_1}^{qq})_{rij}\omega_{1r}^{\alpha\dag}\varepsilon_{\alpha\beta\gamma}\bar q^\beta_{i}i\sig_2q^{c\gamma}_{j}
			+
			(y_{\omega_1}^{eu})_{rij}\omega_{1r}^{\dag}\bar e^c_{i}u_{j}\\&
			+
			(y_{\omega_1}^{du})_{rij}\omega_{1r}^{\alpha\dag}\varepsilon_{\alpha\beta\gamma}\bar d^\beta_{i}u^{c\gamma}_{j}
			+\hermc.\nonumber
		\end{alignedat}
	\end{equation}
	\item Flavor irreps:
	\begin{alignat}{3}
		&\omega_1^{i_q j_\ell} 
		\sim(\bm3_q,\bm3_\ell):
		&\qquad
		&[y_{\omega_1}^{q\ell}]^{i_qk_\ell}_{j_ql_\ell}=y_{\omega_1}^{q\ell}\delta^{i_q}_{j_q}\delta^{k_\ell}_{l_\ell},\nonumber
		\\
		&(\omega_{1})_{ i_q j_q} 
		\sim\bar{\bm6}_q:
		&\qquad &[y_{\omega_1}^{qq}]^{i_qk_q}_{j_ql_q}=\frac{y^{qq}_{\omega_1}}{2}\lzm \delta^{i_q}_{j_q}\delta^{k_q}_{l_q}+\delta^{i_q}_{l_q}\delta^{k_q}_{j_q} \dzm,\nonumber
		\\
		&\omega_1^{i_e j_u} 
		\sim(\bm3_e,\bm3_u):
		&\qquad &[y_{\omega_1}^{eu}]^{i_ek_u}_{j_el_u}=y_{\omega_1}^{eu}\delta^{i_e}_{j_e}\delta^{k_u}_{l_u},\nonumber
		\\
		&(\omega_{1})_{i_d j_u} 
		\sim(\bar{\bm3}_d,\bar{\bm3}_u):
		&\qquad &[ y_{\omega_1}^{du}]^{i_dk_u}_{j_dl_u}= y_{\omega_1}^{du}\delta^{i_d}_{j_d}\delta^{k_u}_{l_u}.\nonumber
	\end{alignat}
	\item Matching:
	\begin{itemize}
	\item $\omega_1^{i_q j_\ell}\sim(\bm3_q,\bm3_\ell)$:
	\begin{equation}
		\begin{alignedat}{2}
			\cL_{\text{SMEFT}}&\supset\nonumber
			\frac{|y_{\omega_1}^{q\ell}|^2}{4M_{\omega_1}^2}\lzm \cO^{(1)}_{\ell q}-\cO^{(3)}_{\ell q} \dzm.
		\end{alignedat}
	\end{equation}
	\item $(\omega_{1})_{ i_q j_q}\sim\bar{\bm6}_q$:
	\begin{equation}
		\begin{alignedat}{2}
			\cL_{\text{SMEFT}}&\supset
			\frac{\lzu y_{\omega_1}^{qq} \dzu^2}{4M_{\omega_1}^2}\lzm \cO_{qq}^{(1)D}-\cO_{qq}^{(3)D}+\cO_{qq}^{(1)E}-\cO_{qq}^{(3)E} \dzm.
			\nonumber
		\end{alignedat}
	\end{equation}
	\item $\omega_1^{i_e j_u}\sim(\bm3_e,\bm3_u)$:
	\begin{equation}
		\begin{alignedat}{2}
			\cL_{\text{SMEFT}}&\supset
			\frac{|y_{\omega_1}^{eu}|^2}{2M_{\omega_1}^2}\cO_{eu}.
			\nonumber
		\end{alignedat}
	\end{equation}
	\item $(\omega_{1})_{i_d j_u}\sim(\bar{\bm3}_d,\bar{\bm3}_u)$:
	\begin{equation}
		\begin{alignedat}{2}
			\cL_{\text{SMEFT}}&\supset
			\frac{\lzu y_{\omega_1}^{du} \dzu^2}{3M_{\omega_1}^2}\lzm \cO_{ud}^{(1)}-3\cO_{ud}^{(8)} \dzm.
			\nonumber
		\end{alignedat}
	\end{equation}
	\end{itemize}
\end{itemize}

\subsubsection*{$\boldsymbol{\omega_2\sim(3,1)_{\frac{2}{3}}}$}
\begin{itemize}
	\item Lagrangian:
	\begin{equation}
		-\cL_S^{(\le4)}\supset 
		(y_{\omega_2})_{rij}\omega_{2r}^{\alpha\dag}\varepsilon_{\alpha\beta\gamma}\bar d_{i}^{\beta}d_{j}^{c\gamma} +\hermc.\nonumber
	\end{equation}
	\item Flavor irreps:
	\begin{alignat}{3}
		\omega_2^{i_d} 
		&\sim \bm3_d:
		&\qquad
		[y_{\omega_2}]^{i_dj_dk_d}&=y_{\omega_2}\varepsilon^{i_dj_dk_d}.\nonumber
	\end{alignat}
	\item Matching:
		\begin{equation}
		\begin{alignedat}{2}
			\cL_{\text{SMEFT}}&\supset
			\frac{\lzu y_{\omega_2} \dzu^2}{M_{\omega_2}^2}\lzm \cO_{dd}^D-\cO_{dd}^E \dzm.
		\nonumber
		\end{alignedat}
	\end{equation}
\end{itemize}

\subsubsection*{$\boldsymbol{\omega_4\sim(3,1)_{-\frac{4}{3}}}$}
\begin{itemize}
	\item Lagrangian:
	\begin{equation}
		-\cL_S^{(\le4)}\supset 
		(y_{\omega_4}^{ed})_{rij}\omega_{4r}^{\dag}\bar e_{i}^cd_{j}
		+
		(y_{\omega_4}^{uu})_{rij}\omega_{4r}^{\alpha\dag}\varepsilon_{\alpha\beta\gamma}\bar u_{i}^\beta u_{j}^{c\gamma}
		+\hermc.\nonumber
	\end{equation}
	\item Flavor irreps:
	\begin{alignat}{3}
		&\omega_4^{i_e j_d} 
		\sim (\bm3_e,\bm3_d):
		&\qquad
		&[y_{\omega_4}^{ed}]^{i_ek_d}_{j_el_d}=y_{\omega_4}^{ed}\delta^{i_e}_{j_e}\delta^{k_d}_{l_d},\nonumber
		\\
		&\omega_4^{i_u} 
		\sim \bm3_u:
		&\qquad 
		&[y_{\omega_4}^{uu}]^{i_uj_uk_u}=y_{\omega_4}^{uu}\varepsilon^{i_uj_uk_u}.\nonumber
	\end{alignat}
	\item Matching:
	\begin{itemize}
	\item $\omega_4^{i_e j_d}\sim (\bm3_e,\bm3_d)$:
		\begin{equation}
		\begin{alignedat}{2}
			\cL_{\text{SMEFT}}&\supset\frac{\lzu y_{\omega_4}^{ed} \dzu^2}{2M_{\omega_4}^2}\cO_{ed}.
		\nonumber
		\end{alignedat}
	\end{equation}
	\item $\omega_4^{i_u}\sim \bm3_u$:
	\begin{equation}
		\begin{alignedat}{2}
			\cL_{\text{SMEFT}}&\supset \frac{\lzu y_{\omega_4}^{uu} \dzu^2}{M_{\omega_4}^2}\lzm \cO_{uu}^D-\cO_{uu}^E \dzm.
		\nonumber
		\end{alignedat}
	\end{equation}
	\end{itemize}
\end{itemize}

\subsubsection*{$\boldsymbol{\Pi_1\sim(3,2)_{\frac{1}{6}}}$}
\begin{itemize}
	\item Lagrangian:
	\begin{equation}
		-\cL_S^{(\le4)}\supset 
		(y_{\Pi_1})_{rij}\Pi_{1r}^\dag i\sig_2 \bar\ell^T_{i}d_{j}
		+\hermc.\nonumber
	\end{equation}
	\item Flavor irreps:
	\begin{alignat}{3}
		(\Pi_{1})^{i_d}_{j_\ell} 
		&\sim (\bar{\bm3}_\ell,\bm3_d):
		&\qquad
		[y_{\Pi_1}]^{i_\ell k_d}_{j_\ell l_d}&=y_{\Pi_1}\delta^{i_\ell}_{j_\ell}\delta^{k_d}_{l_d}.\nonumber
	\end{alignat}
	\item Matching:
	\begin{equation}
		\begin{alignedat}{2}
			\cL_{\text{SMEFT}}&\supset -\frac{\lzu y_{\Pi_1} \dzu^2}{2M_{\Pi_1}^2}\cO_{\ell d}.
		\nonumber
		\end{alignedat}
	\end{equation}
\end{itemize}

\subsubsection*{$\boldsymbol{\Pi_7\sim(3,2)_{\frac{7}{6}}}$}
\begin{itemize}
	\item Lagrangian:
	\begin{equation}
		-\cL_S^{(\le4)}\supset 
		(y_{\Pi_7}^{\ell u})_{rij}\Pi_{7r}^\dag i\sig_2 \bar\ell^T_{i}u_{j}
		+
		(y_{\Pi_7}^{eq})_{rij}\Pi_{7r}^\dag \bar e_{i}q_{j}
		+\hermc.\nonumber
	\end{equation}
	\item Flavor irreps:
	\begin{alignat}{3}
		(\Pi_{7})^{i_u}_{j_\ell} 
		&\sim (\bar{\bm3}_\ell,\bm3_u):
		&\qquad
		[y_{\Pi_7}^{\ell u}]^{i_\ell k_u}_{j_\ell l_u}&=y_{\Pi_7}^{\ell u}\delta^{i_\ell}_{j_\ell}\delta^{k_u}_{l_u},\nonumber
		\\
		(\Pi_{7})^{i_q}_{j_e} 
		&\sim (\bar{\bm3}_e,\bm3_q):
		&\qquad
		[y_{\Pi_7}^{eq}]^{i_e k_q}_{j_e l_q}&=y_{\Pi_7}^{eq}\delta^{i_e}_{j_e}\delta^{k_q}_{l_q}.\nonumber
	\end{alignat}
	\item Matching:
	\begin{itemize}
	\item $(\Pi_{7})^{i_u}_{j_\ell}\sim (\bar{\bm3}_\ell,\bm3_u)$:
	\begin{equation}
		\begin{alignedat}{2}
			\cL_{\text{SMEFT}}&\supset-\frac{\lzu y_{\Pi_7}^{\ell u} \dzu^2}{2M_{\Pi_7}^2}\cO_{\ell u}.
		\nonumber
		\end{alignedat}
	\end{equation}
	\item $(\Pi_{7})^{i_q}_{j_e}\sim (\bar{\bm3}_e,\bm3_q)$:
	\begin{equation}
		\begin{alignedat}{2}
			\cL_{\text{SMEFT}}&\supset-\frac{|y_{\Pi_7}^{qe}|^2}{2M_{\Pi_7}^2}\cO_{qe}.
		\nonumber
		\end{alignedat}
	\end{equation}
	\end{itemize}
\end{itemize}

\subsubsection*{$\boldsymbol{\zeta\sim(3,3)_{-\frac{1}{3}}}$}
\begin{itemize}
	\item Lagrangian:
	\begin{equation}
		-\cL_S^{(\le4)}\supset 
		(y_\zeta^{q\ell})_{rij}\zeta^{a\dag}_{r}\bar q^c_{i}i\sig_2\sig^a\ell_{j}
		+
		(y_\zeta^{qq})_{rij}\zeta^{a\alpha\dag}_r\varepsilon_{\alpha\beta\gamma}\bar q^\beta_{i}\sig^ai\sig_2q^{c\gamma}_{j}
		+\hermc.\nonumber
	\end{equation}
	\item Flavor irreps:
	\begin{alignat}{3}
		&\zeta^{i_q j_\ell} \sim (\bm3_q,\bm3_\ell):
		&\qquad
		&[y_\zeta^{q\ell}]^{i_qk_\ell}_{j_ql_\ell}=y_{\zeta}^{q\ell}\delta^{i_q}_{j_q}\delta^{k_\ell}_{l_\ell},\nonumber
		\\
		&\zeta^{i_q} \sim \bm3_q:
		&\qquad
		&[y_{\zeta}^{qq}]^{i_qj_qk_q}=y_{\zeta}^{qq}\varepsilon^{i_qj_qk_q}.\nonumber
	\end{alignat}
	\item Matching:
	\begin{itemize}
	\item $\zeta^{i_q j_\ell} \sim (\bm3_q,\bm3_\ell)$:
		\begin{equation}
		\begin{alignedat}{2}
			\cL_{\text{SMEFT}}&\supset
			\frac{|y_{\zeta}^{q\ell}|^2}{4M_\zeta^2}\lzm 3\cO^{(1)}_{\ell q}+\cO^{(3)}_{\ell q} \dzm.
		\nonumber
		\end{alignedat}
	\end{equation}
	\item $\zeta^{i_q} \sim \bm3_q$:
		\begin{equation}
		\begin{alignedat}{2}
			\cL_{\text{SMEFT}}&\supset
			\frac{|y_\zeta^{qq}|^2}{2M_\zeta^2}\lzm 3\cO_{qq}^{(1)D}+\cO_{qq}^{(3)D}-3\cO_{qq}^{(1)E}-\cO_{qq}^{(3)E} \dzm.
		\nonumber
		\end{alignedat}
	\end{equation}
	\end{itemize}
\end{itemize}

\subsubsection{Color sextets}

\subsubsection*{$\boldsymbol{\Omega_1\sim(6,1)_{\frac{1}{3}}}$}
\begin{itemize}
	\item Lagrangian:
	\begin{equation}
		-\cL_S^{(\le4)}\supset
		(y_{\Omega_1}^{ud})_{rij}\Omega_{1r}^{\alpha\beta\dag}\bar u_{i}^{c(\alpha|}d_{j}^{|\beta)}
		+
		(y_{\Omega_1}^{qq})_{rij}\Omega_{1r}^{\alpha\beta\dag}\bar q^{c(\alpha|}_{i}i\sig_2 q_{j}^{|\beta)}
		+\hermc.\nonumber
	\end{equation}
	\item Flavor irreps:
	\begin{alignat}{3}
		&\Omega_1^{i_u j_d} \sim (\bm3_u,\bm3_d):
		&\qquad
		&[y_{\Omega_1}^{ud}]^{i_uk_d}_{j_ul_d}=y_{\Omega_1}^{ud}\delta^{i_u}_{j_u}\delta^{k_d}_{l_d},\nonumber
		\\
		&(\Omega_{1})_{i_q} \sim \bar{\bm3}_q:
		&\qquad
		&[y_{\Omega_1}^{qq}]_{i_qj_qk_q}=y_{\Omega_1}^{qq}\varepsilon_{i_qj_qk_q}.\nonumber
	\end{alignat}
	\item Matching:
	\begin{itemize}
	\item $\Omega_1^{i_u j_d} \sim (\bm3_u,\bm3_d)$:
		\begin{equation}
		\begin{alignedat}{2}
			\cL_{\text{SMEFT}}&\supset
			\frac{\lzu y^{ud}_{\Omega_1} \dzu^2}{6M_{\Omega_1}^2}\lzm 2\cO^{(1)}_{ud}+3\cO^{(8)}_{ud} \dzm.
		\nonumber
		\end{alignedat}
	\end{equation}
	\item $(\Omega_{1})_{i_q} \sim \bar{\bm3}_q$:
		\begin{equation}
		\begin{alignedat}{2}
			\cL_{\text{SMEFT}}&\supset
			\frac{|y_{\Omega_1}^{qq}|^2}{4M_{\Omega_1}^2}\lzm \cO_{qq}^{(1)D}-\cO_{qq}^{(3)D}-\cO_{qq}^{(1)E}+\cO_{qq}^{(3)E} \dzm.
		\nonumber
		\end{alignedat}
	\end{equation}
	\end{itemize}
\end{itemize}

\subsubsection*{$\boldsymbol{\Omega_2\sim(6,1)_{-\frac{2}{3}}}$}
\begin{itemize}
	\item Lagrangian:
	\begin{equation}
		-\cL_S^{(\le4)}\supset
		(y_{\Omega_2})_{rij}\Omega_{2r}^{\alpha\beta\dag}\bar d_{i}^{c(\alpha|}d_{j}^{|\beta)}
		+\hermc.\nonumber
	\end{equation}
	\item Flavor irreps:
	\begin{alignat}{3}
		\Omega_2^{i_d j_d} &\sim \bm6_d:
		&\qquad
		[y_{\Omega_2}]^{i_dk_d}_{j_dl_d}&=\frac{y_{\Omega_2}}{2}\lzm \delta^{i_d}_{j_d}\delta^{k_d}_{l_d}+\delta^{i_d}_{l_d}\delta^{k_d}_{j_d} \dzm.\nonumber
	\end{alignat}
	\item Matching:
		\begin{equation}
		\begin{alignedat}{2}
			\cL_{\text{SMEFT}}&\supset \frac{\lzu y_{\Omega_2} \dzu^2}{4M_{\Omega_2}^2}\lzm \cO_{dd}^D+\cO_{dd}^E \dzm.
		\nonumber
		\end{alignedat}
	\end{equation}
\end{itemize}

\subsubsection*{$\boldsymbol{\Omega_4\sim(6,1)_{\frac{4}{3}}}$}
\begin{itemize}
	\item Lagrangian:
	\begin{equation}
		-\cL_S^{(\le4)}\supset
		(y_{\Omega_4})_{rij}\Omega_{4r}^{\alpha\beta\dag}\bar u_{i}^{c(\alpha|}u_{j}^{|\beta)}
		+\hermc.\nonumber
	\end{equation}
	\item Flavor irreps:
	\begin{alignat}{3}
		\Omega_4^{i_u j_u} &\sim \bm6_u:
		&\qquad
		[y_{\Omega_4}]^{i_uk_u}_{j_ul_u}&=\frac{y_{\Omega_4}}{2}\lzm \delta^{i_u}_{j_u}\delta^{k_u}_{l_u}+\delta^{i_u}_{l_u}\delta_{j_u}^{k_u} \dzm.\nonumber
	\end{alignat}
	\item Matching:
		\begin{equation}
		\begin{alignedat}{2}
			\cL_{\text{SMEFT}}&\supset \frac{\lzu y_{\Omega_4} \dzu^2}{4M_{\Omega_4}^2}\lzm \cO_{uu}^D+\cO_{uu}^E \dzm .
		\nonumber
		\end{alignedat}
	\end{equation}
\end{itemize}

\subsubsection*{$\boldsymbol{\Upsilon\sim(6,3)_{\frac{1}{3}}}$}
\begin{itemize}
	\item Lagrangian:
	\begin{equation}
		-\cL_S^{(\le4)}\supset
		(y_\Upsilon)_{rij}\Upsilon_r^{a\alpha\beta\dag}\bar q_{i}^{c(\alpha|}i\sig_2\sig^a q_{j}^{|\beta)}
		+\hermc.\nonumber
	\end{equation}
	\item Flavor irreps:
	\begin{alignat}{3}
		\Upsilon^{i_q j_q} &\sim \bm6_q:
		&\qquad
		[y_\Upsilon]^{i_qk_q}_{j_ql_q}&=\frac{y_{\Upsilon}}{2}\lzm \delta^{i_q}_{j_q}\delta^{k_q}_{l_q}+\delta^{i_q}_{l_q}\delta^{k_q}_{j_q} \dzm.\nonumber
	\end{alignat}
	\item Matching:
	\begin{equation}
		\begin{alignedat}{2}
			\cL_{\text{SMEFT}}&\supset \frac{\lzu y_\Upsilon \dzu^2}{8M_\Upsilon^2}\lzm 3\cO_{qq}^{(1)D}+\cO_{qq}^{(3)D}+3\cO_{qq}^{(1)E}+\cO_{qq}^{(3)E} \dzm.
		\nonumber
		\end{alignedat}
	\end{equation}
\end{itemize}

\subsubsection{Color octets}

\subsubsection*{$\boldsymbol{\Phi\sim(8,2)_{\frac{1}{2}}}$}
\begin{itemize}
	\item Lagrangian:
	\begin{equation}
		-\cL_S^{(\le4)}\supset
		(y_\Phi^{qu})_{rij}\Phi_{r}^{A\dag}i\sig_2\bar q^T_{i}T_Au_{j}
		+
		(y_\Phi^{dq})_{rij}\Phi^{A\dag}_r\bar d_{i}T_A q_{j}
		+\hermc.\nonumber
	\end{equation}
	\item Flavor irreps:
	\begin{alignat}{3}
		\Phi^{i_u}_{j_q} &\sim (\bar{\bm3}_q,\bm3_u):
		&\qquad
		[y_\Phi^{qu}]^{i_qk_u}_{j_ql_u}&=y_\Phi^{qu}\delta^{i_q}_{j_q}\delta^{k_u}_{l_u},\nonumber
		\\
		\Phi^{i_q}_{j_d} &\sim (\bar{\bm3}_d,\bm3_q):
		&\qquad
		[y_\Phi^{dq}]^{i_dk_q}_{j_dl_q}&=y_\Phi^{dq}\delta^{i_d}_{j_d}\delta^{k_q}_{l_q}.\nonumber
	\end{alignat}
	\item Matching:
	\begin{itemize}
	\item $\Phi^{i_u}_{j_q}\sim (\bar{\bm3}_q,\bm3_u)$:
		\begin{equation}
		\begin{alignedat}{2}
			\cL_{\text{SMEFT}}&\supset
			-\frac{\lzu y^{qu}_\Phi \dzu^2}{18M_\Phi^2}\lzm 4\cO_{qu}^{(1)}-3\cO_{qu}^{(8)} \dzm.
		\nonumber
		\end{alignedat}
	\end{equation}
	\item $\Phi^{i_q}_{j_d}\sim (\bar{\bm3}_d,\bm3_q)$:
		\begin{equation}
		\begin{alignedat}{2}
			\cL_{\text{SMEFT}}&\supset
			-\frac{|y^{dq}_\Phi|^2}{18M_\Phi^2}\lzm 4\cO_{qd}^{(1)}-3\cO_{qd}^{(8)} \dzm.
		\nonumber
		\end{alignedat}
	\end{equation}
	\end{itemize}
\end{itemize}


\subsection{Spin 1/2}
\label{app:sm1-spin12}

\subsubsection{Color singlets}
\label{app:sm1-spin12-singlets}

\subsubsection*{$\boldsymbol{N\sim(1,1)_0}$}
\begin{itemize}
	\item Lagrangian:
	\begin{equation}
		\begin{alignedat}{2}
			-\cL_{\text{leptons}}^{(4)}&\supset
			(\lambda_N)_{ri}\bar N_{Rr}\tilde\phi^\dag\ell_{i}
			+\hermc.\nonumber
		\end{alignedat}
	\end{equation}
	\item Flavor irreps:
		\begin{alignat}{3}
		N_{L,R}^{i_\ell} &\sim \bm3_\ell:
		&\qquad
		[\lambda_N]^{i_\ell}_{j_\ell}&=\lambda_N \delta^{i_\ell}_{j_\ell}.\nonumber
	\end{alignat}
	\item Matching:
		\begin{equation}
		\begin{alignedat}{2}
			\cL_{\text{SMEFT}}&\supset 
			\frac{\lzu \lambda_N \dzu^2}{4M_N^2}\lzm \cO_{\phi\ell}^{(1)} -\cO_{\phi\ell}^{(3)} \dzm.
		\nonumber
		\end{alignedat}
	\end{equation}
\end{itemize}

\subsubsection*{$\boldsymbol{E\sim(1,1)_{-1}}$}
\begin{itemize}
	\item Lagrangian:
	\begin{equation}
		\begin{alignedat}{2}
			-\cL_{\text{leptons}}^{(4)}&\supset
			(\lambda_E)_{ri}\bar E_{Rr}\phi^\dag\ell_{i}
			+\hermc.
			\nonumber
		\end{alignedat}
	\end{equation}
	\item Flavor irreps:
		\begin{alignat}{3}
		E_{L,R}^{i_\ell} &\sim \bm3_\ell:
		&\qquad
		[\lambda_E]^{i_\ell}_{j_\ell}&=\lambda_E \delta^{i_\ell}_{j_\ell}.
		\nonumber
	\end{alignat}
	\item Matching:
			\begin{equation}
		\begin{alignedat}{2}
			\cL_{\text{SMEFT}}&\supset 
			-\frac{\lzu\lambda_E\dzu^2}{4M_E^2} \lzm \cO_{\phi\ell}^{(1)} + \cO_{\phi\ell}^{(3)} \dzm+\lzv \frac{\lzu\lambda_E\dzu^2}{2M_E^2} [y_e^*]^{i_\ell}_{j_e}[\cO_{e\phi}]^{j_e}_{i_\ell}+\hermc \dzv.
		\nonumber
		\end{alignedat}
	\end{equation}
\end{itemize}

\subsubsection*{$\boldsymbol{\Delta_1\sim(1,2)_{-\frac{1}{2}}}$}
\begin{itemize}
	\item Lagrangian:
	\begin{equation}
		\begin{alignedat}{2}
			-\cL_{\text{leptons}}^{(4)}&\supset
			(\lambda_{\Delta_1})_{ri}\bar\Delta_{1Lr}\phi e_{i}
			+\hermc.\nonumber
		\end{alignedat}
	\end{equation}
	\item Flavor irreps:
		\begin{alignat}{3}
		\Delta_{1L,R}^{i_e} &\sim \bm3_e:
		&\qquad
		[\lambda_{\Delta_1}]^{i_e}_{j_e}&=\lambda_{\Delta_1} \delta^{i_e}_{j_e}.
		\nonumber
	\end{alignat}
	\item Matching:
			\begin{equation}
		\begin{alignedat}{2}
			\cL_{\text{SMEFT}}&\supset 
			\frac{\lzu \lambda_{\Delta_1} \dzu^2}{2M_{\Delta_1}^2} \cO_{\phi e}+\lzv \frac{\lzu\lambda_{\Delta_1}\dzu^2}{2M_{\Delta_1}^2} [y_e^*]^{i_\ell}_{j_e}[\cO_{e\phi}]^{j_e}_{i_\ell}+\hermc \dzv.
		\nonumber
		\end{alignedat}
	\end{equation}
\end{itemize}

\subsubsection*{$\boldsymbol{\Delta_3\sim(1,2)_{-\frac{3}{2}}}$}
\begin{itemize}
	\item Lagrangian:
	\begin{equation}
		\begin{alignedat}{2}
			-\cL_{\text{leptons}}^{(4)}&\supset
			(\lambda_{\Delta_3})_{ri}\bar\Delta_{3Lr}\tilde\phi e_{i}
			+\hermc.\nonumber
		\end{alignedat}
	\end{equation}
	\item Flavor irreps:
		\begin{alignat}{3}
		\Delta_{3L,R}^{i_e} &\sim \bm3_e:
		&\qquad
		[\lambda_{\Delta_3}]^{i_e}_{j_e}&=\lambda_{\Delta_3} \delta^{i_e}_{j_e}.
		\nonumber
	\end{alignat}
	\item Matching:
			\begin{equation}
		\begin{alignedat}{2}
			\cL_{\text{SMEFT}}&\supset
			-\frac{\lzu \lambda_{\Delta_3} \dzu^2}{2M_{\Delta_3}^2} \cO_{\phi e}+\lzv \frac{\lzu \lambda_{\Delta_3} \dzu^2}{2M_{\Delta_3}^2} [y_e^*]^{i_\ell}_{j_e}[\cO_{e\phi}]^{j_e}_{i_\ell}+\hermc \dzv.
		\nonumber
		\end{alignedat}
	\end{equation}
\end{itemize}

\subsubsection*{$\boldsymbol{\Sigma\sim(1,3)_{0}}$}
\begin{itemize}
	\item Lagrangian:
	\begin{equation}
		\begin{alignedat}{2}
			-\cL_{\text{leptons}}^{(4)}&\supset
			\frac{1}{2}(\lambda_\Sigma)_{ri}\bar\Sigma^a_{Rr}\tilde\phi^\dag\sig^a\ell_{i}
			+\hermc.\nonumber
		\end{alignedat}
	\end{equation}
	\item Flavor irreps:
		\begin{alignat}{3}
		\Sigma_{L,R}^{i_\ell} &\sim \bm3_\ell:
		&\qquad
		[\lambda_\Sigma]^{i_\ell}_{j_\ell}&=\lambda_\Sigma \delta^{i_\ell}_{j_\ell}\nonumber.		
	\end{alignat}
	\item Matching:
			\begin{equation}
		\begin{alignedat}{2}
			\cL_{\text{SMEFT}}&\supset 
			\frac{\lzu\lambda_\Sigma\dzu^2}{16M_\Sigma^2} \lzm 3 \cO_{\phi\ell}^{(1)} +\cO_{\phi\ell}^{(3)}\dzm+\lzv \frac{\lzu \lambda_\Sigma \dzu^2}{4M_\Sigma^2} [y_e^*]^{i_\ell}_{j_e}[\cO_{e\phi}]^{j_e}_{i_\ell}+\hermc \dzv.
		\nonumber
		\end{alignedat}
	\end{equation}
\end{itemize}

\subsubsection*{$\boldsymbol{\Sigma_1\sim(1,3)_{-1}}$}
\begin{itemize}
	\item Lagrangian:
	\begin{equation}
		\begin{alignedat}{2}
			-\cL_{\text{leptons}}^{(4)}&\supset
			\frac{1}{2}(\lambda_{\Sigma_1})_{ri}\bar\Sigma^a_{1Rr}\phi^\dag\sig^a\ell_{i}
			+\hermc.
			\nonumber
		\end{alignedat}
	\end{equation}
	\item Flavor irreps:
		\begin{alignat}{3}
		\Sigma_{1L,R}^{i_\ell} &\sim \bm3_\ell:
		&\qquad
		[\lambda_{\Sigma_1}]^{i_\ell}_{j_\ell}&=\lambda_{\Sigma_1} \delta^{i_\ell}_{j_\ell}.
		\nonumber
		\end{alignat}
	\item Matching:
	\begin{equation}
		\begin{alignedat}{2}
			\cL_{\text{SMEFT}}&\supset
			\frac{\lzu\lambda_{\Sigma_1}\dzu^2}{16M_{\Sigma_1}^2} \lzm \cO_{\phi\ell}^{(3)} -3 \cO_{\phi\ell}^{(1)} \dzm+\lzv \frac{\lzu \lambda_{\Sigma_1} \dzu^2}{8M_{\Sigma_1}^2} [y_e^*]^{i_\ell}_{j_e}[\cO_{e\phi}]^{j_e}_{i_\ell}+\hermc \dzv.
		\nonumber
		\end{alignedat}
	\end{equation}
\end{itemize}


\subsubsection{Color triplets}
\label{app:sm1-spin12-triplets}

\subsubsection*{$\boldsymbol{U\sim(3,1)_{\frac{2}{3}}}$}
\begin{itemize}
	\item Lagrangian:
	\begin{equation}
		\begin{alignedat}{2}
			-\cL^{(4)}_\text{quarks}&\supset
			(\lambda_U)_{ri}\bar U_{Rr}\tilde\phi^\dag q_{i}
			+\hermc.
			\nonumber
		\end{alignedat}
	\end{equation}
	\item Flavor irreps:
		\begin{alignat}{3}
		U_{L,R}^{i_q} &\sim \bm3_q:
		&\qquad
		[\lambda_U]^{i_q}_{j_q}&=\lambda_U \delta^{i_q}_{j_q}.
		\nonumber
	\end{alignat}
	\item Matching:
	\begin{equation}
		\begin{alignedat}{2}
			\cL_{\text{SMEFT}}&\supset 
			\frac{\lzu\lambda_{U}\dzu^2}{4M_{U}^2} \lzm \cO_{\phi q}^{(1)}-\cO_{\phi q}^{(3)} \dzm+\lzv 
			\frac{\lzu \lambda_{U} \dzu^2}{2M_{U}^2} [y_u^*]^{i_q}_{j_u}[\cO_{u\phi}]^{j_u}_{i_q} +\hermc \dzv.
		\nonumber
		\end{alignedat}
	\end{equation}
\end{itemize}

\subsubsection*{$\boldsymbol{D\sim(3,1)_{-\frac{1}{3}}}$}
\begin{itemize}
	\item Lagrangian:
	\begin{equation}
		\begin{alignedat}{2}
			-\cL^{(4)}_\text{quarks}&\supset
			(\lambda_D)_{ri}\bar D_{Rr}\phi^\dag q_{i}
			+\hermc.
			\nonumber
		\end{alignedat}
	\end{equation}
	\item Flavor irreps:
		\begin{alignat}{3}
		D_{L,R}^{i_q} &\sim \bm3_q:
		&\qquad
		[\lambda_D]^{i_q}_{j_q}&=\lambda_D \delta^{i_q}_{j_q}.
		\nonumber
	\end{alignat}
	\item Matching:
	\begin{equation}
		\begin{alignedat}{2}
			\cL_{\text{SMEFT}}&\supset 
			-\frac{\lzu\lambda_{D}\dzu^2}{4M_{D}^2}\lzm \cO_{\phi q}^{(1)} + \cO_{\phi q}^{(3)}\dzm +\lzv 
			\frac{\lzu \lambda_{D} \dzu^2}{2M_{D}^2} [y_d^*]^{i_q}_{j_d}[\cO_{d\phi}]^{j_d}_{i_q}+\hermc\dzv.
		\nonumber
		\end{alignedat}
	\end{equation}
\end{itemize}

\subsubsection*{$\boldsymbol{Q_1\sim(3,2)_{\frac{1}{6}}}$}
\begin{itemize}
	\item Lagrangian:
	\begin{equation}
		\begin{alignedat}{2}
			-\cL^{(4)}_\text{quarks}&\supset
			(\lambda^u_{Q_1})_{ri}\bar Q_{1Lr}\tilde\phi u_{i}
			+
			(\lambda^d_{Q_1})_{ri}\bar Q_{1Lr}\phi d_{i}
			+\hermc.
			\nonumber
		\end{alignedat}
	\end{equation}
	\item Flavor irreps:
		\begin{alignat}{3}
		Q_{1L,R}^{i_u} &\sim \bm3_u:
		&\qquad
		[\lambda^u_{Q_1}]^{i_u}_{j_u}&=\lambda^u_{Q_1} \delta^{i_u}_{j_u},
		\nonumber
		\\
		Q_{1L,R}^{i_d} &\sim \bm3_d:
		&\qquad
		[\lambda^d_{Q_1}]^{i_d}_{j_d}&=\lambda^d_{Q_1} \delta^{i_d}_{j_d}\nonumber.
	\end{alignat}
	\item Matching:
	\begin{itemize}
	\item $Q_{1L,R}^{i_u}\sim \bm3_u$:
	\begin{equation}
		\begin{alignedat}{2}
			\cL_{\text{SMEFT}}&\supset 
			-\frac{|\lambda^u_{Q_1}|^2}{2M_{Q_1}^2} \cO_{\phi u}+\lzv 
			\frac{|\lambda^u_{Q_1}|^2}{2M_{Q_1}^2} [y_u^*]^{i_q}_{j_u}[\cO_{u\phi}]^{j_u}_{i_q}+\hermc\dzv.
		\nonumber
		\end{alignedat}
	\end{equation}
	\item $Q_{1L,R}^{i_d}\sim \bm3_d$:
	\begin{equation}
		\begin{alignedat}{2}
			\cL_{\text{SMEFT}}&\supset 
			\frac{|\lambda^d_{Q_1}|^2}{2M_{Q_1}^2} \cO_{\phi d}+\lzv 
			\frac{|\lambda^d_{Q_1}|^2}{2M_{Q_1}^2} [y_d^*]^{i_q}_{j_d}[\cO_{d\phi}]^{j_d}_{i_q} +\hermc\dzv.
		\nonumber
		\end{alignedat}
	\end{equation}
	\end{itemize}
\end{itemize}

\subsubsection*{$\boldsymbol{Q_5\sim(3,2)_{-\frac{5}{6}}}$}
\begin{itemize}
	\item Lagrangian:
	\begin{equation}
		\begin{alignedat}{2}
			-\cL^{(4)}_\text{quarks}&\supset
			(\lambda_{Q_5})_{ri}\bar Q_{5Lr}\tilde\phi d_{i}
			+\hermc.
			\nonumber
		\end{alignedat}
	\end{equation}
	\item Flavor irreps:
		\begin{alignat}{3}
		Q_{5L,R}^{i_d} &\sim \bm3_d:
		&\qquad
		[\lambda_{Q_5}]^{i_d}_{j_d}&=\lambda_{Q_5} \delta^{i_d}_{j_d}.
		\nonumber
		\end{alignat}
	\item Matching:
	\begin{equation}
		\begin{alignedat}{2}
			\cL_{\text{SMEFT}}&\supset 
			-\frac{\lzu\lambda_{Q_5}\dzu^2}{2M_{Q_5}^2}\cO_{\phi d}+\lzv 
			\frac{\lzu \lambda_{Q_5} \dzu^2}{2M_{Q_5}^2} [ y_d^*]^{i_q}_{j_d}[\cO_{d\phi}]^{j_d}_{i_q}+\hermc\dzv.
		\nonumber
		\end{alignedat}
	\end{equation}
\end{itemize}

\subsubsection*{$\boldsymbol{Q_7\sim(3,2)_{\frac{7}{6}}}$}
\begin{itemize}
	\item Lagrangian:
	\begin{equation}
		\begin{alignedat}{2}
			-\cL^{(4)}_\text{quarks}&\supset
			(\lambda_{Q_7})_{ri}\bar Q_{7Lr}\phi u_{i}
			+\hermc.
			\nonumber
		\end{alignedat}
	\end{equation}
	\item Flavor irreps:
		\begin{alignat}{3}
		Q_{7L,R}^{i_u} &\sim \bm3_u:
		&\qquad
		[\lambda_{Q_7}]^{i_u}_{j_u}&=\lambda_{Q_7} \delta^{i_u}_{j_u}.
		\nonumber
	\end{alignat}
	\item Matching:
\begin{equation}
		\begin{alignedat}{2}
			\cL_{\text{SMEFT}}&\supset 
			\frac{\lzu\lambda_{Q_7}\dzu^2}{2M_{Q_7}^2} \cO_{\phi u}+\lzv \frac{\lzu \lambda_{Q_7} \dzu^2}{2M_{Q_7}^2} [y_u^*]^{i_q}_{j_u}[\cO_{u\phi}]^{j_u}_{i_q}+\hermc\dzv.
		\nonumber
		\end{alignedat}
\end{equation}
\end{itemize}

\subsubsection*{$\boldsymbol{T_1\sim(3,3)_{-\frac{1}{3}}}$}
\begin{itemize}
	\item Lagrangian:
	\begin{equation}
		\begin{alignedat}{2}
			-\cL^{(4)}_\text{quarks}&\supset
			\frac{1}{2}(\lambda_{T_1})_{ri}\bar T^a_{1Rr}\phi^\dag \sig^a q_{i}
			+\hermc.
			\nonumber
		\end{alignedat}
	\end{equation}
	\item Flavor irreps:
		\begin{alignat}{3}
		T_{1L,R}^{i_q} &\sim \bm3_q:
		&\qquad
		[\lambda_{T_1}]^{i_q}_{j_q}&=\lambda_{T_1} \delta^{i_q}_{j_q}.
		\nonumber
		\end{alignat}
	\item Matching:
\begin{equation}
		\begin{alignedat}{2}
			\cL_{\text{SMEFT}}&\supset 
			\frac{\lzu\lambda_{T_1}\dzu^2}{16M_{T_1}^2}
			\lzm \cO_{\phi q}^{(3)}-3\cO_{\phi q}^{(1)} \dzm+\Bigg\{
			\frac{\lzu \lambda_{T_1} \dzu^2}{8M_{T_1}^2} [y_d^*]^{i_q}_{j_d}[\cO_{d\phi}]^{j_d}_{i_q}+\frac{\lzu \lambda_{T_1} \dzu^2}{4M_{T_1}^2} [y_u^*]^{i_q}_{j_u}[\cO_{u\phi}]^{j_u}_{i_q}
			+\hermc
			\Bigg\}.
		\nonumber
		\end{alignedat}
\end{equation}
\end{itemize}

\subsubsection*{$\boldsymbol{T_2\sim(3,3)_{\frac{2}{3}}}$}
\begin{itemize}
	\item Lagrangian:
	\begin{equation}
		\begin{alignedat}{2}
			-\cL^{(4)}_\text{quarks}&\supset
			\frac{1}{2}(\lambda_{T_2})_{ri}\bar T^a_{2Rr}\tilde\phi^\dag \sig^a q_{i}
			+\hermc.
			\nonumber
		\end{alignedat}
	\end{equation}
	\item Flavor irreps:
		\begin{alignat}{3}
		T_{2L,R}^{i_q} &\sim \bm3_q:
		&\qquad
		[\lambda_{T_2}]^{i_q}_{j_q}&=\lambda_{T_2} \delta^{i_q}_{j_q}.
		\nonumber
	\end{alignat}
	\item Matching:
\begin{equation}
		\begin{alignedat}{2}
			\cL_{\text{SMEFT}}&\supset 
			\frac{\lzu\lambda_{T_2}\dzu^2}{16M_{T_2}^2}
			\lzm \cO_{\phi q}^{(3)}+3\cO_{\phi q}^{(1)} \dzm+\Bigg\{ 
			\frac{\lzu \lambda_{T_2} \dzu^2}{4M_{T_2}^2} [y_d^*]^{i_q}_{j_d}[\cO_{d\phi}]^{j_d}_{i_q}+
			\frac{\lzu \lambda_{T_2} \dzu^2}{8M_{T_2}^2} [y_u^*]^{i_q}_{j_u}[\cO_{u\phi}]^{j_u}_{i_q} 
			+\hermc\Bigg\}.
		\nonumber
		\end{alignedat}
\end{equation}
\end{itemize}


\subsection{Spin 1}
\label{app:sm1-spin1}

\subsubsection{Color singlets}
\label{app:sm1-spin1-singlets}

\subsubsection*{$\boldsymbol{\cB\sim(\bm1,\bm1)_0}$}

\begin{itemize}
	\item Lagrangian:
	\begin{equation}
		\begin{alignedat}{2}
			-\cL_V^{(\le4)}&\supset 
			(g_\cB^\ell)_{rij}\cB_r^\mu\bar\ell_{i}\gamma_\mu\ell_{j}
			+
			(g_\cB^q)_{rij}\cB_r^\mu\bar q_{i}\gamma_\mu q_{j}
			+
			(g_\cB^e)_{rij}\cB_r^\mu\bar e_{i}\gamma_\mu e_{j}
			\\&
			+
			(g_\cB^d)_{rij}\cB_r^\mu\bar d_{i}\gamma_\mu d_{j}
			+
			(g_\cB^u)_{rij}\cB_r^\mu\bar u_{i}\gamma_\mu u_{j}
			+\lzs (g_\cB^\phi)_{r}\cB_r^\mu \phi^\dag iD_\mu \phi+\hermc \dzs
			\nonumber.
		\end{alignedat}
	\end{equation}
	\item Flavor irreps:
		\begin{alignat}{3}
		&(B^\mu)\ud{i_\ell}{j_\ell} \sim \bm8_\ell:
		&\qquad
		&[g_\cB^{\ell,A_\ell}]\ud{i_\ell}{j_\ell}=g_\cB^\ell (T^{A_\ell})\ud{i_\ell}{j_\ell}\nonumber,
		\\		
		&(B^\mu)\ud{i_q}{j_q} \sim \bm8_q:
		&\qquad
		&[g_\cB^{q,A_q}]\ud{i_q}{j_q}=g_\cB^q (T^{A_q})\ud{i_q}{j_q},
		\nonumber
		\\
		&(B^\mu)\ud{i_e}{j_e} \sim \bm8_e:
		&\qquad
		&[g_\cB^{e,A_e}]\ud{i_e}{j_e}=g_\cB^e (T^{A_e})\ud{i_e}{j_e},
		\nonumber
		\\
		&(B^\mu)\ud{i_d}{j_d} \sim \bm8_d:
		&\qquad
		&[g_\cB^{d,A_d}]\ud{i_d}{j_d}=g_\cB^d (T^{A_d})\ud{i_d}{j_d},\nonumber
		\\
		&(B^\mu)\ud{i_u}{j_u} \sim \bm8_u:
		&\qquad
		&[g_\cB^{u,A_u}]\ud{i_u}{j_u}=g_\cB^u (T^{A_u})\ud{i_u}{j_u},\nonumber
		\\
		&B^\mu \sim \bm1:
		&\qquad
		&[g_\cB^\ell]^{i_\ell}_{j_\ell}=g_\cB^\ell \delta^{i_\ell}_{j_\ell},
		\quad
		[g_\cB^q]^{i_q}_{j_q}=g_\cB^q \delta^{i_q}_{j_q},
		\quad
		[g_\cB^e]^{i_e}_{j_e}=g_\cB^e \delta^{i_e}_{j_e},
		\nonumber
		\\&&\qquad
		&[g_\cB^d]^{i_d}_{j_d}=g_\cB^d \delta^{i_d}_{j_d},
		\quad
		[g_\cB^u]^{i_u}_{j_u}=g_\cB^u \delta^{i_u}_{j_u},
		\quad g_\cB^\phi
		\nonumber.
	\end{alignat}
	\item Matching:
	\begin{itemize}
	\item $(B^\mu)\ud{i_\ell}{j_\ell}\sim\bm8_\ell$:
			\begin{equation}
		\begin{alignedat}{2}
			\cL_{\text{SMEFT}}&\supset
			-\frac{(g_\cB^\ell)^2}{12M_\cB^2}\lzm
			 3 \cO_{\ell\ell}^E-\cO_{\ell\ell}^D 
			 \dzm.
		\nonumber
		\end{alignedat}
	\end{equation}
	\item $(B^\mu)\ud{i_q}{j_q}\sim\bm8_q$:
			\begin{equation}
		\begin{alignedat}{2}
			\cL_{\text{SMEFT}}&\supset
			-\frac{(g_\cB^q)^2}{12M_\cB^2}\lzm
			 3\cO_{qq}^{(1)E}
			 -\cO_{qq}^{(1)D} \dzm.
		\nonumber
		\end{alignedat}
	\end{equation}
	\item $(B^\mu)\ud{i_e}{j_e}\sim\bm8_e$:
			\begin{equation}
		\begin{alignedat}{2}
			\cL_{\text{SMEFT}}&\supset
			-\frac{(g_\cB^e)^2}{6M_\cB^2} \cO_{ee}.
		\nonumber
		\end{alignedat}
	\end{equation}
	\item $(B^\mu)\ud{i_d}{j_d}\sim\bm8_d$:
			\begin{equation}
		\begin{alignedat}{2}
			\cL_{\text{SMEFT}}&\supset
			-\frac{(g_\cB^d)^2}{12M_\cB^2}\lzm 
			3\cO_{dd}^E
			-\cO_{dd}^D \dzm.
		\nonumber
		\end{alignedat}
	\end{equation}
	\item $(B^\mu)\ud{i_u}{j_u}\sim\bm8_u$:
			\begin{equation}
		\begin{alignedat}{2}
			\cL_{\text{SMEFT}}&\supset
			-\frac{(g_\cB^u)^2}{12M_\cB^2}\lzm 
			3\cO_{uu}^E-\cO_{uu}^D \dzm.
		\nonumber
		\end{alignedat}
	\end{equation}
	\item $B^\mu\sim\bm1$:
			\begin{equation}
		\begin{alignedat}{2}
			\cL_{\text{SMEFT}}\supset
			&-\frac{(g_\cB^\ell)^2}{2M_\cB^2}\cO_{\ell\ell}^D
			-\frac{(g_\cB^q)^2}{2M_\cB^2}\cO_{qq}^{(1)D}
			-\frac{g_\cB^\ell g_\cB^q}{M_\cB^2}\cO_{\ell q}^{(1)}
			-\frac{(g_\cB^e)^2}{2M_\cB^2}\cO_{ee}
			-\frac{(g_\cB^d)^2}{2M_\cB^2}\cO_{dd}^D
			-\frac{(g_\cB^u)^2}{2M_\cB^2}\cO_{uu}^D
			\\&
			-\frac{g_\cB^d g_\cB^e}{M_\cB^2}\cO_{ed}
			-\frac{g_\cB^u g_\cB^e}{M_\cB^2}\cO_{eu}
			-\frac{g_\cB^u g_\cB^d}{M_\cB^2}\cO_{ud}^{(1)}
			-\frac{g_\cB^e g_\cB^\ell}{M_\cB^2}\cO_{\ell e}
			-\frac{g_\cB^d g_\cB^\ell}{M_\cB^2}\cO_{\ell d}
			-\frac{g_\cB^u g_\cB^\ell}{M_\cB^2}\cO_{\ell u}
			\\&
			-\frac{g_\cB^q g_\cB^e}{M_\cB^2}\cO_{qe}
			-\frac{g_\cB^q g_\cB^u}{M_\cB^2}\cO_{qu}^{(1)}
			-\frac{g_\cB^q g_\cB^d}{M_\cB^2}\cO_{qd}^{(1)}
			\\&
			-\lzs \frac{\re (g^\phi_\cB  g^\phi_\cB)}{2M_\cB^2} \dzs \cO_{\phi\Box}
			-\lzs \frac{\re ( g^\phi_\cB  g^\phi_\cB)}{M_\cB^2} +\frac{( g_\cB^\phi)^*  g_\cB^\phi}{M_\cB^2} \dzs \cO_{\phi D}
			-\lzs \frac{\re( g^\phi_\cB)g_\cB^u}{M_\cB^2} \dzs \cO_{\phi u}
			\\&
			-\lzs \frac{\re( g^\phi_\cB)g_\cB^d}{M_\cB^2} \dzs \cO_{\phi d}
			-\lzs \frac{\re( g^\phi_\cB)g_\cB^e}{M_\cB^2} \dzs \cO_{\phi e}
			-\lzs \frac{\re( g^\phi_\cB)g_\cB^\ell}{M_\cB^2} \dzs \cO_{\phi\ell}^{(1)}
			-\lzs \frac{\re( g^\phi_\cB)g_\cB^q}{M_\cB^2} \dzs \cO_{\phi q}^{(1)}
			\\&
			+\Bigg\{\lzs -\frac{i\,\im ( g^\phi_\cB  g^\phi_\cB)}{2M_\cB^2}
			+\frac{i\,\im ( g^\phi_\cB) g_\cB^\ell}{M_\cB^2}
			-\frac{i\,\im ( g^\phi_\cB) g_\cB^e}{M_\cB^2} \dzs [ y_e^*]^{i_\ell}_{j_e}[\cO_{e\phi}]^{j_e}_{i_\ell}
			\\&
			\phantom{aa}+\lzs -\frac{i\,\im ( g^\phi_\cB  g^\phi_\cB)}{2M_\cB^2}
			+\frac{i\,\im ( g^\phi_\cB) g_\cB^q}{M_\cB^2}
			-\frac{i\,\im ( g^\phi_\cB) g_\cB^d}{M_\cB^2} \dzs [ y_d^*]^{i_q}_{j_d}[\cO_{d\phi}]^{j_d}_{i_q}
			\\&
			\phantom{aa}+\lzs -\frac{i\,\im ( g^\phi_\cB  g^\phi_\cB)}{2M_\cB^2}
			+\frac{i\,\im ( g^\phi_\cB) g_\cB^q}{M_\cB^2}
			-\frac{i\,\im ( g^\phi_\cB) g_\cB^u}{M_\cB^2} \dzs [ y_u^*]^{i_q}_{j_u}[\cO_{u\phi}]^{j_u}_{i_q}+\hermc\Bigg\}.
		\nonumber
		\end{alignedat}
	\end{equation}
	\end{itemize}
\end{itemize}

\subsubsection*{$\boldsymbol{\cB_1\sim(\bm1,\bm1)_1}$}

\begin{itemize}
	\item Lagrangian:
	\begin{equation}
		-\cL_V^{(\le4)}\supset (g_{\cB_1}^{du})_{rij}\cB_{1r}^{\mu\dag}\bar d_{i}\gamma_\mu u_{j}
		+(g_{\cB_1}^\phi)_r B_{1r}^{\mu\dag}i(D_\mu \phi)^Ti\sig_2 \phi
		+\hermc.\nonumber
	\end{equation}
	\item Flavor irreps:
		\begin{alignat}{3}
		&(\cB^\mu_{1})^{i_u}_{j_d} \sim (\bar{\bm3}_d,\bm3_u):
		&\qquad
		&[g_{\cB_1}^{du}]^{i_d k_u}_{j_d l_u}=g_{\cB_1}^{du}\delta^{i_d}_{j_d}\delta^{k_u}_{l_u}\nonumber,
		\\
		&\cB^\mu_{1} \sim \bm1:
		&\qquad
		&g_{\cB_1}^\phi\nonumber
		.
	\end{alignat}
	\item Matching:
	\begin{itemize}
	\item $(\cB^\mu_{1})^{i_u}_{j_d} \sim (\bar{\bm3}_d,\bm3_u)$:
			\begin{equation}
		\begin{alignedat}{2}
			\cL_{\text{SMEFT}}&\supset
			-\frac{\lzu g_{\cB_1}^{du} \dzu^2}{3M_{\cB_1}^2}\lzm \cO_{ud}^{(1)}+6\cO_{ud}^{(8)} \dzm.
		\nonumber
		\end{alignedat}
	\end{equation}
	\item $\cB^\mu_{1} \sim \bm1$:
			\begin{equation}
		\begin{alignedat}{2}
			\cL_{\text{SMEFT}}\supset &
			-\frac{2|g_{\cB_1}^\phi|^2}{M_{\cB_1}^2}(\lambda_\phi+C_{\phi4}^{\cB_1})\cO_\phi
			+\frac{|g_{\cB_1}^\phi|^2}{M_{\cB_1}^2}\cO_{\phi D}
			-\frac{|g_{\cB_1}^\phi|^2}{2M_{\cB_1}^2}\cO_{\phi\Box}
			\\&
			-\frac{|g_{\cB_1}^\phi|^2}{2M_{\cB_1}^2}
			\lzv 
			[ y_e^*]^{i_\ell}_{j_e}[\cO_{e\phi}]^{j_e}_{i_\ell}
			+[ y_d^*]^{i_q}_{j_d}[\cO_{d\phi}]^{j_d}_{i_q}
			+[ y_u^*]^{i_q}_{j_u}[\cO_{u\phi}]^{j_u}_{i_q} +\hermc\dzv,
		\nonumber
		\end{alignedat}
	\end{equation}
	where
	\begin{equation}
		C_{\phi4}^{\cB_1}=\frac{\mu_\phi^2 |g_{\cB_1}^\phi|^2}{M_{\cB_1}^2}.
		\nonumber
	\end{equation}
	\end{itemize}
\end{itemize}

\subsubsection*{$\boldsymbol{\cW\sim(\bm1,\bm3)_0}$}

\begin{itemize}
	\item Lagrangian
	\begin{equation}
		-\cL_V^{(\le4)}\supset
		\frac{1}{2}(g_\cW^\ell)_{rij}\cW^{\mu a}_{r}\bar\ell_{i}\sig^a\gamma_\mu\ell_{j}
		+
		\frac{1}{2}(g_\cW^q)_{rij}\cW_r^{\mu a}\bar q_{i}\sig^a\gamma_\mu q_{j}
		+\lzs \frac{1}{2}(g_\cW^\phi)_r \cW_r^{\mu a}\phi^\dag \sig^a iD_\mu \phi+\hermc \dzs.
		\nonumber
	\end{equation}
	\item Flavor irreps:
		\begin{alignat}{3}
		&(\cW^\mu)\ud{i_q}{j_q} \sim \bm8_q:
		&\qquad
		&[g_\cW^{q,A_q}]\ud{i_q}{j_q}=g_\cW^q (T^{A_q})\ud{i_q}{j_q}\nonumber,
		\\		
		&(\cW^\mu)\ud{i_\ell}{j_\ell} \sim \bm8_\ell:
		&\qquad
		&[g_\cW^{\ell,A_\ell}]\ud{i_\ell}{j_\ell}=g_\cW^\ell (T^{A_\ell})\ud{i_\ell}{j_\ell},
		\nonumber
		\\
		&\cW^\mu \sim \bm1:
		&\qquad
		&[g_\cW^{\ell}]^{i_\ell}_{j_\ell}=g_\cW^{\ell} \delta^{i_\ell}_{j_\ell},
		\quad
		[g_\cW^{q}]^{i_q}_{j_q}=g_\cW^{q}\delta^{i_q}_{j_q},\quad g_\cW^\phi
		\nonumber.
	\end{alignat}
	\item Matching:
	\begin{itemize}
	\item $(\cW^\mu)\ud{i_q}{j_q}\sim\bm8_q$:
			\begin{equation}
		\begin{alignedat}{2}
			\cL_{\text{SMEFT}}&\supset
			-\frac{(g_{\cW}^q)^2}{48M_{\cW}^2}\lzm 3\cO_{qq}^{(3)E}-\cO_{qq}^{(3)D} \dzm.
		\nonumber
		\end{alignedat}
	\end{equation}
	\item $(\cW^\mu)\ud{i_\ell}{j_\ell}\sim\bm8_\ell$:
			\begin{equation}
		\begin{alignedat}{2}
			\cL_{\text{SMEFT}}&\supset
			\frac{(g_\cW^\ell)^2}{48M_\cW^2}\lzm 5 \cO_{\ell\ell}^E-7\cO_{\ell\ell}^D \dzm.
		\nonumber
		\end{alignedat}
	\end{equation}
	\item $\cW^\mu\sim\bm1$:
			\begin{equation}
		\begin{alignedat}{2}
			\cL_{\text{SMEFT}}&\supset
			\frac{(g_\cW^\ell)^2}{8M_\cW^2}\lzm \cO_{\ell\ell}^D-2\cO_{\ell\ell}^E \dzm 
			-\frac{(g_\cW^q)^2}{8M_\cW^2}\cO_{qq}^{(3)D}
			-\frac{g_\cW^q g_\cW^\ell}{4M_\cW^2} \cO_{\ell q}^{(3)}
			-\frac{|g_\cW^\phi|^2}{M_\cW^2}(\lambda_\phi+C_{\phi4}^{\cW})\cO_\phi
			\\&
			+\lzs \frac{|g_\cW^\phi|^2}{4M_\cW^2}-\frac{\re( g_\cW^\phi g_\cW^\phi)}{4M_\cW^2}\dzs \cO_{\phi D}
			-\lzs \frac{|g_\cW^\phi|^2}{4M_\cW^2}+\frac{\re( g_\cW^\phi g_\cW^\phi)}{8M_\cW^2}\dzs \cO_{\phi \Box}
			\\&
			-\frac{\re( g_\cW^\phi)g_\cW^\ell}{4M_\cW^2}\cO_{\phi\ell}^{(3)}
			-\frac{\re( g_\cW^\phi)g_\cW^q}{4M_\cW^2}\cO_{\phi q}^{(3)}
			\\&
			+\Bigg\{ \lzs -\frac{i\,\im( g_\cW^\phi  g_\cW^\phi)}{8M_\cW^2} -\frac{|g_\cW^\phi|^2}{4M_\cW^2} +\frac{i\,\im( g_\cW^\phi)g_\cW^\ell}{4M_\cW^2}\dzs [ y_e^*]^{i_\ell}_{j_e}[\cO_{e\phi}]^{j_e}_{i_\ell} 
			\\&
			\phantom{i..}+ \lzs -\frac{i\,\im( g_\cW^\phi  g_\cW^\phi)}{8M_\cW^2} -\frac{|g_\cW^\phi|^2}{4M_\cW^2} +\frac{i\,\im( g_\cW^\phi)g_\cW^q}{4M_\cW^2}\dzs [ y_d^*]^{i_q}_{j_d}[\cO_{d\phi}]^{j_d}_{i_q} 
			\\&
			\phantom{i..}+ \lzs -\frac{i\,\im( g_\cW^\phi  g_\cW^\phi)}{8M_\cW^2} -\frac{|g_\cW^\phi|^2}{4M_\cW^2} -\frac{i\,\im( g_\cW^\phi)g_\cW^q}{4M_\cW^2}\dzs [ y_u^*]^{i_q}_{j_u}[\cO_{u\phi}]^{j_u}_{i_q}+\hermc\Bigg\},
		\nonumber
		\end{alignedat}
	\end{equation}
	where 
	\begin{equation}
		C_{\phi4}^{\cW}=\frac{\mu_\phi^2 |g_\cW^\phi|^2}{2M_\cW^2}.\nonumber
	\end{equation}
	\end{itemize}
\end{itemize}

\subsubsection*{$\boldsymbol{\cW_1\sim(\bm1,\bm3)_1}$}

\begin{itemize}
	\item Lagrangian:
	\begin{equation}
	\begin{alignedat}{2}
		-\cL_V^{(\le4)}&\supset
		\frac{1}{2}(g_{\cW_1})_r \cW_{1r}^{\mu a\dag}i(D_\mu \phi)^T i\sig_2\sig^a\phi+\hermc.
		\nonumber
	\end{alignedat}
\end{equation} 
	\item Flavor irreps:
		\begin{alignat}{3}
		&\cW_1^\mu\sim\bm1:
		&\qquad
		&g_{\cW_1}
		\nonumber.
	\end{alignat}
	\item Matching:
			\begin{equation}
		\begin{alignedat}{2}
			\cL_{\text{SMEFT}}\supset&
			-\frac{ \lzu  g_{\cW_1} \dzu^2}{2M_{\cW_1}^2}(\lambda_\phi+C_{\phi4}^{\cW_1})\cO_\phi
			-\frac{\lzu  g_{\cW_1} \dzu^2}{4M_{\cW_1}^2}\cO_{\phi D}
			-\frac{\lzu  g_{\cW_1} \dzu^2}{8M_{\cW_1}^2}\cO_{\phi\Box}
			\\&
			-\frac{\lzu  g_{\cW_1} \dzu^2}{8M_{\cW_1}^2}\lzv 
			[ y_e^*]^{i_\ell}_{j_e}[\cO_{e\phi}]^{j_e}_{i_\ell}
			+[ y_d^*]^{i_q}_{j_d}[\cO_{d\phi}]^{j_d}_{i_q}
			+[ y_u^*]^{i_q}_{j_u}[\cO_{u\phi}]^{j_u}_{i_q} +\hermc\dzv,
		\nonumber
		\end{alignedat}
	\end{equation}
	where 
	\begin{equation}
		C_{\phi4}^{\cW_1}=\frac{\mu_\phi^2 \lzu g_{\cW_1} \dzu^2}{4M_{\cW_1}^2}.\nonumber
	\end{equation}
\end{itemize}

\subsubsection*{$\boldsymbol{\cL_3\sim(\bm1,\bm2)_{-\frac{3}{2}}}$}

\begin{itemize}
	\item Lagrangian
	\begin{equation}
		-\cL_V^{(\le4)}\supset (g_{\cL_3})_{rij}\cL_{3r}^{\mu\dag}\bar e_{i}^c\gamma_\mu \ell_{j}+\hermc.
		\nonumber
	\end{equation}
	\item Flavor irreps:
		\begin{alignat}{3}
		&(\cL_3^\mu)^{i_e j_\ell} \sim (\bm3_e,\bm3_\ell):
		&\qquad
		&[g_{\cL_3}]^{i_e k_\ell}_{j_e l_\ell}=g_{\cL_3} \delta^{i_e}_{j_e}\delta^{k_\ell}_{l_\ell}\nonumber.
	\end{alignat}
	\item Matching:
			\begin{equation}
		\begin{alignedat}{2}
			\cL_{\text{SMEFT}}&\supset 
			\frac{\lzu g_{\cL_3} \dzu^2}{M_{\cL_3}^2}\cO_{\ell e}.
		\nonumber
		\end{alignedat}
	\end{equation}
\end{itemize}


\subsubsection{Color triplets}

\subsubsection*{$\boldsymbol{\cU_2\sim(\bm3,\bm1)_{\frac{2}{3}}}$}

\begin{itemize}
	\item Lagrangian:
	\begin{equation}
		-\cL_V^{(\le4)}\supset
		(g_{\cU_2}^{ed})_{rij}\cU_{2r}^{\mu\dag}\bar e_{i}\gamma_\mu d_{j}
		+
		(g_{\cU_2}^{\ell q})_{rij}\cU_{2r}^{\mu\dag}\bar\ell_{i}\gamma_\mu q_{j}
		+\hermc.
		\nonumber 
	\end{equation}
	\item Flavor irreps:
		\begin{alignat}{3}
		(\cU_2^\mu)^{i_d}_{j_e} &\sim (\bar{\bm3}_e,\bm3_d):
		&\qquad
		&[g_{\cU_2}^{ed}]^{i_e k_d}_{j_e l_d}=g_{\cU_2}^{ed} \delta^{i_e}_{j_e}\delta^{k_d}_{l_d}\nonumber,
		\\
		(\cU_2^\mu)^{i_q}_{j_\ell} &\sim (\bar{\bm3}_\ell,\bm3_q):
		&\qquad
		&[g_{\cU_2}^{\ell q}]^{i_\ell k_q}_{j_\ell l_q}=g_{\cU_2}^{\ell q} \delta^{i_\ell}_{j_\ell}\delta^{k_q}_{l_q}\nonumber.
	\end{alignat}
	\item Matching:
	\begin{itemize}
	\item $(\cU_2^\mu)^{i_d}_{j_e}\sim (\bar{\bm3}_e,\bm3_d)$:
		\begin{equation}
		\begin{alignedat}{2}
			\cL_{\text{SMEFT}}&\supset
			-\frac{\lzu g_{\cU_2}^{ed} \dzu^2}{M_{\cU_2}^2}\cO_{ed}.
		\nonumber
		\end{alignedat}
	\end{equation}
	\item $(\cU_2^\mu)^{i_q}_{j_\ell}\sim(\bar{\bm3}_\ell,\bm3_q)$:
		\begin{equation}
		\begin{alignedat}{2}
			\cL_{\text{SMEFT}}&\supset
			-\frac{|g_{\cU_2}^{\ell q}|^2}{2M_{\cU_2}^2}\lzm \cO_{\ell q}^{(1)}
			+\cO_{\ell q}^{(3)} \dzm.
		\nonumber
		\end{alignedat}
	\end{equation}
	\end{itemize}
\end{itemize}

\subsubsection*{$\boldsymbol{\cU_5\sim(\bm3,\bm1)_{\frac{5}{3}}}$}

\begin{itemize}
	\item Lagrangian:
	\begin{equation}
		-\cL_V^{(\le4)}\supset
		(g_{\cU_5})_{rij}\cU_{5r}^{\mu\dag}\bar e_{i}\gamma_\mu u_{j}+\hermc.
		\nonumber
	\end{equation}
	\item Flavor irreps:
		\begin{alignat}{3}
		(\cU_5^\mu)^{i_u}_{j_e} &\sim (\bar{\bm3}_e,\bm3_u):
		&\qquad
		&[g_{\cU_5}]^{i_e k_u}_{j_e l_u}=g_{\cU_5} \delta^{i_e}_{j_e}\delta^{k_u}_{l_u}\nonumber.
	\end{alignat}
	\item Matching:
		\begin{equation}
		\begin{alignedat}{2}
			\cL_{\text{SMEFT}}&\supset 
			-\frac{\lzu g_{\cU_5} \dzu^2}{M_{\cU_5}^2}\cO_{eu}.
		\nonumber
		\end{alignedat}
	\end{equation}
\end{itemize}

\subsubsection*{$\boldsymbol{\cQ_1\sim(\bm3,\bm2)_{\frac{1}{6}}}$}

\begin{itemize}
	\item Lagrangian:
	\begin{equation}
		-\cL_V^{(\le4)}\supset
		(g_{\cQ_1}^{u\ell})_{rij} \cQ_{1r}^{\mu\dag}\bar u_{i}^c\gamma_\mu \ell_{j}
		+
		(g_{\cQ_1}^{dq})_{rij}\cQ_{1r}^{\mu \alpha\dag}\varepsilon_{\alpha\beta\gamma}\bar d^\beta_{i}\gamma_\mu i\sig_2q^{c\gamma}_{j}
		+\hermc.
		\nonumber
	\end{equation}
	\item Flavor irreps:
		\begin{alignat}{3}
		(\cQ_1^\mu)^{i_u j_\ell} &\sim (\bm3_u,\bm3_\ell):
		&\qquad
		&[g_{\cQ_1}^{u\ell}]^{i_u k_\ell}_{j_u l_\ell}=g_{\cQ_1}^{u\ell} \delta^{i_u}_{j_u}\delta^{k_\ell}_{l_\ell}\nonumber,
		\\
		(\cQ_1^\mu)_{i_d j_q} &\sim (\bar{\bm3}_d,\bar{\bm3}_q):
		&\qquad
		&[g_{\cQ_1}^{dq}]^{i_d k_q}_{j_d l_q}=g_{\cQ_1}^{dq} \delta^{i_d}_{j_d}\delta^{k_q}_{l_q}\nonumber.
		\nonumber
	\end{alignat}
	\item Matching:
	\begin{itemize}
	\item $(\cQ_1^\mu)^{i_u j_\ell} \sim (\bm3_u,\bm3_\ell)$:
		\begin{equation}
		\begin{alignedat}{2}
			\cL_{\text{SMEFT}}&\supset
			\frac{\lzu g_{\cQ_1}^{u\ell} \dzu^2}{M_{\cQ_1}^2}\cO_{\ell u}.
		\nonumber
		\end{alignedat}
	\end{equation}
	\item $(\cQ_1^\mu)_{i_d j_q}\sim (\bar{\bm3}_d,\bar{\bm3}_q)$:
	\begin{equation}
		\begin{alignedat}{2}
			\cL_{\text{SMEFT}}&\supset
			\frac{2|g_{\cQ_1}^{dq}|^2}{3M_{\cQ_1}^2}\lzm \cO_{qd}^{(1)} -3\cO_{qd}^{(8)} \dzm.
		\nonumber
		\end{alignedat}
	\end{equation}
	\end{itemize}
\end{itemize}

\subsubsection*{$\boldsymbol{\cQ_5\sim(\bm3,\bm2)_{-\frac{5}{6}}}$}

\begin{itemize}
	\item Lagrangian:
	\begin{equation}
		\begin{alignedat}{2}
			-\cL_V^{(\le4)}&\supset
			(g_{\cQ_5}^{d\ell})_{rij}\cQ_{5r}^{\mu\dag}\bar d^c_{i}\gamma_\mu \ell_{j}
			+
			(g_{\cQ_5}^{eq})_{rij}\cQ_{5r}^{\mu\dag}\bar e^c_{i}\gamma_\mu q_{j}
			+
			(g_{\cQ_5}^{uq})_{rij}\cQ_{5r}^{\mu\alpha\dag}\varepsilon_{\alpha\beta\gamma}\bar u^\beta_{i}\gamma_\mu q^{c\gamma}_{j}+\hermc.
			\nonumber
		\end{alignedat}
	\end{equation}
	\item Flavor irreps:
		\begin{alignat}{3}
		(\cQ_5^\mu)^{i_d j_\ell} &\sim (\bm3_d,\bm3_\ell):
		&\qquad
		&[g_{\cQ_5}^{d\ell}]^{i_d k_\ell}_{j_d l_\ell}=g_{\cQ_5}^{d\ell} \delta^{i_d}_{j_d}\delta^{k_\ell}_{l_\ell}\nonumber,
		\\
		(\cQ_5^\mu)^{i_e j_q} &\sim (\bm3_e,\bm3_q):
		&\qquad
		&[g_{\cQ_5}^{eq}]^{i_e k_q}_{j_e l_q}=g_{\cQ_5}^{eq} \delta^{i_e}_{j_e}\delta^{k_q}_{l_q}\nonumber,
		\\
		(\cQ_5^\mu)_{i_u j_q} &\sim (\bar{\bm3}_u,\bar{\bm3}_q):
		&\qquad
		&[g_{\cQ_5}^{uq}]^{i_u k_q}_{j_u l_q}=g_{\cQ_5}^{uq} \delta^{i_u}_{j_u}\delta^{k_q}_{l_q}\nonumber.
	\end{alignat}
	\item Matching:
	\begin{itemize}
	\item $(\cQ_5^\mu)^{i_d j_\ell} \sim (\bm3_d,\bm3_\ell)$:
	\begin{equation}
		\begin{alignedat}{2}
			\cL_{\text{SMEFT}}&\supset
			\frac{\lzu g_{\cQ_5}^{d\ell} \dzu^2}{M_{Q_5}^2}\cO_{\ell d}.
		\nonumber
		\end{alignedat}
	\end{equation}
	\item $(\cQ_5^\mu)^{i_e j_q} \sim (\bm3_e,\bm3_q)$:
	\begin{equation}
		\begin{alignedat}{2}
			\cL_{\text{SMEFT}}&\supset
			\frac{|g_{\cQ_5}^{eq}|^2}{M_{Q_5}^2}\cO_{qe}.
		\nonumber
		\end{alignedat}
	\end{equation}
	\item $(\cQ_5^\mu)_{i_u j_q} \sim (\bar{\bm3}_u,\bar{\bm3}_q)$:
	\begin{equation}
		\begin{alignedat}{2}
			\cL_{\text{SMEFT}}&\supset
			\frac{2|g_{\cQ_5}^{uq}|^2}{3M_{\cQ_5}^2}\lzm \cO_{qu}^{(1)} -3\cO_{qu}^{(8)} \dzm.
		\nonumber
		\end{alignedat}
	\end{equation}
	\end{itemize}
\end{itemize}

\subsubsection*{$\boldsymbol{\cX\sim(\bm3,\bm3)_{\frac{2}{3}}}$}

\begin{itemize}
	\item Lagrangian:
	\begin{equation}
		-\cL_V^{(\le4)}\supset
		\frac{1}{2}(g_\cX)_{rij}\cX_r^{\mu a\dag}\bar\ell_{i}\gamma_\mu\sig^aq_{j}+\hermc.\nonumber
	\end{equation}
	\item Flavor irreps:
		\begin{alignat}{3}
		(\cX^\mu)^{i_q}_{j_\ell} &\sim (\bar{\bm3}_\ell,\bm3_q):
		&\qquad
		&[g_\cX]^{i_\ell k_q}_{j_\ell l_q}=g_\cX \delta^{i_\ell}_{j_\ell}\delta^{k_q}_{l_q}\nonumber.
	\end{alignat}
	\item Matching:
	\begin{equation}
		\begin{alignedat}{2}
			\cL_{\text{SMEFT}}&\supset
			-\frac{\lzu g_\cX \dzu^2}{8M_\cX^2}\lzm 3\cO_{\ell q}^{(1)}-\cO_{\ell q}^{(3)} \dzm.
		\nonumber
		\end{alignedat}
	\end{equation}
\end{itemize}


\subsubsection{Color sextets}

\subsubsection*{$\boldsymbol{\cY_1\sim(\bar{\bm6},\bm2)_{\frac{1}{6}}}$}

\begin{itemize}
	\item Lagrangian:
	\begin{equation}
		-\cL_V^{(\le4)}\supset
		\frac{1}{2}(g_{\cY_1})_{rij}\cY_{1r}^{\mu \alpha\beta\dag}\bar d_{i}^{(\alpha|}\gamma_\mu i\sig_2 q_{j}^{c|\beta)}
		+\hermc.\nonumber
	\end{equation}
	\item Flavor irreps:
		\begin{alignat}{3}
		(\cY_1^\mu)_{i_d j_q} &\sim (\bar{\bm3}_d,\bar{\bm3}_q):
		&\qquad
		&[g_{\cY_1}]^{i_d k_q}_{j_d l_q}=g_{\cY_1} \delta^{i_d}_{j_d}\delta^{k_q}_{l_q}\nonumber.
	\end{alignat}
	\item Matching:
	\begin{equation}
		\begin{alignedat}{2}
			\cL_{\text{SMEFT}}&\supset
			\frac{\lzu g_{\cY_1} \dzu^2}{3M_{\cY_1}^2}\lzm 2\cO_{qd}^{(1)}+3\cO_{qd}^{(8)} \dzm.
		\nonumber
		\end{alignedat}
	\end{equation}
\end{itemize}

\subsubsection*{$\boldsymbol{\cY_5\sim(\bar{\bm6},\bm2)_{-\frac{5}{6}}}$}

\begin{itemize}
	\item Lagrangian:
	\begin{equation}
		-\cL_V^{(\le4)}\supset
		\frac{1}{2}(g_{\cY_5})_{rij}\cY_{5r}^{\mu \alpha\beta \dag}\bar u_{i}^{(\alpha|}\gamma_\mu i\sig_2 q_{j}^{c|\beta)}+\hermc.\nonumber
	\end{equation}
	\item Flavor irreps:
		\begin{alignat}{3}
		(\cY_5^\mu)_{i_u j_q} &\sim (\bar{\bm3}_u,\bar{\bm3}_q):
		&\qquad
		&[g_{\cY_5}]^{i_u k_q}_{j_u l_q}=g_{\cY_5} \delta^{i_u}_{j_u}\delta^{k_q}_{l_q}\nonumber.
	\end{alignat}
	\item Matching:
	\begin{equation}
		\begin{alignedat}{2}
			\cL_{\text{SMEFT}}&\supset
			\frac{\lzu g_{\cY_5} \dzu^2}{3M_{\cY_5}^2}\lzm 2\cO_{qu}^{(1)}+3\cO_{qu}^{(8)} \dzm.
		\nonumber
		\end{alignedat}
	\end{equation}
\end{itemize}

\subsubsection{Color octets}

\subsubsection*{$\boldsymbol{\cG\sim(\bm8,\bm1)_{0}}$}

\begin{itemize}
	\item Lagrangian:
	\begin{equation}
		-\cL_V^{(\le4)}\supset
		(g_\cG^q)_{rij}\cG_r^{\mu A}\bar q_{i}\gamma_\mu T_A q_{j}
		+
		(g_\cG^u)_{rij}\cG_r^{\mu A}\bar u_{i}\gamma_\mu T_A u_{j}
		+
		(g_\cG^d)_{rij}\cG_r^{\mu A}\bar d_{i}\gamma_\mu T_A d_{j}.\nonumber 
	\end{equation}
	\item Flavor irreps:
		\begin{alignat}{3}
		&(\cG^\mu)\ud{i_q}{j_q} \sim \bm8_q:
		&\qquad
		&[g_\cG^{q,A_q}]\ud{i_q}{j_q}=g_\cG^q (T^{A_q})\ud{i_q}{j_q}\nonumber,
		\\
		&(\cG^\mu)\ud{i_u}{j_u} \sim \bm8_u:
				&\qquad
		&[g_\cG^{u,A_u}]\ud{i_u}{j_u}=g_\cG^u (T^{A_u})\ud{i_u}{j_u}\nonumber,
		\\
		&(\cG^\mu)\ud{i_d}{j_d} \sim \bm8_d:
				&\qquad
		&[g_\cG^{d,A_d}]\ud{i_d}{j_d}=g_\cG^d (T^{A_d})\ud{i_d}{j_d}\nonumber,
		\\
		&\cG^\mu \sim \bm1:
				&\qquad
		&[g_\cG^{q}]^{i_q}_{j_q}=g_\cG^q\delta^{i_q}_{j_q},
		\quad 
		[g_\cG^{u}]^{i_u}_{j_u}=g_\cG^u\delta^{i_u}_{j_u},
		\quad 
		[g_\cG^{d}]^{i_d}_{j_d}=g_\cG^d\delta^{i_d}_{j_d}
		\nonumber.
	\end{alignat}
	\item Matching:
	\begin{itemize}
	\item $(\cG^\mu)\ud{i_q}{j_q}\sim\bm8_q$:
		\begin{equation}
		\begin{alignedat}{2}
			\cL_{\text{SMEFT}}&\supset
			-\frac{(g_\cG^q)^2}{144M_\cG^2}\lzm 11\cO_{qq}^{(1)D}-9\cO_{qq}^{(1)E}+9\cO_{qq}^{(3)D}-3\cO_{qq}^{(3)E} \dzm.
		\nonumber
		\end{alignedat}
	\end{equation}
	\item $(\cG^\mu)\ud{i_u}{j_u}\sim\bm8_u$:
		\begin{equation}
		\begin{alignedat}{2}
			\cL_{\text{SMEFT}}&\supset 
			\frac{(g_\cG^u)^2}{36M_\cG^2}\lzm 
			3\cO_{uu}^E
			-5\cO_{uu}^D \dzm.
		\nonumber
		\end{alignedat}
	\end{equation}
	\item $(\cG^\mu)\ud{i_d}{j_d}\sim\bm8_d$:
		\begin{equation}
		\begin{alignedat}{2}
			\cL_{\text{SMEFT}}&\supset 
			\frac{(g_\cG^d)^2}{36M_\cG^2}\lzm 
			3\cO_{dd}^E
			-5\cO_{dd}^D \dzm.
		\nonumber
		\end{alignedat}
	\end{equation}
	\item $\cG^\mu\sim\bm1$:
		\begin{equation}
		\begin{alignedat}{2}
			\cL_{\text{SMEFT}}&\supset
			\frac{(g_\cG^d)^2}{12M_\cG^2}\lzm \cO_{dd}^D-3\cO_{dd}^E \dzm
			+\frac{(g_\cG^u)^2}{12M_\cG^2}\lzm \cO_{uu}^D-3\cO_{uu}^E \dzm
			-\frac{(g_\cG^q)^2}{8M_\cG^2}\cO_{qq}^{(3)E}
			\\&
			-\frac{g_\cG^ug_\cG^q}{M_\cG^2}\cO_{qu}^{(8)}
			-\frac{g_\cG^dg_\cG^q}{M_\cG^2}\cO_{qd}^{(8)}
			+\frac{(g_\cG^q)^2}{24M_\cG^2}\lzm 2\cO_{qq}^{(1)D}-3\cO_{qq}^{(1)E} \dzm-\frac{g_\cG^dg_\cG^u}{M_\cG^2}\cO^{(8)}_{ud}.
		\nonumber
		\end{alignedat}
	\end{equation}
	\end{itemize}
\end{itemize}

\subsubsection*{$\boldsymbol{\cG_1\sim(\bm8,\bm1)_{1}}$}

\begin{itemize}
	\item Lagrangian:
	\begin{equation}
		-\cL_V^{(\le4)}\supset
		(g_{\cG_1})_{rij}\cG_{1r}^{\mu A \dag}\bar d_{i}T_A \gamma_\mu u_{j}+\hermc.
		\nonumber
	\end{equation}
	\item Flavor irreps:
		\begin{alignat}{3}
		(\cG_1^\mu)^{i_u}_{j_d} &\sim (\bar{\bm3}_d,\bm3_u):
		&\qquad
		&[g_{\cG_1}]^{i_d k_u}_{j_d l_u}=g_{\cG_1}\delta^{i_d}_{j_d}\delta^{k_u}_{l_u}
		\nonumber.
	\end{alignat}
	\item Matching:
		\begin{equation}
		\begin{alignedat}{2}
			\cL_{\text{SMEFT}}&\supset
			\frac{\lzu g_{\cG_1} \dzu^2}{9M_{\cG_1}^2}\lzm -4\cO^{(1)}_{ud} +3\cO_{ud}^{(8)} \dzm.
		\nonumber
		\end{alignedat}
	\end{equation}
\end{itemize}

\subsubsection*{$\boldsymbol{\cH\sim(\bm8,\bm3)_{0}}$}

\begin{itemize}
	\item Lagrangian:
	\begin{equation}
		-\cL_V^{(\le4)}\supset
		\frac{1}{2}(g_\cH)_{rij}\cH^{\mu aA}_r\bar q_{i}\gamma_\mu\sig^a T_A q_{j}.\nonumber
	\end{equation}
	\item Flavor irreps:
		\begin{alignat}{3}
		(\cH^\mu)\ud{i_q}{j_q} &\sim \bm8_q:
		&\qquad
		&[g_\cH^{A_q}]\ud{i_q}{j_q}=g_\cH(T^{A_q})\ud{i_q}{j_q},\nonumber
		\\
		\cH^\mu &\sim \bm1:
		&\qquad
		&[g_\cH]^{i_q}_{j_q}=g_\cH \delta^{i_q}_{j_q}.\nonumber
	\end{alignat}
	\item Matching:
	\begin{itemize}
	\item $(\cH^\mu)\ud{i_d}{j_d}\sim\bm8_q$:
		\begin{equation}
		\begin{alignedat}{2}
			\cL_{\text{SMEFT}}&\supset
			-\frac{(g_\cH)^2}{576 M_\cH^2}\lzm 27\cO_{qq}^{(1)D}-9\cO_{qq}^{(1)E}-7\cO_{qq}^{(3)D}-3\cO_{qq}^{(3)E} \dzm.
		\nonumber
		\end{alignedat}
	\end{equation}
	\item $\cH^\mu\sim\bm1$:
		\begin{equation}
		\begin{alignedat}{2}
			\cL_{\text{SMEFT}}&\supset
			\frac{(g_\cH)^2}{96M_\cH^2}\lzm 2\cO_{qq}^{(3)D}+3\cO_{qq}^{(3)E}-9\cO_{qq}^{(1)E} \dzm.
		\nonumber
		\end{alignedat}
	\end{equation}
	\end{itemize}
\end{itemize}






\section{Matching: SM + N fields}
\label{app:smN}

Apart from the trivial contributions that involve summing the WCs of individual fields previously calculated, there are nontrivial contribution when a single diagram in a UV theory has multiple BSM fields. All such cases are outlined in this Appendix.

\subsection{Matching: SM + 2 fields}
\label{app:sm2}

\subsubsection*{$\boldsymbol{\cS\sim(\bm1,\bm1)_{0}}$, $\boldsymbol{\Xi\sim(\bm1,\bm3)_{0}}$}
\begin{itemize}
	\item Lagrangian:
\begin{equation}
	\begin{alignedat}{2}
		-\cL_S^{(\le4)}&\supset 
		(\lambda_{\cS\Xi})_{rs}\cS_r\Xi_s^a(\phi^\dag\sig^a\phi)
		+(\kappa_{\cS\Xi})_{rst}\cS_r\Xi^a_s\Xi^a_t.
		\nonumber
	\end{alignedat}
\end{equation}
	\item Flavor irreps:
		\begin{alignat}{3}
		\cS &\sim \bm1,\quad \Xi\sim\bm1:
		&\qquad
		&\lambda_{\cS\Xi},\quad \kappa_{\cS\Xi},\quad \kappa_\cS,\quad \kappa_\Xi
		.\nonumber
	\end{alignat}
	\item Matching:
		\begin{equation}
		\begin{alignedat}{2}
			\cL_{\text{SMEFT}}&\supset \lzm -\frac{\lambda_{\cS\Xi}\kappa_\cS \kappa_\Xi}{M_\Xi^2 M_\cS^2}+\frac{\kappa_{\cS\Xi}\kappa_\cS \kappa_\Xi^2}{M_\Xi^4 M_\cS^2} \dzm\cO_\phi.
		\nonumber
		\end{alignedat}
	\end{equation}
\end{itemize}

\subsubsection*{$\boldsymbol{\cS\sim(\bm1,\bm1)_{0}}$, $\boldsymbol{\Xi_1\sim(\bm1,\bm3)_{1}}$}
\begin{itemize}
	\item Lagrangian:
\begin{equation}
	\begin{alignedat}{2}
		-\cL_S^{(\le4)}&\supset 
		(\kappa_{\cS\Xi_1})_{rst}\cS_r\Xi_{1s}^{a\dag}\Xi_{1t}^a
		+\lzs (\lambda_{\cS\Xi_1})_{rs}\cS_r\Xi_{1s}^{a\dag}(\tilde\phi^\dag\sig^a\phi) +\hermc \dzs.
		\nonumber
	\end{alignedat}
\end{equation}
	\item Flavor irreps:
		\begin{alignat}{3}
		\cS &\sim \bm1,\quad \Xi_1\sim\bm1:
		&\qquad
		&\lambda_{\cS\Xi_1},\quad \kappa_{\cS\Xi_1},\quad \kappa_\cS,\quad \kappa_{\Xi_1}
		.\nonumber
	\end{alignat}
	\item Matching:
			\begin{equation}
		\begin{alignedat}{2}
			\cL_{\text{SMEFT}}&\supset \lzs \frac{2\kappa_{\cS\Xi_1} \kappa_\cS \lzu \kappa_{\Xi_1} \dzu^2}{M_{\Xi_1}^4 M_\cS^2}-\frac{4\re(\lambda_{\cS\Xi_1}\kappa_{\Xi_1}^*)\kappa_\cS}{M_{\Xi_1}^2M_\cS^2} \dzs\cO_\phi.
		\nonumber
		\end{alignedat}
	\end{equation}
\end{itemize}

\subsubsection*{$\boldsymbol{\cS\sim(\bm1,\bm1)_{0}}$, $\boldsymbol{\varphi\sim(\bm1,\bm2)_{\frac{1}{2}}}$}
\begin{itemize}
	\item Lagrangian:
\begin{equation}
	\begin{alignedat}{2}
		-\cL_S^{(\le4)}&\supset 
		(\kappa_{\cS\varphi})_{rs}\cS_r\varphi_s^\dag\phi +\hermc.
		\nonumber
	\end{alignedat}
\end{equation}
	\item Flavor irreps:
		\begin{alignat}{3}
		\cS &\sim \bm1,\quad \varphi\sim\bm1:
		&\qquad
		&\kappa_{\cS\varphi},\quad \kappa_\cS,\quad \lambda_{\varphi}
		.\nonumber
	\end{alignat}
	\item Matching:
		\begin{equation}
		\begin{alignedat}{2}
			\cL_{\text{SMEFT}}&\supset
                \lzs -\frac{2\re(\kappa_{\cS\varphi} \lambda_\varphi^*)\kappa_\cS}{M_\cS^2 M_\varphi^2}+\frac{\lzu\kappa_{\cS\varphi}\dzu^2 \kappa_\cS^2}{M_\cS^4 M_\varphi^2} \dzs
			\cO_\phi.
		\nonumber
		\end{alignedat}
	\end{equation}
\end{itemize}

\subsubsection*{$\boldsymbol{\Xi\sim(\bm1,\bm3)_{0}}$, $\boldsymbol{\varphi\sim(\bm1,\bm2)_{\frac{1}{2}}}$}
\begin{itemize}
	\item Lagrangian:
\begin{equation}
	\begin{alignedat}{2}
		-\cL_S^{(\le4)}&\supset 
		(\kappa_{\Xi\varphi})_{rs}\Xi_r^a(\varphi_s^\dag\sig^a\phi)+\hermc.
		\nonumber
	\end{alignedat}
\end{equation}
	\item Flavor irreps:
		\begin{alignat}{3}
		\Xi &\sim \bm1,\quad \varphi\sim\bm1:
		&\qquad
		&\kappa_{\Xi\varphi},\quad \kappa_\Xi,\quad \lambda_{\varphi}
		.\nonumber
	\end{alignat}
	\item Matching:
		\begin{equation}
		\begin{alignedat}{2}
			\cL_{\text{SMEFT}}&\supset
			\lzs -\frac{2\re(\kappa_{\Xi\varphi}\lambda_\varphi^*)\kappa_\Xi}{M_\Xi^2M_\varphi^2}+\frac{\lzu \kappa_{\Xi\varphi} \dzu^2 \kappa_\Xi^2}{M_\varphi^2 M_\Xi^4} \dzs\cO_\phi.
		\nonumber
		\end{alignedat}
	\end{equation}
\end{itemize}

\subsubsection*{$\boldsymbol{\Xi_1\sim(\bm1,\bm3)_{1}}$, $\boldsymbol{\varphi\sim(\bm1,\bm2)_{\frac{1}{2}}}$}
\begin{itemize}
	\item Lagrangian:
\begin{equation}
	\begin{alignedat}{2}
		-\cL_S^{(\le4)}&\supset 
		(\kappa_{\Xi_1\varphi})_{rs}\Xi_{1r}^{a\dag}(\tilde\varphi_s^\dag\sig^a\phi)
		+\hermc.
		\nonumber
	\end{alignedat}
\end{equation}
	\item Flavor irreps:
		\begin{alignat}{3}
		\Xi_1 &\sim \bm1,\quad \varphi\sim\bm1:
		&\qquad
		&\kappa_{\Xi_1\varphi},\quad \kappa_{\Xi_1},\quad \lambda_{\varphi}
		.\nonumber
	\end{alignat}
	\item Matching:
		\begin{equation}
		\begin{alignedat}{2}
			\cL_{\text{SMEFT}}&\supset
			 \lzs -\frac{4\re(\kappa_{\Xi_1\varphi}^* \kappa_{\Xi_1} \lambda_\varphi^*)}{M_{\Xi_1}^2 M_\varphi^2}+\frac{4\lzu \kappa_{\Xi_1\varphi}\dzu^2 \lzu \kappa_{\Xi_1} \dzu^2}{M_{\Xi_1}^4 M_\varphi^2} \dzs \cO_\phi.
		\nonumber
		\end{alignedat}
	\end{equation}
\end{itemize}

\subsubsection*{$\boldsymbol{\Xi\sim(\bm1,\bm3)_{0}}$, $\boldsymbol{\Xi_1\sim(\bm1,\bm3)_{1}}$}
\begin{itemize}
	\item Lagrangian:
\begin{equation}
	\begin{alignedat}{2}
		-\cL_S^{(\le4)}&\supset 
		(\kappa_{\Xi\Xi_1})_{rst}f_{abc}\Xi_r^a\Xi_{1s}^{b\dag}\Xi_{1t}^{c}+\lzs (\lambda_{\Xi_1\Xi})_{rs}f_{abc}\Xi_{1r}^{a\dag}\Xi_s^b(\tilde\phi^\dag\sig^c\phi)+\hermc \dzs.
		\nonumber
	\end{alignedat}
\end{equation}
	\item Flavor irreps:
		\begin{alignat}{3}
		\Xi &\sim \bm1,\quad \Xi_1 \sim\bm1:
		&\qquad
		&\kappa_{\Xi\Xi_1},\quad \lambda_{\Xi_1\Xi},\quad \kappa_{\Xi},\quad \kappa_{\Xi_1}
		.\nonumber
	\end{alignat}
	\item Matching:
		\begin{equation}
		\begin{alignedat}{2}
			\cL_{\text{SMEFT}}&\supset
			\lzs -\frac{\sqrt2 \kappa_{\Xi\Xi_1} \kappa_\Xi \lzu\kappa_{\Xi_1}\dzu^2}{M_{\Xi_1}^4 M_\Xi^2}-\frac{2\sqrt2 \re(\lambda_{\Xi_1\Xi} \kappa_{\Xi_1}^*)\kappa_\Xi}{M_\Xi^2 M_{\Xi_1}^2} \dzs \cO_\phi.
		\nonumber
		\end{alignedat}
	\end{equation}
\end{itemize}

\subsubsection*{$\boldsymbol{\Xi\sim(\bm1,\bm3)_{0}}$, $\boldsymbol{\Theta_1\sim(\bm1,\bm4)_{\frac{1}{2}}}$}

\begin{itemize}
	\item Lagrangian:
\begin{equation}
	\begin{alignedat}{2}
		-\cL_S^{(\le4)}&\supset 
		(\kappa_{\Xi\Theta_1})_{rs}\Xi_r^aC_{a\beta}^I\tilde\phi_\beta\varepsilon_{IJ}\Theta_{1s}^J+\hermc.
		\nonumber
	\end{alignedat}
\end{equation}
	\item Flavor irreps:
		\begin{alignat}{3}
		\Xi &\sim \bm1,\quad \Theta_1 \sim\bm1:
		&\qquad
		&\kappa_{\Xi\Theta_1},\quad \lambda_{\Theta_1},\quad \kappa_{\Xi}.\nonumber
	\end{alignat}
	\item Matching:
		\begin{equation}
		\begin{alignedat}{2}
			\cL_{\text{SMEFT}}&\supset
			\lzs -\frac{\re(\kappa_{\Xi\Theta_1} \lambda_{\Theta_1}^*)\kappa_\Xi}{3M_{\Theta_1}^2 M_\Xi^2}+\frac{\lzu\kappa_{\Xi\Theta_1}\dzu^2 \kappa_\Xi^2}{6M_\Xi^4 M_{\Theta_1}^2} \dzs\cO_\phi.
		\nonumber
		\end{alignedat}
	\end{equation}
\end{itemize}

\subsubsection*{$\boldsymbol{\Xi_1\sim(\bm1,\bm3)_{1}}$, $\boldsymbol{\Theta_1\sim(\bm1,\bm4)_{\frac{1}{2}}}$}
\begin{itemize}
	\item Lagrangian:
\begin{equation}
	\begin{alignedat}{2}
		-\cL_S^{(\le4)}&\supset 
		(\kappa_{\Xi_1\Theta_1})_{rs}\Xi_{1r}^{a\dag}C_{a\beta}^I\phi_\beta\varepsilon_{IJ}\Theta_{1s}^J+\hermc.
		\nonumber
	\end{alignedat}
\end{equation}
	\item Flavor irreps:
		\begin{alignat}{3}
		\Xi_1 &\sim \bm1,\quad \Theta_1 \sim\bm1:
		&\qquad
		&\kappa_{\Xi_1\Theta_1},\quad \lambda_{\Theta_1},\quad \kappa_{\Xi_1}.\nonumber
	\end{alignat}
	\item Matching:
		\begin{equation}
		\begin{alignedat}{2}
			\cL_{\text{SMEFT}}&\supset
			 \lzs -\frac{\re(\kappa_{\Xi_1\Theta_1} \kappa_{\Xi_1}^* \lambda_{\Theta_1}^*)}{3M_{\Xi_1}^2 M_{\Theta_1}^2}+\frac{\lzu\kappa_{\Xi_1 \Theta_1}\dzu^2 \lzu\kappa_{\Xi_1}\dzu^2}{6M_{\Theta_1}^2 M_\Xi^4} \dzs\cO_\phi.
		\nonumber
		\end{alignedat}
	\end{equation}
\end{itemize}

\subsubsection*{$\boldsymbol{\Xi_1\sim(\bm1,\bm3)_{1}}$, $\boldsymbol{\Theta_3\sim(\bm1,\bm4)_{\frac{3}{2}}}$}

\begin{itemize}
	\item Lagrangian:
\begin{equation}
	\begin{alignedat}{2}
		-\cL_S^{(\le4)}&\supset 
		(\kappa_{\Xi_1\Theta_3})_{rs}\Xi_{1r}^{a\dag}C_{a\beta}^I\tilde\phi_\beta\varepsilon_{IJ}\Theta_{3s}^J+\hermc.
		\nonumber
	\end{alignedat}
\end{equation}
	\item Flavor irreps:
		\begin{alignat}{3}
		\Xi_1 &\sim \bm1,\quad \Theta_3 \sim\bm1:
		&\qquad
		&\kappa_{\Xi_1\Theta_3},\quad \lambda_{\Theta_3},\quad \kappa_{\Xi_1}.\nonumber
	\end{alignat}
	\item Matching:
		\begin{equation}
		\begin{alignedat}{2}
			\cL_{\text{SMEFT}}&\supset
			 \lzs -\frac{\re(\kappa_{\Xi_1\Theta_3}\kappa_{\Xi_1}^* \lambda_{\Theta_3}^*)}{M_{\Xi_1}^2 M_{\Theta_3}^2}+\frac{\lzu \kappa_{\Xi_1\Theta_3} \dzu^2 \lzu\kappa_{\Xi_1}\dzu^2}{2M_{\Theta_3}^2 M_{\Xi_1}^4} \dzs\cO_\phi.
		\nonumber
		\end{alignedat}
	\end{equation}
\end{itemize}


\subsection{Matching: SM + 3 fields}
\label{app:sm3}

\subsubsection*{$\boldsymbol{\cS\sim(\bm1,\bm1)_{0}}$, $\boldsymbol{\Xi\sim(\bm1,\bm3)_{0}}$, $\boldsymbol{\varphi\sim(\bm1,\bm2)_{\frac{1}{2}}}$}
\begin{itemize}
	\item Flavor irreps:
		\begin{alignat}{3}
		\cS_1 &\sim \bm1,\quad \Xi \sim\bm1,\quad \varphi\sim \bm1:
		&\qquad
		&\kappa_{\Xi\varphi},\quad \kappa_{\cS\varphi},\quad \kappa_\Xi,\quad \kappa_\cS
		.\nonumber
	\end{alignat}
	\item Matching:
			\begin{equation}
		\begin{alignedat}{2}
			\cL_{\text{SMEFT}}&\supset 
			 \frac{2\re(\kappa_{\Xi\varphi}^* \kappa_{\cS\varphi})\kappa_\Xi \kappa_\cS}{M_\Xi^2 M_\cS^2 M_\varphi^2}
			\cO_\phi.
		\nonumber
		\end{alignedat}
	\end{equation}
\end{itemize}

\subsubsection*{$\boldsymbol{\cS\sim(\bm1,\bm1)_{0}}$, $\boldsymbol{\Xi_1\sim(\bm1,\bm3)_{1}}$, $\boldsymbol{\varphi\sim(\bm1,\bm2)_{\frac{1}{2}}}$}
\begin{itemize}
	\item Flavor irreps:
		\begin{alignat}{3}
		\cS_1 &\sim \bm1,\quad \Xi_1 \sim\bm1,\quad \varphi\sim \bm1:
		&\qquad
		&\kappa_{\Xi_1\varphi},\quad \kappa_{\cS\varphi},\quad \kappa_{\Xi_1},\quad \kappa_\cS 
		.\nonumber
	\end{alignat}
	\item Matching:
			\begin{equation}
		\begin{alignedat}{2}
			\cL_{\text{SMEFT}}&\supset
			 \frac{4\re(\kappa_{\Xi_1\varphi} \kappa_{\Xi_1}^* \kappa_{\cS\varphi})\kappa_\cS}{M_\varphi^2 M_{\Xi_1}^2 M_\cS^2}\cO_\phi.
		\nonumber
		\end{alignedat}
	\end{equation}
\end{itemize}

\subsubsection*{$\boldsymbol{\Xi\sim(\bm1,\bm3)_{0}}$, $\boldsymbol{\Xi_1\sim(\bm1,\bm3)_{1}}$, $\boldsymbol{\varphi\sim(\bm1,\bm2)_{\frac{1}{2}}}$}
\begin{itemize}
	\item Flavor irreps:
		\begin{alignat}{3}
		\Xi &\sim \bm1,\quad \Xi_1 \sim\bm1,\quad \varphi\sim \bm1:
		&\qquad
		&\kappa_{\Xi\varphi},\quad \kappa_{\Xi\varphi},\quad \kappa_\Xi,\quad \kappa_{\Xi_1}
		.\nonumber
	\end{alignat}
	\item Matching:
		\begin{equation}
		\begin{alignedat}{2}
			\cL_{\text{SMEFT}}&\supset
			 \frac{4\re(\kappa_{\Xi_1\varphi}^* \kappa_{\Xi_1} \kappa_{\Xi\varphi}^*)\kappa_\Xi}{M_\varphi^2 M_{\Xi_1}^2 M_\Xi^2} \cO_\phi.
		\nonumber
		\end{alignedat}
	\end{equation}
\end{itemize}

\subsubsection*{$\boldsymbol{\Xi\sim(\bm1,\bm3)_{0}}$, $\boldsymbol{\Xi_1\sim(\bm1,\bm3)_{1}}$, $\boldsymbol{\Theta_1\sim(\bm1,\bm4)_{\frac{1}{2}}}$}
\begin{itemize}
	\item Flavor irreps:
		\begin{alignat}{3}
		\Xi &\sim \bm1,\quad \Xi_1 \sim\bm1,\quad \Theta_1\sim \bm1:
		&\qquad
		&\kappa_{\Xi\Theta_1},\quad \kappa_{\Xi_1\Theta_1},\quad \kappa_\Xi,\quad \kappa_{\Xi_1}
		.\nonumber
	\end{alignat}
	\item Matching:
		\begin{equation}
		\begin{alignedat}{2}
			\cL_{\text{SMEFT}}&\supset
			 \frac{\re(\kappa_{\Xi_1\Theta_1}^* \kappa_{\Xi_1} \kappa_{\Xi\Theta_1})\kappa_\Xi}{3M_\Xi^2 M_{\Xi_1}^2 M_{\Theta_1}^2} \cO_\phi.
		\nonumber
		\end{alignedat}
	\end{equation}
\end{itemize}

\section{Non-renormalizable interactions}
\label{app:nrint}
Lastly, for completeness, in this section, we list the dimension-5 interaction terms of BSM mediators with the SM fields allowed by the flavor symmetry together with the corresponding coupling tensors, and we present their contributions to the matching. Scale $f$ is introduced on dimensional grounds. We also note that the contributions to the matching are always proportional to the product of the form $\tilde g_f\times g_f$, where $\tilde g_f$ denotes the generic dimension-5 coupling and $g_f$ denotes either dimension-3 or dimension-4 coupling, so that, in order to have the non-vanishing contributions to the matching from these terms, the corresponding $g_f$ couplings need to be activated as well (see Appendix \ref{app:sm1-spin0-singlets} and \ref{app:sm1-spin12} for the relevant renormalizable terms and couplings).  For simplicity, we omit the contribution from the field $\mathcal L_1 \sim(1,2)_{\frac{1}{2}}$~\cite{deBlas:2017xtg}.

\subsection{Spin 0}
\subsubsection*{$\boldsymbol{\mathcal S\sim(1,1)_0}$}
\begin{itemize}
	\item Lagrangian:
	\begin{equation}
		\begin{alignedat}{2}
		-\cL_S^{(5)}&\supset\frac{1}{f}
		\bigg[
		(\tilde k_\cS^\phi)_r\cS_r (D_\mu\phi)^\dag D^\mu\phi
		+(\tilde\lambda_\cS)_r\cS_r|\phi|^4
		+(\tilde k^B_\cS)_r\cS_r B_{\mu\nu}B^{\mu\nu}
		+(\tilde k^{\tilde B}_\cS)_r\cS_r B_{\mu\nu}\tilde B^{\mu\nu}
		\\&
		+(\tilde k_\cS^W)_r\cS_r W^a_{\mu\nu} W^{a\,\mu\nu}
		+(\tilde k_\cS^{\tilde W})_r\cS_r W^a_{\mu\nu} \tilde W^{a\,\mu\nu}
		+(\tilde k_\cS^G)_r\cS_r G^A_{\mu\nu}G^{A\,\mu\nu}
		+(\tilde k_\cS^{\tilde G})_r\cS_r G^A_{\mu\nu}\tilde G^{A\,\mu\nu}
		\bigg].
		\nonumber
		\end{alignedat}
	\end{equation}
	\item Flavor irreps:
	\begin{alignat}{3}
		S&\sim\bm1:
		&\qquad
		\tilde k_\cS^\phi,\quad\tilde\lambda_\cS,\quad\tilde k_\cS^B,\quad\tilde k^{\tilde B}_\cS,\quad\tilde k_\cS^W,\quad\tilde k_\cS^{\tilde W},\quad\tilde k_\cS^G,\quad\tilde k_\cS^{\tilde G}.\nonumber
	\end{alignat}
	\item Matching:
			\begin{equation}
		\begin{alignedat}{2}
			\cL_{\text{SMEFT}}&\supset
			\frac{1}{f}\frac{\tilde k_\cS^\phi \kappa_\cS}{2M_\cS^2}\cO_{\phi\Box}
			+\frac{1}{f}\frac{\kappa_\cS}{M_\cS^2}\lzs 2\tilde k_\cS^\phi(\lambda_\phi+\tilde C_{\phi4}^\cS)  +\tilde\lambda_\cS \dzs\cO_\phi
			\\&
			+\frac{1}{f}\frac{\kappa_\cS}{M_\cS^2} \lzm \tilde k_\cS^B\cO_{\phi B}+\tilde k_\cS^{\tilde B}\cO_{\phi \tilde B}+\tilde k_\cS^{W}\cO_{\phi W}+\tilde k_\cS^{\tilde W}\cO_{\phi \tilde W}+\tilde k_\cS^G\cO_{\phi G}+\tilde k_\cS^{\tilde G}\cO_{\phi \tilde G} \dzm
			\\&
			+\lzv \frac{1}{f}\frac{\tilde k_\cS^\phi \kappa_\cS}{2M_\cS^2} \lzm [y_e^*]^{j_\ell}_{i_e}[\cO_{e\phi}]^{i_e}_{j_\ell}+[ y_d^*]^{j_q}_{i_d}[\cO_{d\phi}]^{i_d}_{j_q}+[ y_u^*]^{j_q}_{i_u}[\cO_{u\phi}]^{i_u}_{j_q} \dzm+\hermc \dzv,
		\nonumber
		\end{alignedat}
	\end{equation}
	where
	\begin{equation}
		\tilde C_{\phi4}^\cS=-\frac{1}{f}\frac{\mu_\phi^2 \tilde k_\cS^\phi \kappa_\cS}{M_\cS^2}.\nonumber
	\end{equation}
\end{itemize}

\subsubsection*{$\boldsymbol{\Xi\sim(1,3)_{0}}$}
\begin{itemize}
	\item Lagrangian:
	\begin{equation}
	\begin{alignedat}{2}
		-\cL_S^{(5)}&\supset 
		\frac{1}{f}
		\bigg\{
		(\tilde k^\phi_\Xi)_r\Xi^a_r D_\mu \phi^\dag \sig^a D^\mu \phi
		+(\tilde\lambda_\Xi)_r\Xi^a_r |\phi|^2\phi^\dag \sig^a\phi
		+(\tilde k^{WB}_\Xi)_r \Xi^a_r W^a_{\mu\nu}B^{\mu\nu}
		\\&
		+(\tilde k^{W\tilde B}_\Xi)_r \Xi^a_r W^a_{\mu\nu}\tilde B^{\mu\nu}
		\bigg\}.
		\nonumber
	\end{alignedat}
\end{equation}
	\item Flavor irreps:
	\begin{alignat}{3}
		\Xi &\sim\bm1:
		&\qquad
		\tilde k^\phi_\Xi,\quad\tilde\lambda_\Xi,\quad\tilde k^{WB}_\Xi,\quad\tilde k^{W\tilde B}_\Xi
		\nonumber.
	\end{alignat}
\item Matching:
		\begin{equation}
		\begin{alignedat}{2}
			\cL_{\text{SMEFT}}&\supset 
			\frac{1}{f}\frac{2\tilde k_\Xi^\phi \kappa_\Xi}{M_\Xi^2} \cO_{\phi D}
			-\frac{1}{f}\frac{\tilde k_\Xi^\phi \kappa_\Xi}{2M_\Xi^2}\cO_{\phi\Box}
			+\frac{1}{f}\frac{\kappa_\Xi}{M_\Xi^2}\lzm \tilde k_\Xi^{WB}\cO_{\phi WB}+\tilde k_\Xi^{W\tilde B}\cO_{\phi W\tilde B} \dzm
			\\&
			-\Bigg\{  \frac{1}{f}\frac{\tilde k_\Xi^\phi \kappa_\Xi}{2M_\Xi^2} \lzm [ y_e^*]^{j_\ell}_{i_e}[\cO_{e\phi}]^{i_e}_{j_\ell}+[ y_d^*]^{j_q}_{i_d}[\cO_{d\phi}]^{i_d}_{j_q}+[ y_u^*]^{j_q}_{i_u}[\cO_{u\phi}]^{i_u}_{j_q} \dzm+\hermc \Bigg\}
			\\&
			+\frac{1}{f}\frac{\kappa_\Xi}{M_\Xi^2}\lzs \tilde\lambda_\Xi-2\tilde k_\Xi^\phi(\lambda_\phi+\tilde C_{\phi4}^\Xi)  \dzs \cO_\phi,
		\nonumber
		\end{alignedat}
	\end{equation}
	where
	\begin{equation}
		\tilde C_{\phi4}^\Xi=\frac{1}{f}\frac{\mu_\phi^2 \tilde k_\Xi^\phi \kappa_\Xi}{M_\Xi^2}.\nonumber
	\end{equation}
\end{itemize}

\subsubsection*{$\boldsymbol{\Xi_1\sim(1,3)_{1}}$}
\begin{itemize}
	\item Lagrangian:
	\begin{equation}
		\begin{alignedat}{2}
		-\cL_S^{(5)}&\supset
			\frac{1}{f}\Big[(\tilde k_{\Xi_1})_r \Xi_{1r}^{a\dag} D_\mu \tilde\phi^\dag \sig^a D^\mu\phi
			+(\tilde\lambda_{\Xi_1})_r\Xi_{1r}^{a\dag}|\phi|^2\tilde\phi^\dag\sig^a\phi\Big]+\hermc.
			\nonumber
		\end{alignedat}
	\end{equation}
	\item Flavor irreps:
		\begin{alignat}{3}
		&\Xi_{1}
		\sim\bm1:
		&\qquad &\tilde k_{\Xi_1},\quad\tilde\lambda_{\Xi_1}\nonumber.
		\nonumber
	\end{alignat}
	\item Matching:
		\begin{equation}
		\small
		\begin{alignedat}{2}
			\cL_{\text{SMEFT}}&\supset 
			\frac{1}{f}\frac{4\re(\tilde\lambda_{\Xi_1}\kappa_{\Xi_1}^*)}{M_{\Xi_1}^2} \cO_{\phi}
			-\frac{1}{f}\frac{4\re(\tilde k_{\Xi_1} \kappa_{\Xi_1}^*)}{M_{\Xi_1}^2} \cO_{\phi D}
			-\frac{1}{f}\frac{2\re(\tilde k_{\Xi_1} \kappa_{\Xi_1}^*)}{M_{\Xi_1}^2} \cO_{\phi\Box}\\&
			-\Bigg\{\frac{1}{f}\frac{2i\,\im(\tilde k_{\Xi_1} \kappa_{\Xi_1}^*)}{M_{\Xi_1}^2}\lzm [ y_e^*]^{j_\ell}_{i_e}[\cO_{e\phi}]^{i_e}_{j_\ell}+[ y_d^*]^{j_q}_{i_d}[\cO_{d\phi}]^{i_d}_{j_q}+[ y_u^*]^{j_q}_{i_u}[\cO_{u\phi}]^{i_u}_{j_q} \dzm 
			+\hermc \Bigg\}.
		\nonumber
		\end{alignedat}
	\end{equation}
\end{itemize}

\subsection{Spin 1/2}
\subsubsection*{Color singlets}

\subsubsection*{$\boldsymbol{E\sim(1,1)_{-1}}$}
\begin{itemize}
	\item Lagrangian:
	\begin{equation}
		\begin{alignedat}{2}
		-\cL_{\text{leptons}}^{(5)}&\supset\frac{1}{f}
			(\tilde\lambda^\ell_E)_{ri}\bar E_{Li}\gamma^\mu(D_\mu\phi)^\dag\ell_{i}
			+\hermc.
			\nonumber
		\end{alignedat}
	\end{equation}
	\item Flavor irreps:
		\begin{alignat}{3}
		E_{L,R}^{i_\ell} &\sim \bm3_\ell:
		&\qquad
		[\tilde\lambda^\ell_E]^{i_\ell}_{j_\ell}=\tilde\lambda^\ell_E \delta^{i_\ell}_{j_\ell}.
		\nonumber
	\end{alignat}
	\item Matching:
			\begin{equation}
		\begin{alignedat}{2}
			\cL_{\text{SMEFT}}&\supset 
			\lzm \frac{i}{f}\frac{\tilde\lambda_E^\ell \lambda_E^*}{4M_E} -\frac{i}{f}\frac{(\tilde\lambda_E^\ell)^*\lambda_E}{4M_E} \dzm \lzm \cO_{\phi\ell}^{(1)} + \cO_{\phi\ell}^{(3)} \dzm
			\\&
			+\lzv \lzm \frac{i}{f}\frac{\tilde\lambda_E^\ell \lambda_E^*}{2M_E}+\frac{i}{f}\frac{(\tilde\lambda_E^\ell)^* \lambda_E}{2M_E} \dzm [y_e^*]^{i_\ell}_{j_e}[\cO_{e\phi}]^{j_e}_{i_\ell}+\hermc \dzv.
		\nonumber
		\end{alignedat}
	\end{equation}
\end{itemize}

\subsubsection*{$\boldsymbol{\Delta_1\sim(1,2)_{-\frac{1}{2}}}$}
\begin{itemize}
	\item Lagrangian:
	\begin{equation}
		\begin{alignedat}{2}
		-\cL_{\text{leptons}}^{(5)}&\supset\frac{1}{f}(\tilde\lambda^e_{\Delta_1})_{ri}\bar\Delta_{1Rr}\slashed{D}\phi e_{i}+\hermc.\nonumber
		\end{alignedat}
	\end{equation}
	\item Flavor irreps:
		\begin{alignat}{3}
		\Delta_{1L,R}^{i_e} &\sim \bm3_e:
		&\qquad
		[\tilde\lambda^e_{\Delta_1}]^{i_e}_{j_e}=\tilde\lambda^e_{\Delta_1} \delta^{i_e}_{j_e}.
		\nonumber
	\end{alignat}
	\item Matching:
			\begin{equation}
		\begin{alignedat}{2}
			\cL_{\text{SMEFT}}&\supset 
			\lzm\frac{i}{f}\frac{(\tilde\lambda^e_{\Delta_1})^* \lambda_{\Delta_1}}{2M_{\Delta_1}}-\frac{i}{f}\frac{\tilde\lambda^e_{\Delta_1}\lambda^*_{\Delta_1}}{2M_{\Delta_1}}\dzm \cO_{\phi e}
			\\&
			-\lzv \lzm \frac{i}{f}\frac{(\tilde\lambda^e_{\Delta_1})^* \lambda_{\Delta_1}}{2M_{\Delta_1}} +\frac{i}{f}\frac{\tilde\lambda^e_{\Delta_1}\lambda^*_{\Delta_1}}{2M_{\Delta_1}}\dzm [ y_e^*]^{i_\ell}_{j_e}[\cO_{e\phi}]^{j_e}_{i_\ell}+\hermc \dzv.
		\nonumber
		\end{alignedat}
	\end{equation}
\end{itemize}

\subsubsection*{$\boldsymbol{\Delta_3\sim(1,2)_{-\frac{3}{2}}}$}
\begin{itemize}
	\item Lagrangian:
	\begin{equation}
		\begin{alignedat}{2}
		-\cL_{\text{leptons}}^{(5)}&\supset\frac{1}{f}
			(\tilde\lambda^e_{\Delta_3})_{ri}\bar\Delta_{3Rr}\slashed{D}\tilde\phi e_{i}+\hermc.\nonumber
		\end{alignedat}
	\end{equation}
	\item Flavor irreps:
		\begin{alignat}{3}
		\Delta_{3L,R}^{i_e} &\sim \bm3_e:
		&\qquad
		[\tilde\lambda^e_{\Delta_3}]^{i_e}_{j_e}=\tilde\lambda^e_{\Delta_3} \delta^{i_e}_{j_e}.
		\nonumber
	\end{alignat}
	\item Matching:
			\begin{equation}
		\begin{alignedat}{2}
			\cL_{\text{SMEFT}}&\supset
			\lzm 
			\frac{i}{f}\frac{\tilde\lambda^e_{\Delta_3} \lambda_{\Delta_3}^*}{2M_{\Delta_3} }-\frac{i}{f}\frac{(\tilde\lambda^e_{\Delta_3})^* \lambda_{\Delta_3}}{2M_{\Delta_3}}
			\dzm \cO_{\phi e}
			\\&
			-\lzv \lzm 
			\frac{i}{f}\frac{(\tilde\lambda^e_{\Delta_3})^*\lambda_{\Delta_3}}{2M_{\Delta_3}}
			+\frac{i}{f}\frac{\tilde\lambda^e_{\Delta_3}\lambda_{\Delta_3}^*}{2M_{\Delta_3}} \dzm [y_e^*]^{i_\ell}_{j_e}[\cO_{e\phi}]^{j_e}_{i_\ell}+\hermc \dzv.
		\nonumber
		\end{alignedat}
	\end{equation}
\end{itemize}

\subsubsection*{$\boldsymbol{\Sigma\sim(1,3)_{0}}$}
\begin{itemize}
	\item Lagrangian:
	\begin{equation}
		\begin{alignedat}{2}
		-\cL_{\text{leptons}}^{(5)}&\supset\frac{1}{f}
			(\tilde\lambda_\Sigma^\ell)_{ri}\bar\Sigma^{ca}_{Rr}\gamma^\mu(D_\mu\tilde\phi)^\dag\sig^a\ell_{i}+\hermc.\nonumber
		\end{alignedat}
	\end{equation}
	\item Flavor irreps:
		\begin{alignat}{3}
		\Sigma_{L,R}^{i_\ell} &\sim \bm3_\ell:
		&\qquad
		[\tilde\lambda_\Sigma^\ell]^{i_\ell}_{j_\ell}=\tilde\lambda_\Sigma^\ell \delta^{i_\ell}_{j_\ell}\nonumber.		
	\end{alignat}
	\item Matching:
			\begin{equation}
		\begin{alignedat}{2}
			\cL_{\text{SMEFT}}&\supset 
			\lzm 
			\frac{i}{f}\frac{(\tilde\lambda^\ell_\Sigma)^* \lambda_\Sigma}{8M_\Sigma}
			-\frac{i}{f}\frac{\tilde\lambda^\ell_\Sigma \lambda_\Sigma^*}{8M_\Sigma} \dzm			\lzm 3 \cO_{\phi\ell}^{(1)} +\cO_{\phi\ell}^{(3)}\dzm
			\\&
			+\lzv \lzm 
			\frac{i}{f}\frac{(\tilde\lambda^\ell_\Sigma)^*\lambda_\Sigma}{2M_\Sigma}
			+\frac{i}{f}\frac{\tilde\lambda^\ell_\Sigma \lambda_\Sigma^*}{2M_\Sigma}
			\dzm [y_e^*]^{i_\ell}_{j_e}[\cO_{e\phi}]^{j_e}_{i_\ell}+\hermc \dzv.
		\nonumber
		\end{alignedat}
	\end{equation}
\end{itemize}

\subsubsection*{$\boldsymbol{\Sigma_1\sim(1,3)_{-1}}$}
\begin{itemize}
	\item Lagrangian:
	\begin{equation}
		\begin{alignedat}{2}
		-\cL_{\text{leptons}}^{(5)}&\supset\frac{1}{f}
			(\tilde\lambda^\ell_{\Sigma_1})_{ri}\bar\Sigma^a_{1Lr}\gamma^\mu(D_\mu\phi)^\dag\sig^a\ell_{i}+\hermc.
			\nonumber
		\end{alignedat}
	\end{equation}
	\item Flavor irreps:
		\begin{alignat}{3}
		\Sigma_{1L,R}^{i_\ell} &\sim \bm3_\ell:
		&\qquad
		[\tilde\lambda^\ell_{\Sigma_1}]^{i_\ell}_{j_\ell}=\tilde\lambda^\ell_{\Sigma_1} \delta^{i_\ell}_{j_\ell}.
		\nonumber
		\end{alignat}
	\item Matching:
	\begin{equation}
		\begin{alignedat}{2}
			\cL_{\text{SMEFT}}&\supset
			\lzm 
			\frac{i}{f}\frac{ (\tilde\lambda^\ell_{\Sigma_1})^*\lambda_{\Sigma_1}}{8M_{\Sigma_1}}-\frac{i}{f}\frac{\tilde\lambda^\ell_{\Sigma_1}\lambda^*_{\Sigma_1}}{8M_{\Sigma_1}}
			 \dzm \lzm \cO_{\phi\ell}^{(3)} -3 \cO_{\phi\ell}^{(1)} \dzm
			\\&
			+\lzv \lzm 
			\frac{i}{f}\frac{\tilde\lambda^\ell_{\Sigma_1} \lambda^*_{\Sigma_1}}{4M_{\Sigma_1}}
			+\frac{i}{f}\frac{(\tilde\lambda^\ell_{\Sigma_1})^* \lambda_{\Sigma_1}}{4M_{\Sigma_1}} \dzm [y_e^*]^{i_\ell}_{j_e}[\cO_{e\phi}]^{j_e}_{i_\ell}+\hermc \dzv.
		\nonumber
		\end{alignedat}
	\end{equation}
\end{itemize}

\subsubsection*{Color triplets}
\subsubsection*{$\boldsymbol{U\sim(3,1)_{\frac{2}{3}}}$}
\begin{itemize}
	\item Lagrangian:
	\begin{equation}
		\begin{alignedat}{2}
		-\cL^{(5)}_\text{quarks}&\supset\frac{1}{f}
			(\tilde\lambda^q_U)_{ri}\bar U_{Lr}\gamma^\mu(D_\mu\tilde\phi)^\dag q_{i}+\hermc.
			\nonumber
		\end{alignedat}
	\end{equation}
	\item Flavor irreps:
		\begin{alignat}{3}
		U_{L,R}^{i_q} &\sim \bm3_q:
		&\qquad
		[\tilde\lambda^q_U]^{i_q	}_{j_q}=\tilde\lambda^q_U \delta^{i_q}_{j_q}.
		\nonumber
	\end{alignat}
	\item Matching:
	\begin{equation}
		\begin{alignedat}{2}
			\cL_{\text{SMEFT}}&\supset 
			\lzm 			
			\frac{i}{f}\frac{(\tilde\lambda^q_{U})^*\lambda_{U}}{4M_{U}}-\frac{i}{f}\frac{\tilde\lambda^q_{U}\lambda^*_{U}}{4M_{U}}
			 \dzm \lzm \cO_{\phi q}^{(1)}-\cO_{\phi q}^{(3)} \dzm 
			\\&
			+\lzv 
			\lzm 
			\frac{i}{f}\frac{\tilde\lambda^q_{U} \lambda^*_{U}}{2M_{U}}
			+\frac{i}{f}\frac{(\tilde\lambda^q_{U})^* \lambda_{U}}{2M_{U}} \dzm [ y_u^*]^{i_q}_{j_u}[\cO_{u\phi}]^{j_u}_{i_q} +\hermc \dzv.
		\nonumber
		\end{alignedat}
	\end{equation}
\end{itemize}

\subsubsection*{$\boldsymbol{D\sim(3,1)_{-\frac{1}{3}}}$}
\begin{itemize}
	\item Lagrangian:
	\begin{equation}
		\begin{alignedat}{2}
		-\cL^{(5)}_\text{quarks}&\supset\frac{1}{f}
			(\tilde\lambda^q_D)_{ri}\bar D_{Lr}\gamma^\mu(D_\mu\phi)^\dag q_{i}+\hermc.
			\nonumber
		\end{alignedat}
	\end{equation}
	\item Flavor irreps:
		\begin{alignat}{3}
		D_{L,R}^{i_q} &\sim \bm3_q:
		&\qquad
		[\tilde\lambda^q_D]^{i_q	}_{j_q}=\tilde\lambda^q_D \delta^{i_q}_{j_q}.
		\nonumber
	\end{alignat}
	\item Matching:
	\begin{equation}
		\begin{alignedat}{2}
			\cL_{\text{SMEFT}}&\supset 
			\lzm 
			\frac{i}{f}\frac{\tilde\lambda^q_{D}\lambda^*_{D}}{4M_{D}}
			-\frac{i}{f}\frac{(\tilde\lambda^q_{D})^*\lambda_{D}}{4M_{D}} \dzm\lzm \cO_{\phi q}^{(1)} + \cO_{\phi q}^{(3)}\dzm 
			\\&
			+\lzv 
			\lzm 
			\frac{i}{f}\frac{\tilde\lambda^q_{D} \lambda^*_{D}}{2M_{D}}
			+\frac{i}{f}\frac{(\tilde\lambda^q_{D})^* \lambda_{D}}{2M_{D}} \dzm [y_d^*]^{i_q}_{j_d}[\cO_{d\phi}]^{j_d}_{i_q}+\hermc\dzv.
		\nonumber
		\end{alignedat}
	\end{equation}
\end{itemize}

\subsubsection*{$\boldsymbol{Q_1\sim(3,2)_{\frac{1}{6}}}$}
\begin{itemize}
	\item Lagrangian:
	\begin{equation}
		\begin{alignedat}{2}
		-\cL^{(5)}_\text{quarks}&\supset\frac{1}{f}
			\Big[
			(\tilde\lambda^u_{Q_1})_{ri}\bar Q_{1Rr}\slashed D \tilde\phi u_{i}
			+
			(\tilde\lambda^d_{Q_1})_{ri}\bar Q_{1Rr}\slashed D\phi d_{i}
			\Big]
			+\hermc.
			\nonumber
		\end{alignedat}
	\end{equation}
	\item Flavor irreps:
		\begin{alignat}{3}
		Q_{1L,R}^{i_u} &\sim \bm3_u:
		&\qquad
		[\tilde\lambda^u_{Q_1}]^{i_u}_{j_u}=\tilde\lambda^u_{Q_1} \delta^{i_u}_{j_u},
		\nonumber
		\\
		Q_{1L,R}^{i_d} &\sim \bm3_d:
		&\qquad
		[\tilde\lambda^d_{Q_1}]^{i_d}_{j_d}=\tilde\lambda^d_{Q_1} \delta^{i_d}_{j_d}
		\nonumber.
	\end{alignat}
	\item Matching:
	\begin{itemize}
	\item $Q_{1L,R}^{i_u}\sim \bm3_u$:
	\begin{equation}
		\begin{alignedat}{2}
			\cL_{\text{SMEFT}}&\supset 
			\lzm 
			\frac{i}{f}\frac{\tilde\lambda^u_{Q_1} (\lambda^u_{Q_1})^*}{2M_{Q_1}}-\frac{i}{f}\frac{(\tilde\lambda^u_{Q_1})^* \lambda^u_{Q_1}}{2M_{Q_1}}
			 \dzm \cO_{\phi u}
			\\&
			-\lzv 
			\lzm 
			\frac{i}{f}\frac{(\tilde\lambda^u_{Q_1})^* \lambda^u_{Q_1}}{2M_{Q_1}}
			+\frac{i}{f}\frac{\tilde\lambda^u_{Q_1} (\lambda^u_{Q_1})^*}{2M_{Q_1}} \dzm [y_u^*]^{i_q}_{j_u}[\cO_{u\phi}]^{j_u}_{i_q}+\hermc\dzv.
		\nonumber
		\end{alignedat}
	\end{equation}
	\item $Q_{1L,R}^{i_d}\sim \bm3_d$:
	\begin{equation}
		\begin{alignedat}{2}
			\cL_{\text{SMEFT}}&\supset 
			\lzm 
			\frac{i}{f}\frac{(\tilde\lambda^d_{Q_1})^* \lambda^d_{Q_1}}{2M_{Q_1}}
			-\frac{i}{f}\frac{\tilde\lambda^d_{Q_1} (\lambda^d_{Q_1})^*}{2M_{Q_1}} \dzm \cO_{\phi d}
			\\&
			-\lzv 
			\lzm 
			\frac{i}{f}\frac{(\tilde\lambda^d_{Q_1})^* \lambda^d_{Q_1}}{2M_{Q_1}}
			+\frac{i}{f}\frac{\tilde\lambda^d_{Q_1} (\lambda^d_{Q_1})^*}{2M_{Q_1}} \dzm [y_d^*]^{i_q}_{j_d}[\cO_{d\phi}]^{j_d}_{i_q} +\hermc\dzv.
		\nonumber
		\end{alignedat}
	\end{equation}
	\end{itemize}
\end{itemize}

\subsubsection*{$\boldsymbol{Q_5\sim(3,2)_{-\frac{5}{6}}}$}
\begin{itemize}
	\item Lagrangian:
	\begin{equation}
		\begin{alignedat}{2}
		-\cL^{(5)}_\text{quarks}&\supset\frac{1}{f}
			(\tilde\lambda^d_{Q_5})_{ri}\bar Q_{5Rr}\slashed D\tilde\phi d_{i}
			+\hermc.
			\nonumber
		\end{alignedat}
	\end{equation}
	\item Flavor irreps:
		\begin{alignat}{3}
		Q_{5L,R}^{i_d} &\sim \bm3_d:
		&\qquad
		[\tilde\lambda^d_{Q_5}]^{i_d}_{j_d}=\tilde\lambda^d_{Q_5} \delta^{i_d}_{j_d}.
		\nonumber
		\end{alignat}
	\item Matching:
	\begin{equation}
		\begin{alignedat}{2}
			\cL_{\text{SMEFT}}&\supset 
			\lzm 
			\frac{i}{f}\frac{\tilde\lambda^d_{Q_5} (\lambda_{Q_5})^*}{2M_{Q_5}}-\frac{i}{f}\frac{(\tilde\lambda^d_{Q_5})^* \lambda_{Q_5}}{2M_{Q_5}}
			 \dzm \cO_{\phi d}
			\\&
			-\lzv 
			\lzm 
			\frac{i}{f}\frac{(\tilde\lambda^d_{Q_5})^* \lambda_{Q_5}}{2M_{Q_5}}
			+\frac{i}{f}\frac{\tilde\lambda^d_{Q_5} (\lambda_{Q_5})^*}{2M_{Q_5}} \dzm [y_d^*]^{i_q}_{j_d}[\cO_{d\phi}]^{j_d}_{i_q}+\hermc\dzv.
		\nonumber
		\end{alignedat}
	\end{equation}
\end{itemize}

\subsubsection*{$\boldsymbol{Q_7\sim(3,2)_{\frac{7}{6}}}$}
\begin{itemize}
	\item Lagrangian:
	\begin{equation}
		\begin{alignedat}{2}
		-\cL^{(5)}_\text{quarks}&\supset\frac{1}{f}
			(\tilde\lambda^u_{Q_7})_{ri}\bar Q_{7Rr}\slashed D\phi u_{i}
			+\hermc.
			\nonumber
		\end{alignedat}
	\end{equation}
	\item Flavor irreps:
		\begin{alignat}{3}
		Q_{7L,R}^{i_u} &\sim \bm3_u:
		&\qquad
		[\tilde\lambda^u_{Q_7}]^{i_u}_{j_u}=\tilde\lambda^u_{Q_7} \delta^{i_u}_{j_u}.
		\nonumber
	\end{alignat}
	\item Matching:
\begin{equation}
		\begin{alignedat}{2}
			\cL_{\text{SMEFT}}&\supset
			\lzm 
			\frac{i}{f}\frac{(\tilde\lambda^u_{Q_7})^* \lambda_{Q_7}}{2M_{Q_7}}
			-\frac{i}{f}\frac{\tilde\lambda^u_{Q_7} (\lambda_{Q_7})^*}{2M_{Q_7}} \dzm \cO_{\phi u}
			\\&
			-\lzv 
			\lzm 
			\frac{i}{f}\frac{(\tilde\lambda^u_{Q_7})^* \lambda_{Q_7}}{2M_{Q_7}}
			+\frac{i}{f}\frac{\tilde\lambda^u_{Q_7} (\lambda_{Q_7})^*}{2M_{Q_7}} \dzm [y_u^*]^{i_q}_{j_u}[\cO_{u\phi}]^{j_u}_{i_q}+\hermc\dzv.
		\nonumber
		\end{alignedat}
\end{equation}
\end{itemize}

\subsubsection*{$\boldsymbol{T_1\sim(3,3)_{-\frac{1}{3}}}$}
\begin{itemize}
	\item Lagrangian:
	\begin{equation}
		\begin{alignedat}{2}
		-\cL^{(5)}_\text{quarks}&\supset\frac{1}{f}
			(\tilde\lambda^q_{T_1})_{ri}\bar T_{1Lr}^a\gamma^\mu(D_\mu\phi)^\dag \sig^a q_{i}+\hermc.
			\nonumber
		\end{alignedat}
	\end{equation}
	\item Flavor irreps:
		\begin{alignat}{3}
		T_{1L,R}^{i_q} &\sim \bm3_q:
		&\qquad
		[\tilde\lambda^q_{T_1}]^{i_q	}_{j_q}=\tilde\lambda^q_{T_1} \delta^{i_q}_{j_q}.
		\nonumber
		\end{alignat}
	\item Matching:
\begin{equation}
		\begin{alignedat}{2}
			\cL_{\text{SMEFT}}&\supset 
			\lzm 
			\frac{i}{f}\frac{(\tilde\lambda^q_{T_1})^* \lambda_{T_1}}{8M_{T_1}}-\frac{i}{f}\frac{\tilde\lambda^q_{T_1} \lambda_{T_1}^*}{8M_{T_1}}
			 \dzm
			\lzm \cO_{\phi q}^{(3)}-3\cO_{\phi q}^{(1)} \dzm
			\\&
			+\Bigg\{
			\lzm 
			\frac{i}{f}\frac{\tilde\lambda^q_{T_1} \lambda_{T_1}^*}{4M_{T_1}}
			+\frac{i}{f}\frac{(\tilde\lambda^q_{T_1})^* \lambda_{T_1}}{4M_{T_1}} \dzm [ y_d^*]^{i_q}_{j_d}[\cO_{d\phi}]^{j_d}_{i_q}
			\\&
			\phantom{+}+
			\lzm 
			\frac{i}{f}\frac{\tilde\lambda^q_{T_1} \lambda_{T_1}^*}{2M_{T_1}}
			+\frac{i}{f}\frac{(\tilde\lambda^q_{T_1})^* \lambda_{T_1}}{2M_{T_1}} \dzm [y_u^*]^{i_q}_{j_u}[\cO_{u\phi}]^{j_u}_{i_q}
			+\hermc
			\Bigg\}.
		\nonumber
		\end{alignedat}
\end{equation}
\end{itemize}

\subsubsection*{$\boldsymbol{T_2\sim(3,3)_{\frac{2}{3}}}$}
\begin{itemize}
	\item Lagrangian:
	\begin{equation}
		\begin{alignedat}{2}
		-\cL^{(5)}_\text{quarks}&\supset\frac{1}{f}
			(\tilde\lambda^q_{T_2})_{ri}\bar T_{2Lr}^a\gamma^\mu(D_\mu\tilde\phi)^\dag \sig^a q_{i}+\hermc.
			\nonumber
		\end{alignedat}
	\end{equation}
	\item Flavor irreps:
		\begin{alignat}{3}
		T_{2L,R}^{i_q} &\sim \bm3_q:
		&\qquad
		[\tilde\lambda^q_{T_2}]^{i_q	}_{j_q}=\tilde\lambda^q_{T_2} \delta^{i_q}_{j_q}.
		\nonumber
	\end{alignat}
	\item Matching:
\begin{equation}
		\begin{alignedat}{2}
			\cL_{\text{SMEFT}}&\supset 
			\lzm 
			\frac{i}{f}\frac{(\tilde\lambda^q_{T_2})^* \lambda_{T_2}}{8M_{T_2}}-\frac{i}{f}\frac{\tilde\lambda^q_{T_2} \lambda_{T_2}^*}{8M_{T_2}}
			 \dzm 
			\lzm \cO_{\phi q}^{(3)}+3\cO_{\phi q}^{(1)} \dzm
			\\&
			+\Bigg\{ 
			\lzm 
			\frac{i}{f}\frac{\tilde\lambda^q_{T_2} \lambda_{T_2}^*}{2M_{T_2}}
			+\frac{i}{f}\frac{(\tilde\lambda^q_{T_2})^* \lambda_{T_2}}{2M_{T_2}} \dzm [y_d^*]^{i_q}_{j_d}[\cO_{d\phi}]^{j_d}_{i_q}
			\\&
			\phantom{+}+
			\lzm 
			\frac{i}{f}\frac{\tilde\lambda^q_{T_2} \lambda_{T_2}^*}{4M_{T_2}}
			+\frac{i}{f}\frac{(\tilde\lambda^q_{T_2})^* \lambda_{T_2}}{4M_{T_2}} \dzm [y_u^*]^{i_q}_{j_u}[\cO_{u\phi}]^{j_u}_{i_q} 
			+\hermc\Bigg\}.
		\nonumber
		\end{alignedat}
\end{equation}
\end{itemize}

\newpage

\bibliography{references}
\bibliographystyle{JHEP}

\end{document}